\documentclass[preprint,12pt,authoryear]{elsarticle}

\usepackage{amssymb}
\usepackage{amsmath}

\usepackage{tabularx}   
\usepackage{makecell}   
\usepackage{booktabs}   
\usepackage{float}      
\usepackage{pdflscape}
\usepackage{url}        

\journal{Social Networks}

\begin{document}

\begin{frontmatter}



\title{Refining Relational Event Models: Bayesian Penalization and Variable Selection in REMs}


\author[inst1]{Jonathan Koop} 

\affiliation[1]{organization={Department of Methodology and Statistics, Tilburg University},
            addressline={Prof. Cobbenhagenlaan 125}, 
            city={Tilburg},
            postcode={5037 DB}, 
            state={North Brabant},
            country={Netherlands}}

\author[2]{Sara van Erp}            
\author[2, 3]{Mahdi Shafiee Kamalabad}      
            
\affiliation[2]{organization={Department of Methodology and Statistics, Utrecht University},
            addressline={Padualaan 14}, 
            city={Utrecht},
            postcode={3584 CH}, 
            state={Utrecht},
            country={Netherlands}}
\affiliation[3]{organization={Centre for Complex Systems Studies, Utrecht University},
            addressline={Leuvenlaan 4}, 
            city={Utrecht},
            postcode={3584 CE}, 
            state={Utrecht},
            country={Netherlands}}  

\begin{abstract}
Relational Event Models (REMs) provide valuable insights into the dynamics of longitudinal social networks. Yet, the vast availability of potential predictors for a dyad's event rate poses the risk of selecting irrelevant variables and specifying an overfitted model that does not generalize to new data. Despite the recent popularity of Bayesian regularization methods to address this, there has not been a systematic evaluation of the performance of Exact Bayesian Regularization (EBR) and Approximate Bayesian Regularization (ABR) against standard Maximum Likelihood (ML) procedures commonly used for REMs. To address this, we conduct a simulation study in which directed Relational Event History (REH) data is generated with endogenous and exogenous effects of varying strength and compare the performance of (a) unregularized ML estimation, (b) ABR through normal‐approximations of the likelihood with Ridge and Horseshoe priors, and (c) EBR with Horseshoe priors. We find that threshold-based criteria applied to ABR and EBR with Horseshoe priors outperform univariate selection using p-values from ML estimation in variable selection accuracy. Regarding predictive performance, ABR with Ridge priors and EBR with Horseshoe priors outperform ML REMs, particularly for small sample sizes. Given only small gains with large computational burdens when using EBR, we therefore advise researchers to use ABR with a suitable prior.
\end{abstract}




\end{frontmatter}



\section{Introduction}

As everyday social and organizational processes generate increasingly rich digital footprints, relational event history (REH) data have become more widely available. These data, which record social interactions over time and include at minimum a sender, a receiver, and the timing or ordering of interactions, offer unprecedented opportunities to study the dynamics of complex networks. In contexts ranging from animal behavior \citep{Tranmer2015}, organizational communication such as email exchange \citep{Quintane2013}, to acts of violence within criminal networks \citep{Niezink2022}, REH data prompt researchers to examine the factors that drive these relational events. To this end, relational event models (REMs)\citep[REMs;][]{Butts2008} have emerged as the gold standard for analyzing REH data and uncovering the dynamic patterns that shape network interactions over time. REMs explicitly account for the temporal and sequential nature of the data by modeling the interaction rate as a function of network statistics that capture the evolving structural and relational features of the network.

Given the relatively limited theoretical guidance on the drivers of relational events \citep{Karimova2023}, selecting appropriate variables from the many potential predictors poses a significant challenge for applied researchers employing REMs. Many node or edge attributes, termed exogenous variables (e.g., age or gender), can be used as predictors to explain the rate of interactions among actors \citep{MeijerinkBosman2022}. Specific to REH data is the existence of endogenous variables summarizing characteristics of previous relational events at a specific time point. Examples of these include inertia, reflecting the frequency of prior interactions (with the same direction) between the same actors or reciprocity, indicating the frequency of previous interactions in the opposite direction between the same pair of actors \citep{Leenders2016}. Numerous endogenous variables, in addition to inertia, complicate the identification of key variables, resulting in a large number of potential predictors to include in a model.

To efficiently identify a model and avoid overfitting, different regularized regression techniques were introduced for linear models \citep{Hastie2015} promising the selection of important predictors. In classical frameworks, techniques such as Ridge regression \citep[$\ell_2$ norm;][]{Hoerl1970} and LASSO  \citep[$\ell_1$ norm;][]{Tibshirani1996} are prominently applied. Their Bayesian counterpart, Bayesian regularization, has been shown to perform equally or better by incorporating shrinkage through zero-centered prior distributions, providing more reliable parameter uncertainty estimates and allowing simultaneous estimation of model and penalty parameters \citep{vanErp2019}. Through the use of specific prior distributions on the parameters of interest, Bayesian approaches can lead to equivalent approaches as commonly used frequentist penalties. Ridge regression, for instance, corresponds to a Gaussian prior where the penalty parameter $\lambda$ influences the prior variance, and LASSO is equivalent to a Bayesian model with a Laplace prior centered around zero. Beyond these, Bayesian frameworks also allow for more flexible and hierarchical prior structures that do not have a direct one-to-one frequentist equivalent, such as the Horseshoe prior or spike-and-slab priors. These approaches can provide improved variable selection and shrinkage behavior, particularly in sparse settings \citep{Carvalho2010, vanErp2019, Piironen2017}.

While \citet{Karimova2025} demonstrated similar performance of Approximate Bayesian Regularization (ABR) and Exact Bayesian Regularization (EBR) for modeling REH data in a single empirical case, our study provides the first systematic comparison across a broad range of sample sizes ($M = 100$--$6{,}400$) and effect configurations. This broader and more controlled design offers a key advantage over a single-case comparison, as it allows us to identify the conditions under which each approach performs best and to provide more generalizable, evidence-based guidance for applied researchers working with REH data. Accordingly, this article addresses the following research questions: (a) Under what conditions, in terms of sample size and effect size, does Bayesian regularization improve variable selection in REMs relative to standard MLE? (b) Does Bayesian regularization improve predictive performance over MLE? (c) Do the computational trade-offs justify the use of ABR over EBR in practice?

To address these questions, we systematically examine the use of  Bayesian regularization in REMs. More specifically, we conduct simulation studies using synthetic REH data with the sample sizes ranging from $M=100$ to $M=6,400$ relational events in order to assess the effectiveness of these regularization techniques across different settings. In light of findings by \citet{Karimova2025}, Building on the findings of \citet{Karimova2025}, we compare the performance of ABR and EBR with approaches based solely on MLE, focusing on both variable selection and predictive performance across a range of settings. Furthermore, we demonstrate their practical usefulness by applying the best performing models on Spotify data.

The article is structured as follows: First, we provide the necessary statistical underpinnings by expanding on the statistical foundations of REMs, the rationale for the use of Bayesian regularization methods, as well as the statistical foundations of exact Bayesian Regularization and its normal approximation. We then describe the properties of the chosen prior distributions, accomplishing the shrinkage, and expand on the chosen selection criteria for the regularized models. Following this, the simulation procedure is outlined. We consequently present the results from the simulation study and apply the procedure to real-world data from Spotify, illustrating how the method can help identify the most important drivers of collaborations between popular artists on the platform. Finally, these results as well as potential limitations are discussed, and guidance for future research is given.

\section{Methodology}

\subsection{The Relational Event Model} \label{remsection}

 Relational event models (REMs) provide a flexible framework for analyzing fine-grained REH data, allowing for the estimation of the likelihood of relational events over time \citep{Butts2008}. They build on the assumption that time intervals between events follow an exponential distribution, where the rate parameter $\lambda$ varies between dyads as a function of endogenous and exogenous network statistics. At any given time point $t$, the overall event rate can be written as $\lambda_t' = \sum_{s,r} \lambda^{\mathrm{REM}}(s,r,t),$ where $\lambda^{\mathrm{REM}}(s,r,t)$ represents the instantaneous rate of an event occurring from sender $s$ to receiver $r$ at time point $t$. The summation includes all pairs within the riskset $\mathcal{R}_t$ , summarizing every plausible combination of senders and receivers at time $t$. Consequently, the time intervals between events are assumed to be distributed as
\begin{equation}
    \Delta t \sim \text{Exponential}(\lambda_t') \label{eq:rem_exp}.
\end{equation}

The specific event rates are determined through a log-linear model of covariates such that
\begin{equation}
\log \lambda^{\text{REM}}(s,r,t) = \boldsymbol{X}_{s,r}(t)'\boldsymbol{\beta}, \label{eq:rem_equation}
\end{equation}

where $\boldsymbol{X}_{s,r}(t)$ denotes the vector of statistics, including both exogenous and endogenous statistics, and $\boldsymbol{\beta}$ represents the vector of regression coefficients. The probability of a given event, for which sender $s'$ interacts with receiver $r'$, is defined as a multinomial probability distribution, with the probability mass function
\begin{equation}
p((s, r) = (s', r') \mid \mathcal{A}_t) = \frac{\lambda^{\text{REM}}(s', r', t)}{\sum_{s, r} \lambda^{\text{REM}}(s, r, t)}, \label{eq:rem_multinomial}
\end{equation}

where $\mathcal{A}_t$ represents the relational event history up to time $t$. This formulation allows REMs to account for the full interaction history, providing a framework for dynamic relational processes.

The wide availability of both exogenous and endogenous variables \citep{Butts2008} increases the risk of overfitting, can hinder model interpretability, and may substantially increase computational time when too many predictors and interaction effects are included. Therefore, it is crucial to identify a sparse model that captures the relevant predictors while excluding spurious effects.

\subsection{Bayesian Regularization}

In identifying such a sparse model, penalized regression techniques, specifically LASSO and Ridge regression, enjoy widespread popularity \citep{Hoerl1970, Tibshirani1996}. Given drawbacks such as the LASSO’s underestimation of standard errors \citep{Casella2010} and computational time, Bayesian regularization has gained recent popularity. Applying Bayes' theorem, coefficients are shrunken, with a posterior distribution influenced by the likelihood function and a prior distribution centered around zero. Markov Chain Monte Carlo (MCMC) sampling consequently allows simultaneously estimating coefficients and penalty parameters which determine the amount of shrinkage \citep{vanErp2019}. 

Beneficial to Bayesian regularization is particularly its ability to include several prior distributions, going beyond Bayesian equivalents to LASSO and Ridge regression in Laplace and normal prior distributions \citep{Park2008, Loesgen1990}. Particularly, nonconcave Horseshoe priors \citep{Carvalho2010} have demonstrated strong performance \citep{vanErp2019}, as their shape, characterized by a high probability mass near zero combined with heavy tails, enables strong shrinkage of weak coefficients towards zero while maintaining strong effects of larger coefficients.

\subsubsection{Exact Bayesian Regularization}

Given the limited availability of software packages for the application of Bayesian regularization on classical REMs introduced in Section \ref{remsection}, we make use of a reparameterization of the model to apply exact shrinkage where the likelihood is not approximated.  Let $\mu_{s,r,m}$ denote the expected number of events for dyad $(s,r)$ in interval $m$, and let $\Delta_m$ be the length of that interval. Since a rate is defined as the expected number of events per unit of time, the corresponding relational event rate is obtained by dividing the expected count by the interval length: $\lambda^{\mathrm{REM}}(s,r,t) = \frac{\mu_{s,r,m}}{\Delta_m},$ for all $t$ in interval $m$. As shown by \citet{Vieira2024}, this allows for a reparameterization of the classical REM to a Poisson regression

\begin{equation}
\log(\mu_{s,r,m})=\log(\Delta_m) + \boldsymbol{X}_{s,r}(t_m)' \boldsymbol{\beta} ,
\end{equation}

where the natural logarithm of the inter-event time from the previous to the current event, $\log(\Delta_m)$, is included as offset. As in Equation \ref{eq:rem_equation}, $\boldsymbol{X}_{s,r}(t)$ represents the vector of statistics, including both exogenous and endogenous statistics, and $\boldsymbol{\beta}$ is the vector of regression coefficients. The full reparameterization can be found in the Appendix (\ref{eq:remtopois}). This transformation then enables the use of widely used R packages allowing for MCMC sampling for Bayesian generalized linear models, such as \texttt{brms} \citep{brms}. 

The likelihood function for this reformulated Poisson model is given by

\begin{equation}
p(\boldsymbol{y} \mid \boldsymbol{X}, \boldsymbol{\beta}, \boldsymbol{\Delta})
= \prod_{m=1}^{M} \prod_{(s,r) \in \mathcal{R}_m}
\frac{\mu_{s,r,m}^{\,y_{s,r,m}} \exp\{-\mu_{s,r,m}\}}{y_{s,r,m}!},
\end{equation}

with $\mu_{s,r,m} = \Delta_m\,\exp\{\boldsymbol{X}_{s,r}(t_m)' \boldsymbol{\beta}\}$. Combining this likelihood with a shrinkage prior \(p_{\text{Bayes}}(\boldsymbol{\beta} \mid \lambda)\) (and, where applicable, a hyperprior \(p(\lambda)\)) yields the joint posterior distribution
\begin{equation}
p_{\text{Bayes}}(\boldsymbol{\beta}, \lambda_{\mathrm{shrink}} \mid \boldsymbol{y}, \boldsymbol{X}, \boldsymbol{\Delta})
\propto p(\boldsymbol{y} \mid \boldsymbol{X}, \boldsymbol{\beta}, \boldsymbol{\Delta})\,
p_{\text{Bayes}}(\boldsymbol{\beta} \mid \lambda_{\mathrm{shrink}})\, p(\lambda_{\mathrm{shrink}}),
\end{equation}

which can be estimated by the application of MCMC sampling. 

Problematically, however, the estimation of the Poisson model relies on a transformed data set that can be large, since for every realized event $m$ it includes all unrealized events for the time point $t_m$ as well. For large risk sets, this can lead to computational difficulties, to which particularly directed REMs are prone. For instance, in a case where $M=1000$ events with $N=50$ actors are observed, this would lead to a risk set of size $50 \times 49=2,450$ and thus a data frame of $2,450 \times 1000 = 2,450,000$ rows. Although case-control sampling has recently emerged as a promising approach to mitigate the substantial computational demands in larger networks \citep{Lerner2019}, its integration with REMs and EBR lies beyond the scope of the present study and remains an avenue for future research.

\subsection{Approximate Bayesian Regularization}

To address this computational disadvantage, \citet{Karimova2025} introduced Approximate Bayesian Regularization (ABR), approximating the likelihood function of $\boldsymbol{\beta}$ by a normal distribution centered around the MLE estimates $\hat{\boldsymbol{\beta}}_{MLE}$ with the error covariance matrix $\hat{\Sigma}_{\boldsymbol{\beta}}$ as covariance matrix:

\begin{equation}
p(\boldsymbol{y} \mid \boldsymbol{X}, \boldsymbol{\beta}) \sim \mathcal{N}(\boldsymbol{\beta} \mid \hat{\boldsymbol{\beta}}_{\text{MLE}}, \hat{\Sigma}_{\boldsymbol{\beta}}).
\end{equation}

Combining this with a shrinkage prior $p_{Bayes}(\boldsymbol{\beta} | \lambda) \, p(\lambda)$ leads to the posterior distribution

\begin{equation}
    \hat{p}_{Bayes}(\boldsymbol{\beta}, \lambda_{\mathrm{shrink}} | \boldsymbol{y}, \boldsymbol{X}) \propto \mathcal{N}(\boldsymbol{\beta} | \hat{\boldsymbol{\beta}}_{MLE}, \hat{\Sigma}_{\boldsymbol{\beta}}) \, p_{Bayes}(\boldsymbol{\beta} | \lambda_{\mathrm{shrink}}) \, p(\lambda_{\mathrm{shrink}}).
    \label{eq:ABR}
\end{equation}

Applying this to REH data, a REM is estimated using MLE to derive unregularized estimates $\hat{\boldsymbol{\beta}}_{MLE}$ and their associated error covariance matrix $\hat{\Sigma}_{\boldsymbol{\beta}}$. These serve as the distribution parameters for a normal distribution to approximate the likelihood function of $\boldsymbol{\beta}$. Regularization is consequently introduced by combining this normal approximation with a prior distribution centered around zero. By MCMC sampling, a posterior distribution is obtained.

In practice, Bayesian regularization, whether using the exact likelihood function or this approximation, has been shown to enhance the predictive performance of REMs. Applying exact Bayesian regularization, \citet{Karimova2023} evaluated the performance of different models by examining the proportion of observed events that fell within the top 5\%, 10\%, and 20\% most probable events predicted by the models. Regularized Bayesian REMs with Ridge, LASSO, and Horseshoe priors were found to perform comparably to standard MLE-based REMs when predicting in-sample observations (events used for training the REM) from voice loops during the Apollo-13 mission. Moreover, they observed the regularized models to outperform the former with regard to out-of-sample observations, indicating that regularization may indeed reduce overfitting.

Demonstrating the use of ABR for REMs, \citet{Karimova2025} observed ABR with Ridge, LASSO, and Horseshoe priors to perform similarly to exact Bayesian methods. An application on Apollo-13 voice loop data demonstrated similar in-sample predictive performance and slightly lower out-of-sample performance compared to the exact approach. While these results are encouraging, the performance of ABR and EBR in REMs across different numbers of events $M$ and effect types (i.e., endogenous and exogenous) in variable selection and prediction remains to be systematically explored.

\subsection{Prior Specification}

In Bayesian regularization, the choice of the prior distribution represents a key decision as it governs the extent to which parameters are shrunken towards zero. In this realm, various choices are discussed in the literature \citep[see e.g.][]{vanErp2019}. Due to its feasibility in implementation and analytical simplicity, we choose to explore the behavior of the Ridge prior here. Additionally, the Horseshoe prior is evaluated, given its capacity to perform strong shrinkage on small coefficients while retaining truly influential predictors with large effects, making it particularly suitable for high-dimensional problems where variable selection is of interest.

\paragraph{The Ridge Prior} Within the framework of Bayesian regularization, the Ridge prior draws each regression coefficient independently towards zero via a zero-centered normal distribution. More formally, for any chosen regression parameter $\beta_i$, this prior can be mathematically expressed as

\begin{equation}
\label{eq:prior_ridge}
\beta_i \mid \lambda,\sigma^2 \sim \mathcal{N}\bigl(0, \frac{\sigma^2}{\lambda})
\end{equation}

where the shrinkage parameter $\lambda$ affects the prior variance, and thus in turn influences the strength of the shrinkage. That is, the higher $\lambda$, the stronger the shrinkage towards zero. If no hyperprior for $\lambda$ is specified (i.e., if $\lambda$ is not sampled, but instead constrained to a certain value), this formulation leads to an $\ell_2$ penalty on the regression coefficients equivalent to frequentist Ridge regularization. With the Ridge prior, all coefficients are subject to the same degree of shrinkage toward zero, without any of them being shrunken to be exactly zero. In contrast to its frequentist counterpart, Bayesian Ridge regression allows for treating $\lambda$ as a hyperparameter with its own prior distribution (i.e., the hyperprior), which is popularly defined as a half-Cauchy distribution. Given that the difference between the estimated and true effect grows with larger true effects \citep{vanErp2019}, the Ridge prior can be expected to perform poorly in sparse models, i.e., when only few variables have strong true effects while most others do not exert influence on the event rate. Since social networks are often driven by many predictors with small but non-negligible effects instead of only a few dominant predictors, we can expect the Ridge prior to perform well.

\paragraph{The Horseshoe Prior} While the Ridge prior applies the same shrinkage to all coefficients, the Horseshoe prior applies a combination of a global shrinkage similar to that in Bayesian Ridge with local shrinkage specifically designed to adaptively regularize regression coefficients in high-dimensional and potentially sparse data settings \citep{Carvalho2010}. Its hierarchical structure assigns each coefficient $\beta_i$ a conditional prior distribution of the following form:
\begin{equation}
\begin{split}
(\beta_i \mid \tau_i^2) &\sim \mathcal{N}\bigl(0,\, \tau_i^2\bigr), \\
\tau_i^2 &\sim C^+(0,\lambda_{\mathrm{shrink}}), \\
\lambda &\sim C^+(0,\sigma).
\end{split}
\end{equation}

where $\lambda$ functions as a global scale parameter similar to $\lambda$ in Bayesian Ridge, and $\tau_i$ serves as a local scale parameter, which is assigned a half-Cauchy prior distribution for each parameter. This particular construction results in a marginal prior distribution for each $\beta_i$ that is characterized by a high concentration of probability mass around zero, together with heavy tails. The Horseshoe prior's unique shape allows it to effectively conduct strong shrinkage on small coefficients, while leaving predictors with strong effects relatively unaffected \citep{Carvalho2010}. Findings from simulation studies \citep{vanErp2019, Carvalho2010, Piironen2017} indicate that this behavior enables the Horseshoe prior to outperform shrinkage methods applying the same degree of shrinkage to all coefficients.

\subsection{Selecting Variables From Regularized Models}

Because regularized models with Ridge and Horseshoe priors do not shrink parameters to exactly zero, explicit criteria are required to identify which variables to retain. To achieve this, we evaluate two variable selection strategies: criteria based on credible intervals and those relying on absolute thresholds.

\paragraph{Credible Interval Criteria}  
Credible intervals provide a Bayesian analogue to frequentist confidence intervals. They provide information about the parameter range that is in accordance with a certain degree of belief. Similarly to procedures in frequentist null hypothesis significance testing, credible intervals can be used to assess whether there is significant evidence for a parameter being different from zero. This property can also be made use of in the context of variable selection through the application of credible interval criteria: Specifically, a variable is selected if its highest‐posterior‐density interval, i.e., the smallest range that contains a specified proportion of the posterior distribution, does not include zero \citep{Du2021}.

\paragraph{Threshold-Based Criteria}  
An alternative to using credible intervals for variable selection is based on thresholding posterior estimates directly. Under this scheme, variables are selected if a summary statistic of their posterior distribution, such as the posterior mean, median, or mode, is larger in absolute value than a pre-defined threshold  \citep[see e.g.][]{Ishwaran2005}. This threshold can be chosen based on practical considerations like the minimum effect size deemed relevant in a given application, or derived from theoretical considerations regarding the distribution of noise in the model.

\section{Simulation Study}

\subsection{Study Setup}

\subsubsection{Data Generation and Conditions}

To evaluate the performance of Bayesian regularization in selecting variables and making predictions within REMs, a simulation study is conducted comparing it to simply selecting predictors with statistically significant effects from frequentist analyses at $\alpha=0.05$. Consequently, directed REH data is generated with the event rate being explained by the inclusion of three endogenous effects identified on Apollo-13 data \citep{ShafieeKamalabad2023} , and six exogenous variables, three of which were continuous ($Z_1$, $Z_2$ and $Z_3$) and three binary ($Z_4$, $Z_5$ and $Z_6$). Effect sizes of the natural logarithms of 1.22 for weak, 1.86 for moderate, and 3.00 for strong effects were assigned to one variable each, following the conventions used in Cox proportional hazards models which can be applied to REMs \citep{Olivier2017, Schecter2020}. The remaining 29 endogenous statistics for directed relational events in the \texttt{remstats} R package \citep{remstats} and an additional 10 exogenous variables were specified to have no effect, leading to the data-generating model
\begin{equation}
\begin{aligned}
    \log \lambda^{\text{REM}}(s,r,t) = & -8 + 0.7 \, \text{IndegreeSender}(s,r,t) + 0.15 \, \text{Reciprocity}(s,r,t) \\
    & - 0.45 \, \text{OutdegreeReceiver}(s,r,t) \\
    & + \log(1.22) \, \text{Min}(Z_1) + \log(1.22) \, \text{Max}(Z_1) \\
    & + \log(1.86) \, \text{Min}(Z_2) + \log(1.86) \, \text{Max}(Z_2) \\
    & + \log(3.00) \, \text{Min}(Z_3) + \log(3.00) \, \text{Max}(Z_3) \\
    & + \log(1.22) \, \text{Same}(Z_4) + \log(1.86) \, \text{Same}(Z_5) \\
    & + \log(3.00) \, \text{Same}(Z_6),
\end{aligned}
\label{eq:rem_model}
\end{equation}

where $\text{Min}(Z_i)$ is the minimum value of the time-invariant exogenous variable $Z_i$ within sender $s$ and receiver $r$, and $\text{Max}(Z_i)$ is the maximum.

In accordance with \citet{Lakdawala2025}, REH data were generated by iteratively computing the REM event rate, \(\lambda^{\mathrm{REM}}(s,r,t)\), drawing the inter-event time \(\Delta t\) from the exponential distribution in Equation~\eqref{eq:rem_exp}, and sampling the next sender--receiver dyad from the multinomial distribution in Equation~\eqref{eq:rem_multinomial}. Each simulated data set contained \(M = 6{,}400\) events among \(N = 50\) potential actors. Excluding self-ties, all \(N(N-1)=2{,}450\) directed dyads were included in the risk set \(\mathcal{R}_t\) at all times \(t\). In total, 100 independent data sets were generated.

To assess the effect of the number of events $M$ on variable selection performance, each data set was subsetted to include the first 100, 200, 400, 800, 1,600, 3,200, and finally all 6,400 events, resulting in 700 data sets in total.

In addition to the subsetting approach, we generated independent data sets for each sample size to assess the robustness of the results and to determine whether the observed performance patterns generalized across independently simulated relational event histories.

\subsubsection{Evaluation of the Performance}

To evaluate the performance in discovering the true data-generating model, four REMs were estimated in a first step: A standard REM relying simply on the MLE estimates, a model resulting from ABR with a Ridge prior, an ABR model with a Horseshoe prior, and an EBR model with a Horseshoe prior, which was only computed for $M=100$ and $M=200$ due to computational limitations. For MLE estimates, univariate selection at $\alpha = 0.05$ (i.e., variables with statistically significant effects in the full model) served as a rather naive selection criterion. For the three models resulting from ABR and EBR, the 95\% highest density interval criterion (HDI; i.e., choosing variables with intervals that exclude zero) as well as threshold-based criteria with the absolute value of the posterior mean and mode using 0.1 as threshold (i.e., $|\hat\beta|\geq0.1$) were applied. We excluded the posterior median from the reported results, as it yielded a highly similar performance to that of the posterior mean. Detailed results, including those from the posterior median criterion, are provided in the Appendix (Tables \ref{tab:tdr} to \ref{tab:mcc}).

For each selection criterion, true discovery rates, defined as the proportion of truly non-zero effects that is selected, were computed separately for endogenous and exogenous statistics, due to their different nature. Endogenous rates were averaged over the three effects, while exogenous rates were reported per effect size. This split follows from the priors' shrinkage profiles with the Horseshoe combining a spike at zero with heavy tails compared to the Ridge that shrinks all effects more proportionally. False discovery rates describing the proportion of variables with no truly nonzero effect being selected were calculated separately for endogenous and exogenous effects. The two were calculated as:

\begin{equation}
\begin{split}
\mathrm{TDR}
&= \frac{\displaystyle\sum_{i=1}^p 1\{\hat\beta_i \neq 0,\;\beta_i \neq 0\}}
       {\displaystyle\sum_{i=1}^p 1\{\beta_i \neq 0\}}, \\[1ex]
\mathrm{FDR}
&= \frac{\displaystyle\sum_{i=1}^p 1\{\hat\beta_i \neq 0,\;\beta_i = 0\}}
       {\displaystyle\sum_{i=1}^p 1\{\beta_i = 0\}},
\end{split}
\end{equation}

where $\hat{\beta}_i$ is constrained to zero if a variable is not selected by a criterion.

To quantify the trade-off between true and false discoveries in a single metric, we defined a composite Euclidean distance measure commonly used for balancing similar metrics \citep{Nahm2022}

\begin{equation}
d = \sqrt{(1-TDR)^2+(FDR)^2}.
\end{equation}

Furthermore, the classification into non-zero and zero effects conducted by the variable selection procedures was evaluated using Matthews' correlation coefficient \citep{Matthews1975}, taking on the value -1 if in complete disagreement and 1 in the case of complete agreement of true effects and selected effects.

Moreover, the capacity to recover the true parameters was assessed by analyzing the estimates' bias and variance. Bias was quantified as the average deviation of the estimates from the true effects over the 100 iterations, and variance as the variance of this deviation across replications. To assess the predictive performance of the three models, both in-sample and out-of-sample metrics were used. Event rates were calculated at each time point $t$, and the proportion of cases in which the actual event is included in the 5\%, 10\%, and 20\% highest probability events was evaluated. For in-sample performance, this was applied to all events that were used to train the REM. For out-of-sample performance, an additional 1,000 out-of-sample events that were generated on top were evaluated. We assessed the predictive performance for the full REMs as well as for REMs only including predictors selected through the assessed criteria.

The models, the applied selection criteria, selection evaluation, and predictive performance measures are displayed in Table \ref{tab:model_comparison}.

\begin{table}[h]
\centering
\renewcommand{\arraystretch}{1.3} 
\caption{Overview of variable selection criteria, variable selection evaluation, and predictive performance measures used in the four models: Maximum Likelihood Estimation (MLE), Approximate Bayesian Regularization with Ridge prior (ABR Ridge), Approximate Bayesian Regularization with Horseshoe prior (ABR HS), and Exact Bayesian Regularization with Horseshoe prior (EBR HS). HDI abbreviates Highest Density Interval. $\overline{\beta}$ represents the posterior mean of $\beta$, $\beta_{\text{mdn}}$ the posterior median and $\beta_{\text{mod}}$ the posterior mode.}
{\footnotesize
\begin{tabularx}{\textwidth}{p{3cm}X X X X}
\hline
\textbf{Model} 
  & \textbf{MLE} 
  & \textbf{ABR (Ridge)} 
  & \textbf{ABR (HS)} 
  & \textbf{EBR (HS)} \\ 
\hline
\textbf{Selection Criterion} 
  & \makecell[l]{%
      Univariate\\
      Selection\\
      ($\alpha=0.05$)
    }
  & \makecell[l]{%
      95\% HDI\\
      $|\overline{\beta}|\ge 0.1$\\
      $|\beta_{\text{mdn}}|\ge 0.1$\\
      $|\beta_{\text{mod}}|\ge 0.1$
    }
  & \makecell[l]{%
      95\% HDI\\
      $|\overline{\beta}|\ge 0.1$\\
      $|\beta_{\text{mdn}}|\ge 0.1$\\
      $|\beta_{\text{mod}}|\ge 0.1$
    }
  & \makecell[l]{%
      95\% HDI\\
      $|\overline{\beta}|\ge 0.1$\\
      $|\beta_{\text{mdn}}|\ge 0.1$\\
      $|\beta_{\text{mod}}|\ge 0.1$
    } \\ 
    \noalign{\vskip 0.5em} 
\hline
\noalign{\vskip 0.5em} 
\textbf{Variable Selection Evaluation} 
  & \multicolumn{4}{l}{\makecell[l]{True and False Discovery Rates, Distance,\\ Matthews' Correlation Coefficient}} \\ 
\hline
\noalign{\vskip 0.5em} 
\textbf{Predictive Performance Measure} 
  & \multicolumn{4}{l}{\makecell[l]{Proportion of events in which observed event falls in \\top 5\%, 10\%, and 20\% most probable events for \\sparse and full models (in-sample and out-of-sample)}} \\ 
\hline
\end{tabularx}
}
\label{tab:model_comparison}
\end{table}

Analyses were conducted in \textit{R} 4.4.2 \citep{R}, using \texttt{remify} \citep{remify}, \texttt{remstats} \citep{remstats}, \texttt{remstimate} \citep{remstimate}, \texttt{shrinkem} \citep{shrinkem}, and \texttt{brms} \citep{brms}. The code is available at \url{https://github.com/jonathankoop/BayesianRegularization-REMs}.

\subsection{Results} \label{Results}

\subsubsection{Variable Selection} \label{variableselection}

\begin{figure}[H]
    \centering
    \includegraphics[width=\textwidth]{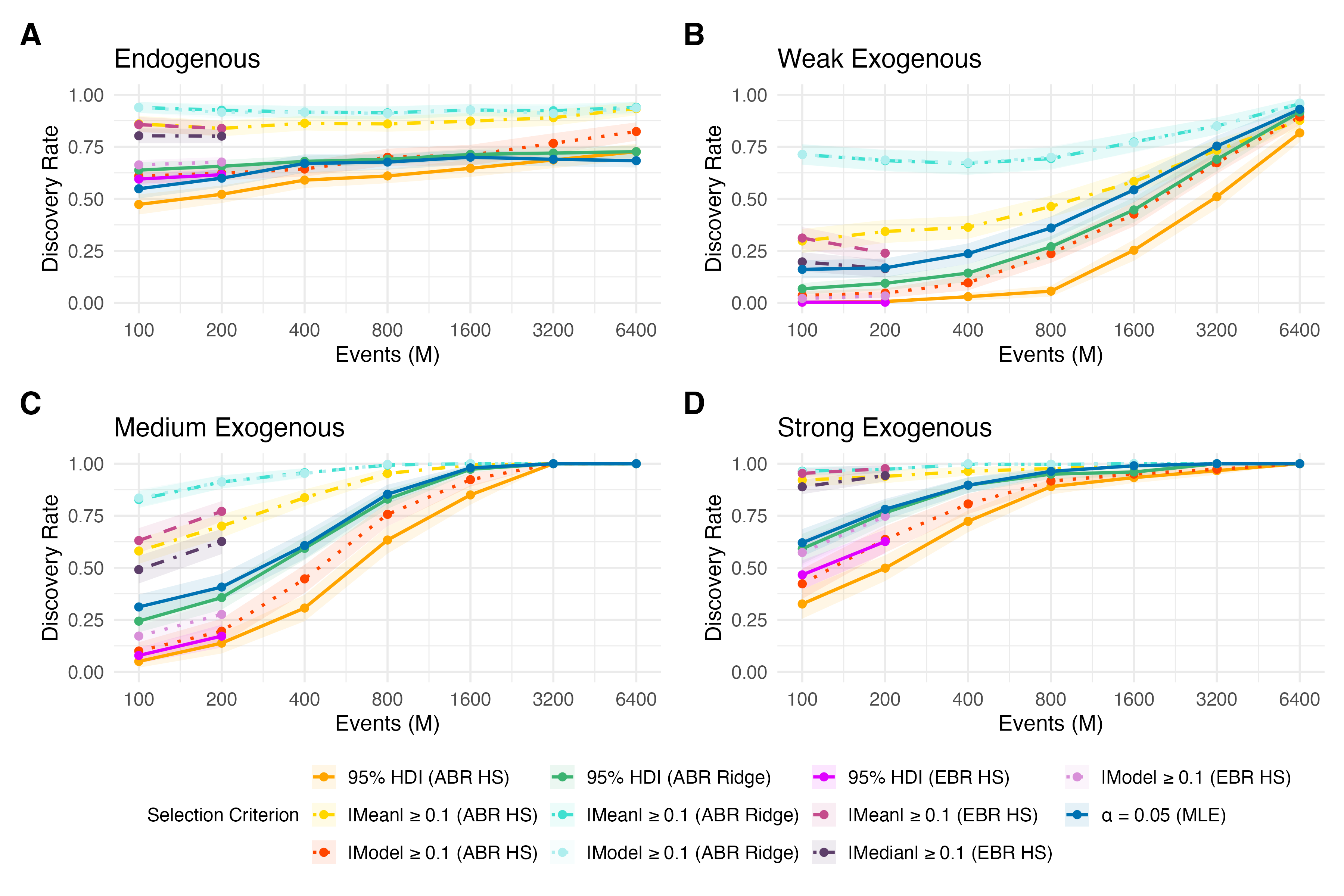}
    \caption{\textbf{True Discovery Rates Across Sample Sizes and Effect Types and Sizes} \\ True Discovery Rates (TDR) under varying effect sizes and numbers of events $M$ for endogenous and exogenous effects. Panels~(A)--(D) correspond to four scenarios: (A)~Endogenous Effects, (B)~Weak Exogenous Effects, (C)~Moderate Exogenous Effects, and (D)~Strong Exogenous Effects. Horizontal axes show the number of events $M$ from 100 to 6,400 and are log-scaled. Vertical axes represent the mean proportion of truly nonzero effects that are correctly identified (TDR) with a 95\% confidence interval across 100 iterations for each $M$. Eleven selection criteria are compared: Maximum Likelihood Estimation (MLE) using univariate selection at $\alpha=0.05$ (solid blue); Approximate Bayesian Regularization (ABR) with a Horseshoe prior using the 95\% HDI (solid orange), and thresholds on the mean (dashed yellow) and mode (dotted red); ABR with a Ridge prior using the 95\% HDI (solid green), and thresholds on the mean (dashed turquoise) and mode (dotted cyan); and Exact Bayesian Regularization (EBR) with a Horseshoe prior using the 95\% HDI (solid magenta), and thresholds on the mean (dashed rose), median (dot-dashed purple) and mode (dotted violet).}
    \label{fig:true_discovery_rates}
\end{figure}

Figure \ref{fig:true_discovery_rates} visualizes the mean true discovery rates over the 100 generated data sets and estimated models by the number of events $M$ for the three endogenous effects, and the exogenous effects by effect size with 95\% confidence intervals across the simulated data sets. Overall, the results reveal a remarkably different behavior in the discovery between non-zero endogenous effects displayed in Panel A and exogenous effects in Panel B, C, and D. 

Regarding the endogenous effects, threshold-based selection criteria for ABR models perform best. Here, particularly those using the posterior mean from ABR with Ridge and Horseshoe priors and - where calculated \footnote{For computational reasons we only estimated EBR models for $M=100$ and $M=200$}- EBR with Horseshoe prior, as well as the posterior mode from ABR with a Ridge prior (see the dotted cyan line), perform well with discovery rates above 0.8 for all sample sizes. Compared to that, the posterior mode from ABR with a Horseshoe prior performs worse with rates comparable to the HDI-based selection primarily for small $M$s ($TDR_{100}=0.609$ to $TDR_{6400}=0.823$). Using $p<0.05$ for MLE estimates performs similarly to the 95\% highest density credible interval from ABR with Ridge and Horseshoe priors with TDRs between 0.5 and 0.7 for most sample sizes. Overall, true discovery rates remain at a moderate level across all numbers of events ($M$) with no particular increase with higher sample sizes. Interestingly, for variables selected using statistical significance in the MLE model, the true discovery rate shows a slight decrease as the number of events surpasses 1,600.

For exogenous predictors, the TDR increases as expected with larger sample sizes, indicating improved recovery of true effects. Once again, threshold-based selection criteria based on the posterior mean show the strongest performance. This is the case across all effect and event sizes with the Ridge prior performing best. MLE using univariate selection at $\alpha=0.05$ and HDI-based selection using the Ridge prior perform worse with TDRs being very similar for all settings. In contrast to that, HDI-based selection after applying shrinkage via the Horseshoe prior shows the worst performance overall. This can be explained by the addition of shrinkage, adding more probability mass in the posterior distribution near estimates of zero, thus leading to more HDIs that include zero. This effect is strongest for weak exogenous effects using ABR with a Horseshoe prior as it performs particularly strong shrinkage on weak effects compared to stronger ones due to its strong concentration of probability mass around zero. Specific TDRs with their Monte Carlo standard errors across iterations can be found in Table \ref{tab:tdr}.

\begin{figure}[H]
    \centering
    \includegraphics[width=\textwidth]{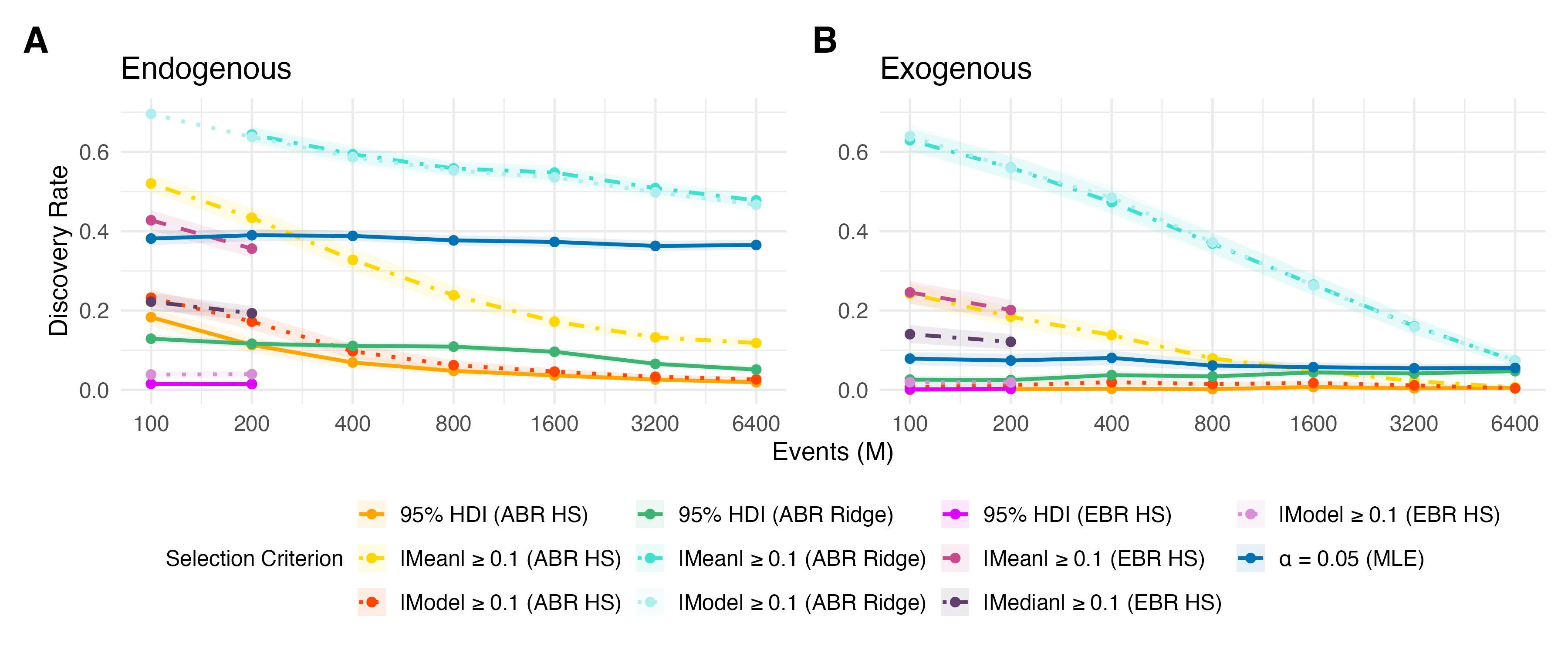}
    \caption{\textbf{False Discovery Rates Across Sample Sizes and Effect Types} \\ False Discovery Rates (FDR) under varying numbers of events $M$ for endogenous and exogenous effects. Panels~(A) and (B) show two scenarios: (A)~Endogenous Effects, and (B)~Exogenous Effects. Horizontal axes display the number of events $M$ from 100 to 6,400 and are log-scaled. Vertical axes represent the mean proportion of truly zero effects that are incorrectly identified as nonzero (FDR) with a  95\% confidence interval across 100 iterations for each $M$. Eleven selection criteria are compared: Maximum Likelihood Estimation (MLE) using univariate selection at $\alpha=0.05$ (solid blue); Approximate Bayesian Regularization (ABR) with a Horseshoe prior using the 95\% HDI (solid orange), and thresholds on the mean (dashed yellow) and mode (dotted red); ABR with a Ridge prior using the 95\% HDI (solid green), and thresholds on the mean (dashed turquoise) and mode (dotted cyan); and Exact Bayesian Regularization (EBR) with a Horseshoe prior using the 95\% HDI (solid magenta), and thresholds on the mean (dashed rose), median (dot-dashed purple) and mode (dotted violet).}
    \label{fig:false_discovery_rates}
\end{figure}

Figure \ref{fig:false_discovery_rates} displays the performance of the selection criteria regarding false discovery rates for endogenous and exogenous effects by the number of events $M$. Similar to the true discovery rates, threshold-based criteria using the posterior mean, leads to the highest rates of false discoveries. With type-I errors controlled at $\alpha=0.05$, the false discovery rates for exogenous effects align with expectations, being around 0.05 across all $M$s. In contrast, the higher rates for endogenous effects are surprising and suggest the need for further investigation. A plausible explanation may be high collinearity between endogenous statistics. By contrast, selection from the 95\% HDI-criterion using ABR with Horseshoe and Ridge priors displays the lowest false discovery rate. Particularly, those resulting from shrinkage with a Horseshoe prior, including the threshold-based criterion using the posterior mode, perform best with false discovery rates particularly low for small $M$s. Given the high discovery rates, univariate selection from MLE and threshold-based criteria using the posterior mean from ABR generally select most variables, thus leading to both high true and false discovery rates. In contrast, HDI-based criteria using ABR estimates are more conservative with fewer false discoveries at the cost of missing out on discovering true effects. Detailed results can be found in Table \ref{tab:fdr} in the Appendix.

\begin{figure}[H]
    \centering
    \includegraphics[width=\textwidth]{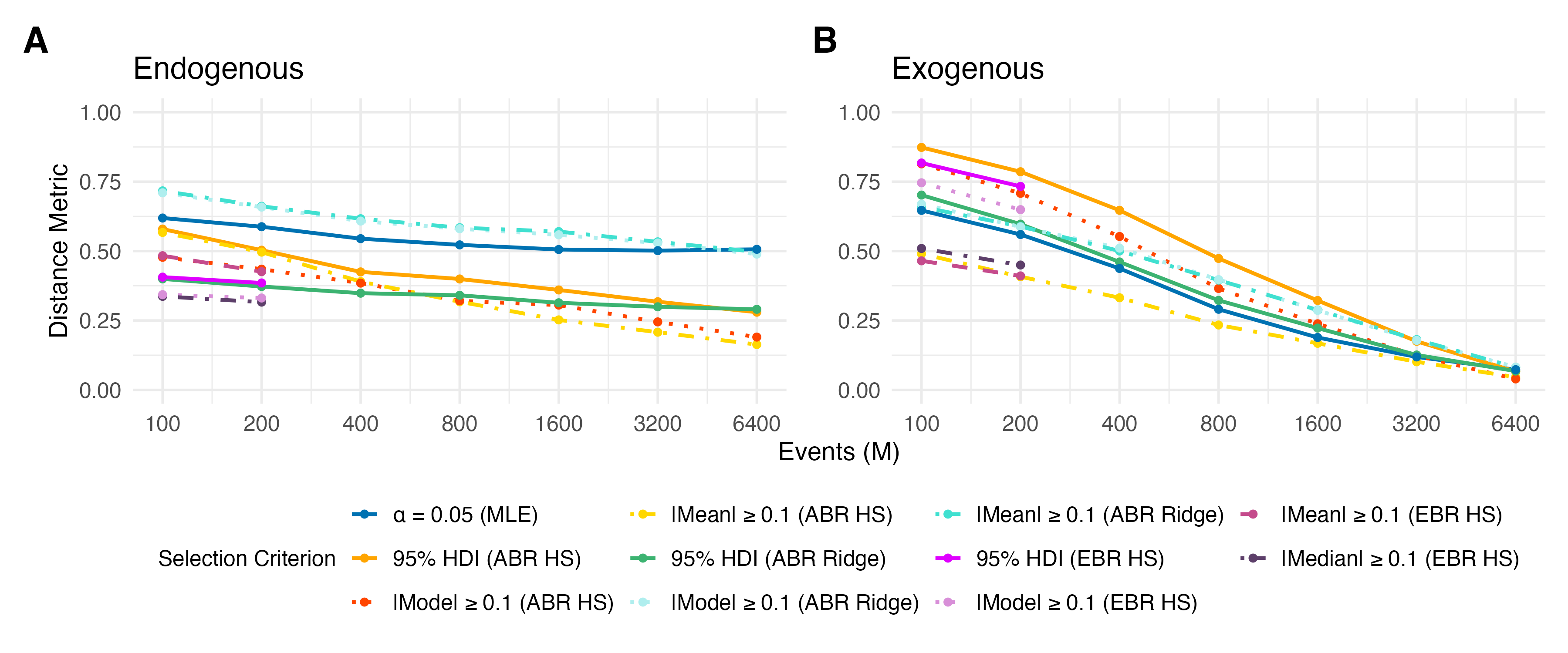}
    \caption{\textbf{Distance Metrics Across Sample Sizes and Effect Types} \\ Distance metrics under varying numbers of events $M$ for endogenous and exogenous effects. Panels~(A) and (B) show two scenarios: (A)~Endogenous Effects, and (B)~Exogenous Effects. Horizontal axes display the number of events $M$ from 100 to 6,400 and are log-scaled. Vertical axes represent the mean distance-based error metric across 100 iterations. Eleven selection criteria are compared: Maximum Likelihood Estimation (MLE) using univariate selection at $\alpha=0.05$ (solid blue); Approximate Bayesian Regularization (ABR) with a Horseshoe prior using the 95\% HDI (solid orange), and thresholds on the mean (dashed yellow) and mode (dotted red); ABR with a Ridge prior using the 95\% HDI (solid green), and thresholds on the mean (dashed turquoise) and mode (dotted cyan); and Exact Bayesian Regularization (EBR) with a Horseshoe prior using the 95\% HDI (solid magenta), and thresholds on the mean (dashed rose), median (dot-dashed purple) and mode (dotted violet).}

    \label{fig:distance_plots}
\end{figure}

To counterbalance true and false discoveries, distances were evaluated for endogenous and exogenous effects, as displayed in Figure \ref{fig:distance_plots} (see Table \ref{tab:dist} for details). Generally, lower distance measures imply better performance of a selection criterion in balancing low false discovery rates and high true discovery rates. As is to be expected, distances decrease as the number of events increases, implying better performances with higher sample sizes. This trend is, however, less pronounced for endogenous effects, where distances show only limited reduction, particularly for univariate selection from MLE where they remain consistently high regardless of the number of events. For endogenous effects, HDI-based selection from ABR Ridge shows a relatively low distance, only being outperformed by threshold-based selection from ABR Horseshoe for $M \geq 1600$. For endogenous effects, univariate selection from MLE performs poorly for all $M$. For exogenous effects, threshold based criteria on coefficents regularized with a Horseshoe prior perform best. Univariate selection from MLE performs markedly worse for small $M$s with diminishing differences with more events.

\begin{figure}[H]
    \centering
    \includegraphics[width=\textwidth]{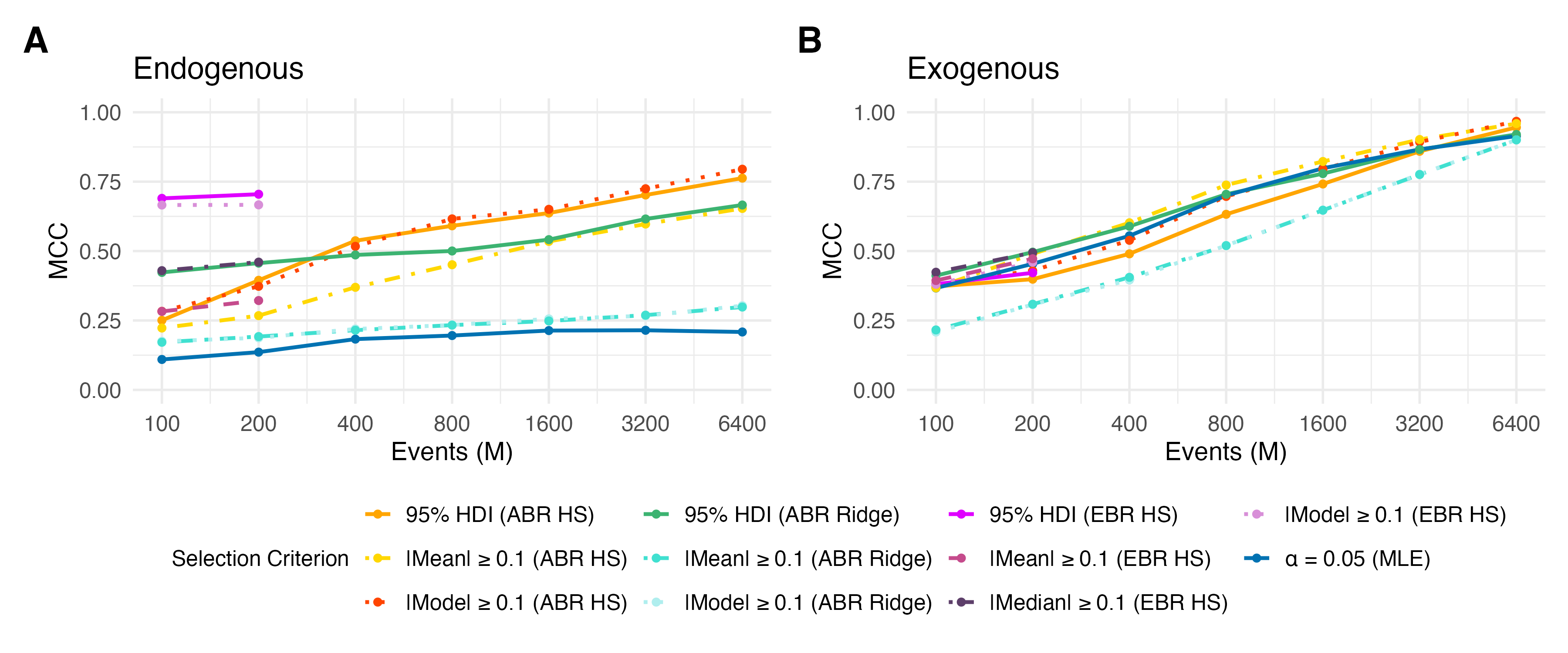}
    \caption{\textbf{Matthews' Correlation Coefficient Across Sample Sizes and Effect Types} \\ Matthews' Correlation Coefficient (MCC) under varying numbers of events $M$ for endogenous and exogenous effects. Panels~(A) and (B) show two scenarios: (A)~Endogenous Effects, and (B)~Exogenous Effects. Horizontal axes display the number of events $M$ from 100 to 6,400 and are log-scaled. Vertical axes represent the mean MCC across 100 iterations. Eleven selection criteria are compared: Maximum Likelihood Estimation (MLE) using univariate selection at $\alpha=0.05$ (solid blue); Approximate Bayesian Regularization (ABR) with a Horseshoe prior using the 95\% HDI (solid orange), and thresholds on the mean (dashed yellow) and mode (dotted red); ABR with a Ridge prior using the 95\% HDI (solid green), and thresholds on the mean (dashed turquoise) and mode (dotted cyan); and Exact Bayesian Regularization (EBR) with a Horseshoe prior using the 95\% HDI (solid magenta), and thresholds on the mean (dashed rose), median (dot-dashed purple) and mode (dotted violet).}
    \label{fig:mcc_plots}
\end{figure}

Figure~\ref{fig:mcc_plots} presents Matthews' correlation coefficient across selection criteria and numbers of events. In the case of perfect classification, i.e., if all true non-zero effects and none of the zero effects are selected, it takes on the value 1. When classification is entirely incorrect, MCC equals -1. Unlike the aforementioned distance measures, MCC assesses the accuracy with which effects are classified as zero or non-zero, independently of the previously calculated discovery rates. As a result, the MCC takes into account the number of true non-zero and zero effects. This means that in the case of a sparse true data-generating model, implying few true non-zero effects and many zero effects, the two are not weighted equally, but rather each variable affects the MCC similarly. Here, this implies that more conservative selection criteria, i.e., criteria with lower false discovery rates, are expected to perform better with regard to the MCC. 

Selection from regularized estimates performs much better than univariate selection from MLE for endogenous effect with increasing advantages when more events are considered. Due to their less conservative nature, the previously superior threshold-based criteria on the posterior mean on ABR estimates now exhibit worse performance than their HDI-based counterparts across most sample sizes. For exogenous effects, the differences in MCCs are smaller with only thresholding the posterior mean from ABR Ridge performing worse than all others. Overall, MCCs increase much stronger with sample sizes for exogenous than for endogenous effects. MCCs and their Monte Carlo standard errors across iterations can be found in Table \ref{tab:mcc}.

Results from the analysis of the independently generated data match those discussed above. They also reveal that threshold-based selection criteria on ABR with a Horseshoe prior perform best in balancing high TDRs with low FDRs. Detailed results can be found in Tables \ref{tab:tdr_ind}, \ref{tab:fdr_ind}, \ref{tab:dist_ind}, \ref{tab:mcc_ind} in the Appendix.

\subsubsection{Bias and Variance} \label{biasvariance}

Overall, estimates from all models show considerate levels of bias, with the bias averaged across all $\hat{\beta}_i$ for MLE, $Bias(\hat{\boldsymbol{\beta}}_{MLE})=1.39$, being highest, and ABR models being lower ($Bias(\hat{\boldsymbol{\beta}}_{ABR_{HS}})=0.30$; $Bias(\hat{\boldsymbol{\beta}}_{ABR_{Ridge}})=0.19$) over all sample sizes $M$. EBR with a Horseshoe prior, which was only computed for $M=100$ and $M=200$, displayed an average bias of $Bias(\hat{\boldsymbol{\beta}}_{ABR_{HS}})=0.15$. Considering the bias of the individual coefficients, however, reveals that this bias is heavily rooted in the estimates for the participation shifts. Over all $M$, these estimated coefficients are extremely large with average values over 11 for the most extreme cases, which largely contributes to this bias. When excluding them the bias drastically decreases. Average estimates for all coefficients except those for participation shifts are displayed in \ref{fig:plots_mle}. Participation shifts are displayed separately due to their extreme bias in \ref{fig:plots_mle_ps}. Mean biases and variances of the models by number of events are included in Table \ref{tab:bias_variance} in the Appendix.

Results from the independent analysis are in line with these results and are displayed in Table \ref{tab:biasvar_independent} in the Appendix.

\subsubsection{Predictive Performance}

\begin{figure}[H]
    \centering
    \includegraphics[width=\textwidth]{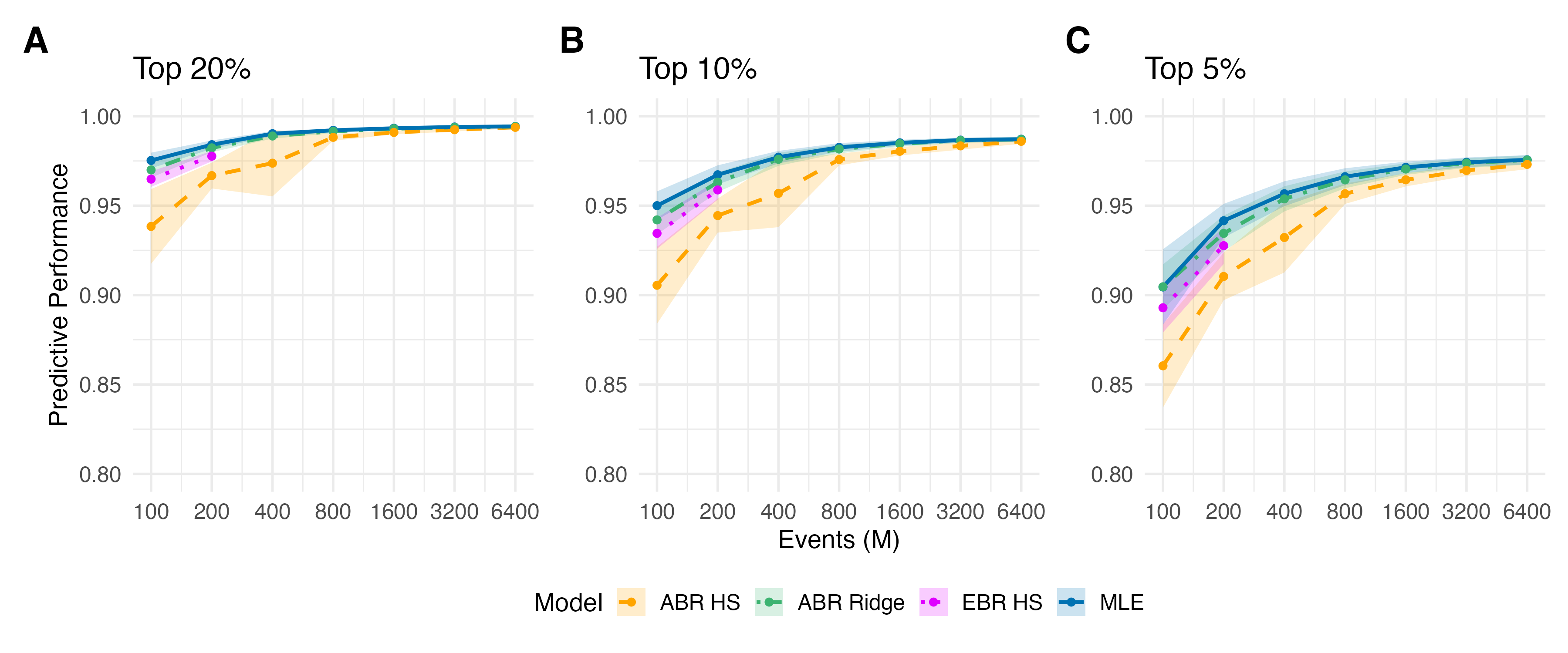}
    \caption{\textbf{In-Sample Predictive Performance of Full Models} \\ In-sample predictive performance assessed as the proportion of in-sample events ranking among the top 20\%, 10\%, and 5\% most probable events under three modeling approaches including all predictor variables. Panel~(A) shows the top 20\%; Panel~(B) the top 10\%, and Panel~(C) the top 5\%. Horizontal axes display the number of events $M$ from 100 to 6,400 and are log-scaled. Vertical axes represent the mean proportion of observed events in the specified top percentiles with a  95\% confidence interval across 100 iterations for each $M$. Models include Maximum Likelihood Estimation (MLE, solid blue line), Approximate Bayesian Regularization with a Horseshoe prior (ABR HS, dashed orange line), Approximate Bayesian Regularization with a Ridge prior (ABR Ridge, dot-dashed green line), and Exact Bayesian Regularization with a Horseshoe prior (EBR HS, dotted purple line).}
    \label{fig:predictive_performance_is}
\end{figure}

\paragraph{Full Models}
In a first step, we assess the models' predictive performance, including all variables, as displayed in Figure \ref{fig:predictive_performance_is}. Overall, in-sample performance, i.e., a model's performance in predicting dyads that were used to train the REM, improves as the sample size grows. REMs from MLE generally perform best with ABR with a Ridge prior showing nearly equal performance across sample sizes and percentiles. REMs with Horseshoe priors display a comparable performance only when exact regularization is applied. ABR performs considerably worse than its exact counterpart where estimated. Given that in MLE coefficients are estimated to maximize the training data's likelihood these findings are as expected.

\begin{figure}[H]
    \centering
    \includegraphics[width=\textwidth]{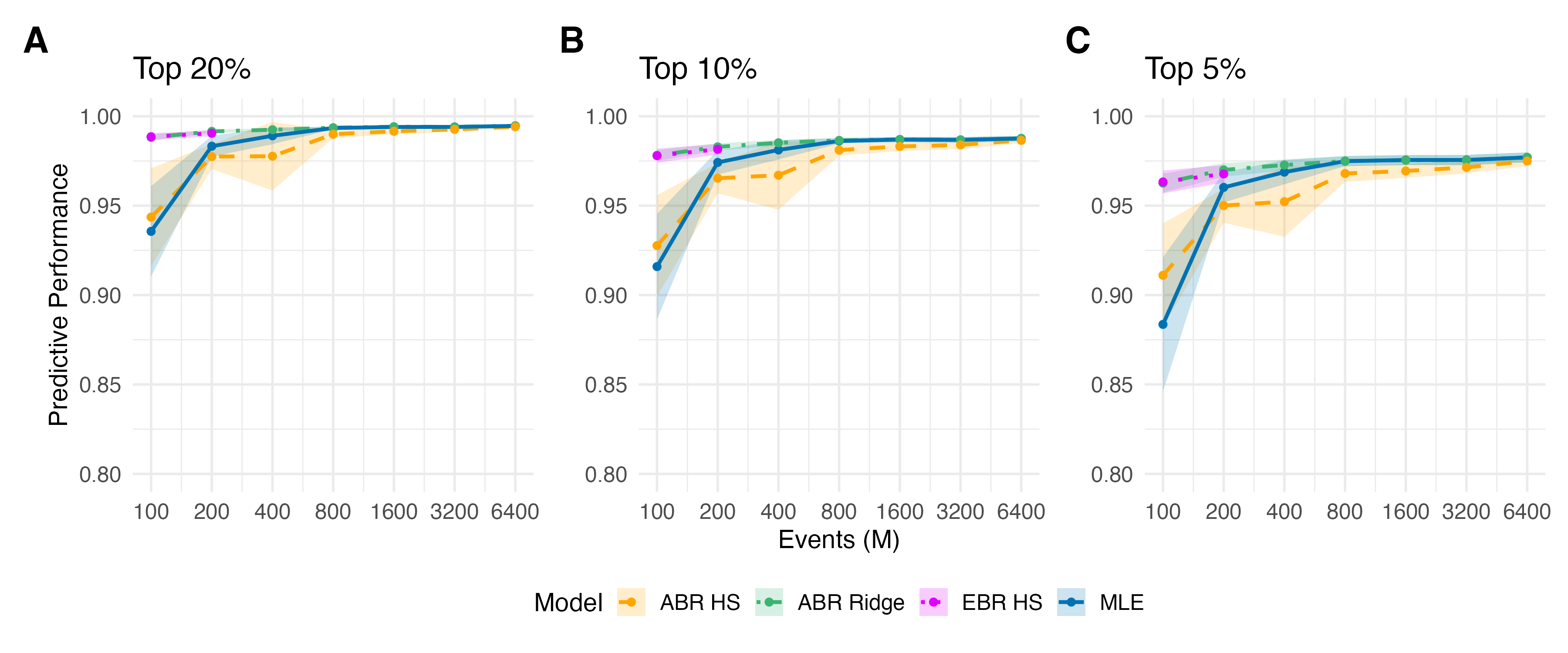}
    \caption{\textbf{Out-Of-Sample Predictive Performance of Full Models} \\ Out-of-sample predictive performance assessed as the proportion of 1000 out-of-sample events ranking among the top 20\%, 10\%, and 5\% most probable events under three modeling approaches, including all predictor variables. Panel~(A) shows the top 20\%; Panel~(B) the top 10\%, and Panel~(C) the top 5\%. Horizontal axes display the number of events $M$ from 100 to 6,400 and are log-scaled. Vertical axes represent the mean proportion of observed events in the specified top percentiles with a  95\% confidence interval across 100 iterations for each $M$. Models include Maximum Likelihood Estimation (MLE, solid blue line), Approximate Bayesian Regularization with a Horseshoe prior (ABR HS, dashed orange line), Approximate Bayesian Regularization with a Ridge prior (ABR Ridge, dot-dashed green line), and Exact Bayesian Regularization with a Horseshoe prior (EBR HS, dotted purple line).}
    \label{fig:predictive_performance}
\end{figure}

The evaluation of out-of-sample predictive performance displayed in Figure \ref{fig:predictive_performance}, assessing the models' ability to generalize to new data, reveals a different narrative, showcasing the main advantage of regularization. Here, independent of the sample size ABR models with a Ridge prior as well as EBR Horseshoe show good predictive performance, while MLE and ABR Horseshoe are much lower, particularly for small $M$s. In contrast to in-sample performance, ABR Ridge models perform similarly well regardless of the sample sizes, implying that already small $M$s are sufficient for making good predictions in case regularized models are used. The addition of new data does not improve the quality of predictions in that case. If sample sizes are sufficiently large (specifically if $M \geq 400$), MLE performs similarly well to ABR Ridge models.

Analyses where REH data was generated independently generally yield comparable results. Here, however, MLE performs best for in-sample performance, which can be explained by the different risk sets implied by the data sets: While the dependent analyses consider all potential dyads in most cases (i.e., they assign all 2,450 potential dyads a probability), the risk set of the dependent data often only contains the actors of the realized events in the risk set. Tables for their predictive performance can be found in the Appendix (Tables \ref{tab:pp_is_independent} and \ref{tab:pp_oos_independent}).

\begin{figure}[H]
    \centering
    \includegraphics[width=\textwidth]{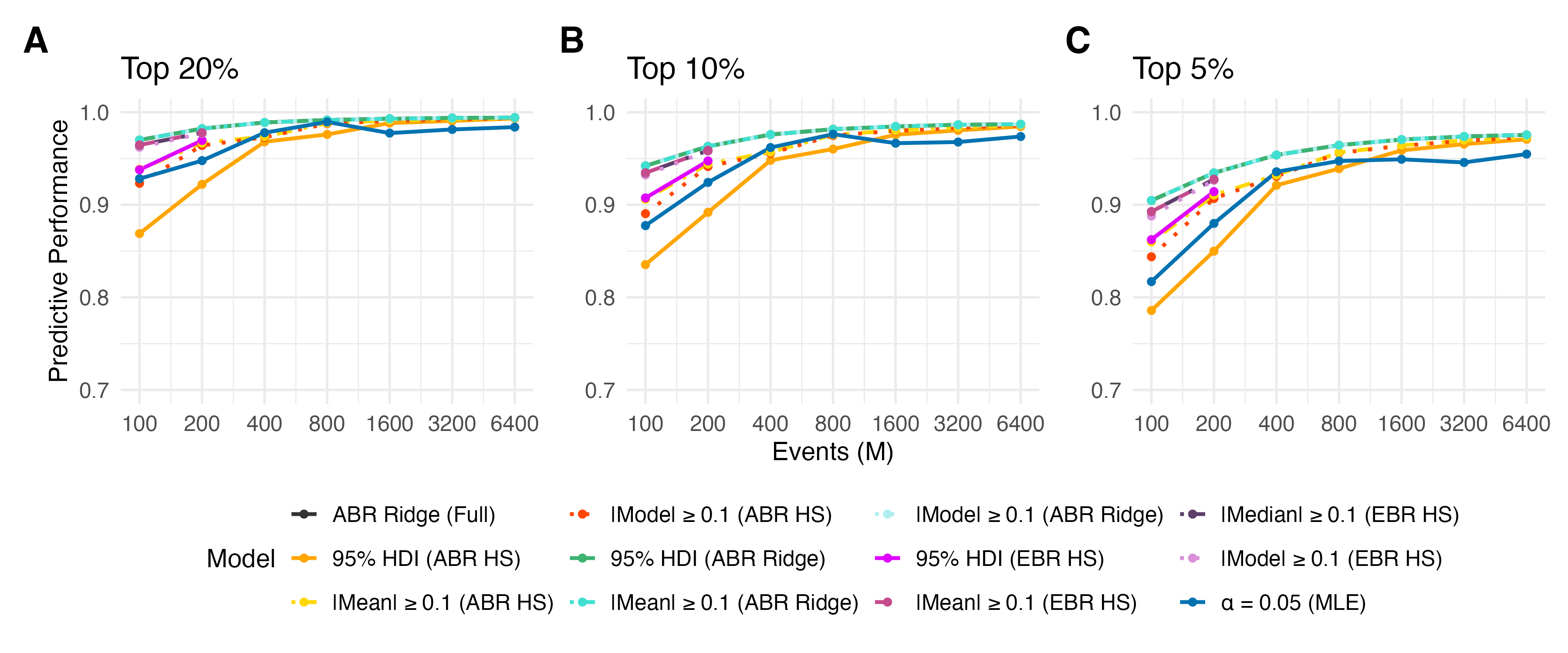}
    \caption{\textbf{In-Sample Predictive Performance of Sparse Models} \\ In-sample predictive performance assessed as the proportion of in-sample events ranking among the top 20\%, 10\%, and 5\% most probable events under three modeling approaches including only selected predictor variables from various selection criteria. Panel~(A) shows the top 20\%; Panel~(B) the top 10\%, and Panel~(C) the top 5\%. Horizontal axes display the number of events $M$ from 100 to 6,400 and are log-scaled. Vertical axes represent the mean proportion of observed events in the specified top percentiles across 100 iterations for each $M$. Models include Maximum Likelihood Estimation (MLE) using univariate selection at $\alpha=0.05$ (solid blue); Approximate Bayesian Regularization (ABR) with a Horseshoe prior using the 95\% HDI (solid orange), and thresholds on the mean (dashed yellow) and mode (dotted red); ABR with a Ridge prior using the 95\% HDI (solid green), and thresholds on the mean (dashed turquoise) and mode (dotted cyan); and Exact Bayesian Regularization (EBR) with a Horseshoe prior using the 95\% HDI (solid magenta), and thresholds on the mean (dashed rose), median (dot-dashed purple) and mode (dotted violet).}
    \label{fig:predictive_performance_is_sparse}
\end{figure}

\paragraph{Sparse Models}
In a second step, we apply the variable selection criteria assessed in Section \ref{variableselection} to the respective models. Figure \ref{fig:predictive_performance_is_sparse} displays the predictive performance of these sparser models. Besides those models, the full ABR Ridge model with all predictors is included as reference.

For in-sample prediction, ABR Ridge models only including those variables that were selected by the threshold criterion on the posterior mode and mean perform equally well to their full counterpart over all sample sizes despite including fewer variables. For those $M$s that were explored for EBR with Horseshoe prior, sparse models after selection using threshold-based critera show similar performance. In contrast, models from ABR Horseshoe perform considerably worse, particularly for small $M$s.

Regarding out-of-sample performance, results are similar. As above, sparser models after applying threshold-based criteria on ABR with Ridge prior perform very similarly to the full ABR Ridge model. Again, the EBR model only including those effects whose absolute value exceeds or equals 0.1 is among the well-performing models. The sparser models resulting from ABR with a Horseshoe prior perform considerably worse, particularly for small sample sizes. In summary, the simulation thus demonstrates that employing ABR on REMs yields more parsimonious models with enhanced predictive performance only with a Ridge prior, particularly for out-of-sample predictions. Against theoretical expectations, the Horseshoe prior performed considerably worse. EBR's better performance for small sample sizes may indicate, however, that this could be rooted in the Normal approximation of the likelihood.

The analyses of the independent data demonstrate that these findings are robust. Similarly, REMs applying ABR with shrinkage through Ridge priors, only including those predictors whose posterior mean is larger than or equal to 0.1, show comparable performance to the full models. Tables of the performance metrics can be found in the Appendix (Tables \ref{tab:pp_is_independent} and \ref{tab:pp_oos_independent}).

\begin{figure}[H]
    \centering
    \includegraphics[width=\textwidth]{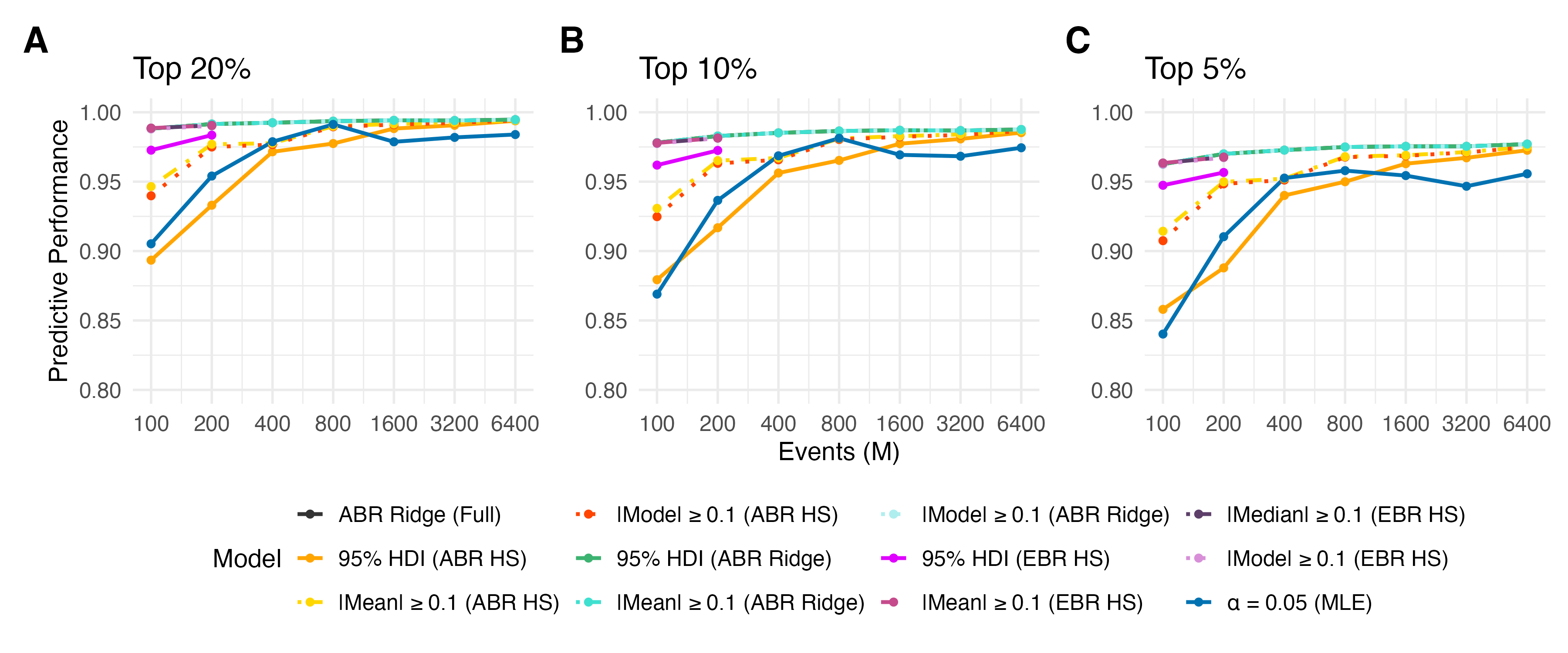}
    \caption{\textbf{Out-Of-Sample Predictive Performance of Sparse Models} \\ Out-of-sample predictive performance assessed as the proportion of 1000 out-of-sample events ranking among the top 20\%, 10\%, and 5\% most probable events under three modeling approaches, including only selected predictor variables from various selection criteria. Panel~(A) shows the top 20\%; Panel~(B) the top 10\%, and Panel~(C) the top 5\%. Horizontal axes display the number of events $M$ from 100 to 6,400 and are log-scaled. Vertical axes represent the mean proportion of observed events in the specified top percentiles across 100 iterations for each $M$. Models include Maximum Likelihood Estimation (MLE) using univariate selection at $\alpha=0.05$ (solid blue); Approximate Bayesian Regularization (ABR) with a Horseshoe prior using the 95\% HDI (solid orange), and thresholds on the mean (dashed yellow) and mode (dotted red); ABR with a Ridge prior using the 95\% HDI (solid green), and thresholds on the mean (dashed turquoise) and mode (dotted cyan); and Exact Bayesian Regularization (EBR) with a Horseshoe prior using the 95\% HDI (solid magenta), and thresholds on the mean (dashed rose), median (dot-dashed purple) and mode (dotted violet).}
    \label{fig:predictive_performance_oos_sparse}
\end{figure}

\section{Empirical Application}

We illustrate the use of Bayesian regularization on a real-world data set containing collaborations between artists on Spotify. Data were collected from the Spotify API \citep{spotify2025webapi}, focusing on the 100 most popular artists according to Spotify’s popularity index, out of which those were subsetted who had collaborated at least twice with other artists. This resulted in a data set of 244 collaboration events (songs) involving 62 unique actors (artists) that occurred between January 2010 and March 2023. For each artist, attributes such as \textit{Gender}, \textit{Age}, \textit{Country}, and \textit{Popularity} were collected from Last.fm \citep{lastfm}. Exogenous dyadic statistics based on these attributes, as listed in Table \ref{tab:variables}, were computed. In addition, all endogenous statistics for directed relational event models available in the \texttt{remstats} R package \citep{remstats} were included.

\begin{table}[h!]
\centering
\caption{Actor attributes and exogenous statistics included in the empirical application}
\label{tab:variables}
\begin{tabularx}{\textwidth}{l X l}
\toprule
\textbf{Attribute} & \textbf{Characteristics of actor $i$} & \textbf{Exogenous statistic} \\
\midrule
$\text{Gender}_i$     & Gender of Artist $i$                                        & $\text{same}(\text{Gender})$ \\
$\text{Age}_i$        & Age of Artist $i$ on December 31, 2024                      & $\min(\text{Age}),\,\max(\text{Age})$ \\
$\text{Country}_i$    & Country of Birth of Artist $i$                              & $\text{same}(\text{Country})$ \\
$\text{Popularity}_i$ & Popularity of Artist $i$ as published by Spotify (up to 100) & $\min(\text{Popularity}),\,\max(\text{Popularity})$ \\
\bottomrule
\end{tabularx}
\end{table}

\subsection{Variable Selection}

\begin{figure}[H]
    \centering
    \includegraphics[width=\textwidth]{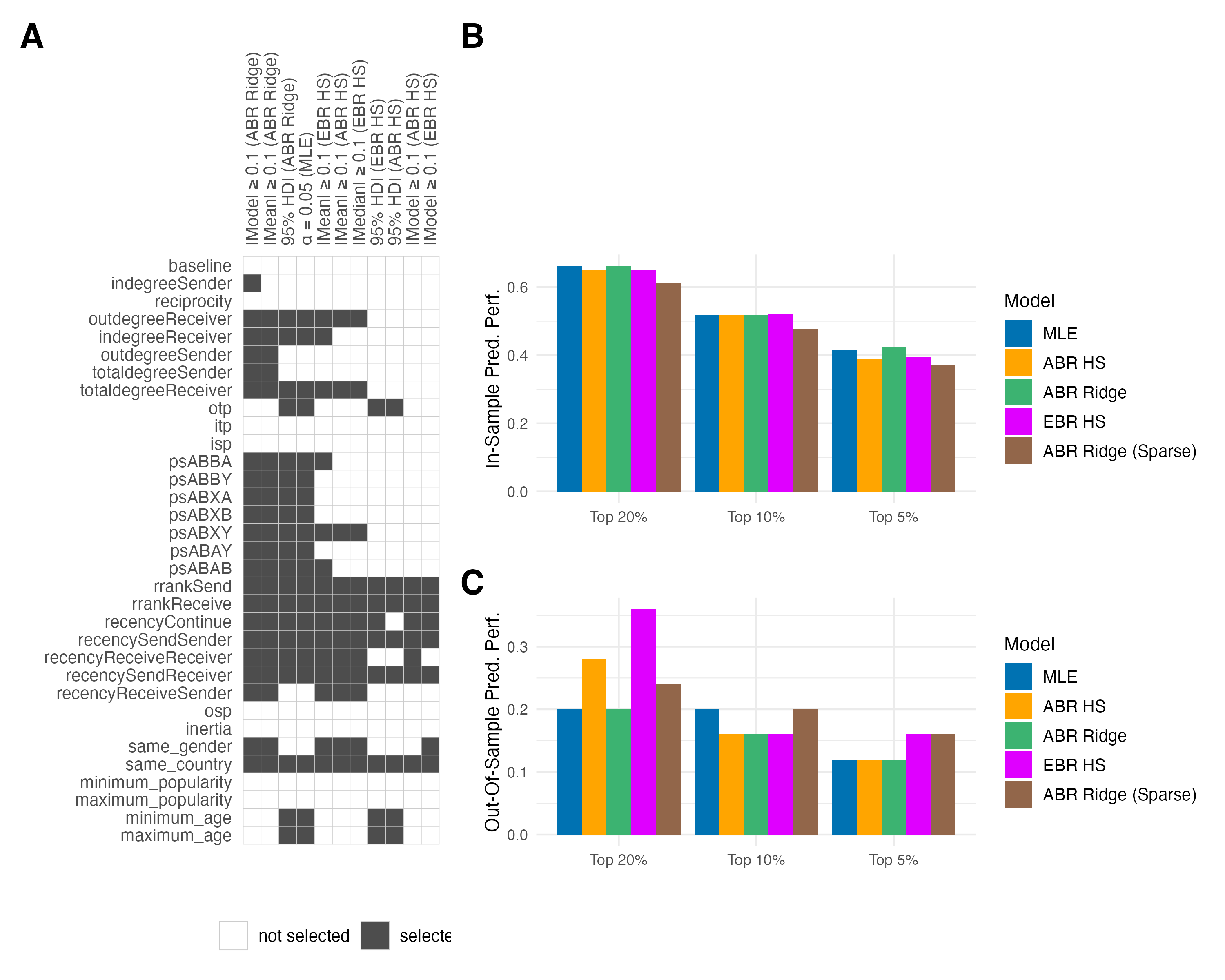}
    \caption{\textbf{Variable Selection and Predictive Performance on Spotify Data} \\ Model selection and predictive performance of the empirical application. Panels~(A)–(C) show: (A) Variable selection across 10 selection criteria, with rows listing covariates and columns denoting criteria. Filled squares indicate selected effects. (B) In-sample predictive performance measured as the proportion of observed events ranked within the top 20\%, 10\%, and 5\% most probable events. (C) Out-of-sample predictive performance using the same metrics for the last 10\% of events after training a Relational Event Model with the first 90\% of events. Horizontal axes in (B) and (C) display the top probability quantiles, and vertical axes represent predictive performance.}
    \label{fig:application}
\end{figure}

Panel A of Figure \ref{fig:application} presents the results of variable selection across the different selection criteria, following the approaches compared in the simulation study. As in the synthetic data sets, threshold-based criteria result in the selection of a larger number of predictors compared to univariate selection via MLE and HDI-based approaches from ABR. The estimation of the standard REM from MLE was the fastest, taking approximately 12 seconds on a laptop computer with an Apple M1 Pro processor. ABR with horseshoe and ridge priors (ABR) took around 30 seconds. In contrast, EBR was considerably more computationally intensive, requiring approximately 6 hours.

In \textit{rrankSend}, \textit{rrankeceive}, \textit{recencySendSender}, and \textit{recencySendReceiver}, four endogenous statistics are selected in all the methods presented, providing strong evidence that these effects are relevant to explain the dynamics of collaboration between artists. Among exogenous statistics, the effect of artists sharing the same country of origin showed the strongest evidence, being selected by all criteria. These results substantively suggest that both endogenous network effects and cultural similarity play an important role in shaping collaborative ties between top Spotify artists.

\subsection{Predictive Performance}

Panel B of Figure \ref{fig:application} displays in-sample predictive performances. Here, the four full models (i.e., MLE, ABR Ridge, ABR Horseshoe and EBR Horseshoe) are assessed before variable selection and compared to a sparser ABR Ridge model, resulting from variables selected through using the threshold-based criterion on the posterior mean. Overall, predictive performance is very similar across the evaluated methods, with only minor differences in performance metrics. 

Regarding out-of-sample predictive performance (visualized in Panel C of Figure \ref{fig:application}), assessed on the final 10\% of events ($M=25$) after training a REM on the first 90\% of events ($M=218$), the advantages of Bayesian regularizations in preventing overfitting show, with slightly better performances of the full EBR model with Horseshoe prior and the sparse model from ABR Ridge. Similar to the in-sample performance, the four models display similar metrics over the different percentiles, confirming the finding that regularized models can actually prevent overfitting.

\section{Discussion}

This paper evaluated the performance of Approximate (ABR) and Exact Bayesian Regularization (EBR) at selecting variables and making predictions in Relational Event Models (REMs), comparing both to a naive approach utilizing maximum likelihood estimation (MLE). A simulation study was conducted where 100 relational event history (REH) data sets of $M=6400$ events were generated and subsetted to different sizes to assess the behavior of three approaches across differently sized edgelists. 

Our simulation results revealed a clear divergence between endogenous and exogenous effects in terms of selection performance. Threshold-based selection using ABR (posterior mean/mode $\geq$ 0.1) identified the most variables, yielding the highest true discovery rates but also elevated false discovery rates, especially for endogenous effects. In contrast, criteria using the 95\% highest density credible interval were more conservative in variable selection. Overall, threshold-based selection using the posterior mean from ABR and EBR with a Horseshoe prior achieved the best trade-off between true and false discoveries, as indicated by the lowest combined distance metric and high Matthews' Correlation Coefficients across effect types and sample sizes. As expected, shrinkage with Horseshoe priors performed particularly well at recovering moderate and strong effects. Where calculated, exact approaches performed only slightly better than the approximations, underscoring ABR's advantages described by \citet{Karimova2025}. Overall, the MLE coefficients exhibited the least bias, which was strongly distorted by largely biased estimates for participation shifts. Regarding predictive performance, ABR with a Ridge prior performed best for both in-sample and out-of-sample data, particularly for small sample sizes, matching previous findings by \citet{Karimova2023}. An empirical example illustrated the application of Bayesian regularization procedures, while underscoring the computational burden of using EBR.

The findings are limited given the following: First, our simulation only considered endogenous statistics that are available in the \texttt{remstats} package. Other statistics, such as geodesic distances or between-centrality, may behave differently when exposed to shrinkage through ABR and EBR. Second, the generated synthetic data here assumes linear effects, whereas endogenous statistics may in reality exhibit nonlinearity \citep[see e.g.][]{Bauer2021, Fritz2021}, which may in turn affect the behavior of shrinkage. Third, the chosen setting with three endogenous effects of \textit{indegreeSender}, \textit{reciprocity}, and \textit{outdegreeReceiver} identified on Apollo-13 data is very specific, potentially limiting generalizability to settings where other data-generating mechanisms are expected. Fourth, REMs from MLE performed unexpectedly in selecting endogenous effects, displaying False Discovery Rates far above the expected $\alpha=0.05$ under univariate selection, likely due to strong multicollinearity among endogenous statistics in the \texttt{remstats} package. In light of additional analyses on collinearity \footnote{available at \url{https://jonathankoop.eu/blog/collinearity-rem/}}, particularly degree statistics tend to correlate highly with each other. In fact, many endogenous statistics measure very similar constructs: For example, the statistic \textit{recencySendSender}, measuring the time passed since a potential sender sent a tie, can be highly correlated with the number of ties a potential sender has sent (i.e., \textit{outdegreeSender}). The existence of such collinearity may also explain the biased coefficients observed under MLE. Consequently, future research should explore (1) the extent to which endogenous statistics correlate and (2) how this multicollinearity relates to the capacity to identify true non-zero effects.

In light of these findings, future research should investigate methods specifically tailored to handling correlated predictors. One potential approach is the use of dimension reduction techniques such as Principal Component Analysis \citep[as applied to communication networks by][]{Pilny2020} or Independent Component Analysis. Since these methods summarize all endogenous predictors into uncorrelated components, we consider this approach to be a potentially viable solution for combining related endogenous statistics. Alternatively, QR decomposition, as offered in the \texttt{brms} package \citep{brms}, could be applied. We consider another means to combat correlated predictors to be the application of grouped shrinkage to predictors by first clustering highly correlated statistics (e.g., all degree‐based or recency statistics) and then applying a group‐wise penalty.

In addition, we recommend exploring different thresholds and credible intervals: Although the threshold-based criterion outperformed both univariate feature selection via maximum-likelihood estimation and the 95\% highest-density credible interval in our simulations, it remains plausible that a less stringent HDI cutoff, such as the 90\% or 85\% HDI, may in reality be superior to $\lvert\beta\rvert\ge0.1$. In turn, other thresholds for posterior means and medians may also demonstrate better performance than 0.1. A potential solution to this may be an approach similar to that applied by \citet{vanErp2019} who investigated HDIs ranging from 0 to 100\% in steps of 10\% and chose an interval that minimized the distance. A similar strategy could then also be explored for threshold-based criteria.

Furthermore, considering recent works \citep{Amati2024, Boschi2025}, alternative metrics for assessing predictive performance should be explored, given that the current method overlooks the role of time points $t$ in coefficient estimation. This would be particularly useful in light of the negligible difference between performances for the synthetic data with large $M$s and the real-world REH data.

As regularization is particularly superior to MLE in high-dimensional settings (i.e., where the ratio of observations to predictors is low), an increased number of potential actors could be examined, given that MLE demonstrates lower statistical power with larger risk sets \citep{Schecter2020}. Moreover, examining the inclusion of interaction effects between endogenous and exogenous variables, which would further increase dimensionality, remains to be explored.

Future work could combine Bayesian regularization with changepoint models to account for non-stationarity in REMs. The relational event history could be segmented into periods with approximately stable dynamics, and regularized Bayesian REMs could be fitted within each segment. Posterior information from one segment, such as coefficient distributions or, could then be used as informative prior information for the next segment, allowing the model to couple onformation and and borrow strength over time while still capturing changes in predictor relevance \citep{ShafieeKamalabad2018, ShafieeKamalabad2019}.

\section{Conclusion}

This study provides the first systematic, simulation-based comparison of exact and approximate Bayesian regularization in Relational Event Models, demonstrating that threshold-based selection via approximate Bayesian regularization delivers a computationally efficient and statistically robust framework for variable selection when Horseshoe priors are used and for prediction when Ridge priors are applied. Applied researchers working with relational event history data should, therefore, fit their full relational event model using approximate Bayesian regularization retaining only those predictors whose posterior means exceed or is equal to $\lvert0.1\rvert$ with a prior choice depending on their primary goal: If it is explanatory, we recommend the use of a Horseshoe prior, in the case it is purely predictive we advise using a Ridge prior. This ascertains good identification of substantive effects while preventing overfitting and unstable estimates, particularly in settings with small sample sizes.

\section*{Acknowledgement}
This research was supported by a NWO Veni Grant (Vl.Veni.221G.005) to SvE.

\section*{Declaration of Generative AI and AI-assisted technologies in the writing process}
During the preparation of this work the author(s) used ChatGPT in order to improve language and readability, to edit tables, and to improve code. After using this tool/service, the author(s) reviewed and edited the content as needed and take(s) full responsibility for the content of the publication.



\bibliographystyle{elsarticle-harv}
\bibliography{literature}

\newpage

\appendix

\section{Reparameterization of REMs to Poisson Regressions}
\label{app1}

Since $\lambda^{REM}(s,r,t) = \frac{\mu_{s,r,m}}{\Delta_m}$,

\begin{equation} \label{eq:remtopois}
    \begin{aligned}
        \log \lambda^{\text{REM}}(s,r,t) &= \boldsymbol{X}_{s,r}(t)'\boldsymbol{\beta} \\
        \log \frac{\mu_{s,r,m}}{\Delta_m} &= \boldsymbol{X}_{s,r}(t)'\boldsymbol{\beta} \\
        \log \mu_{s,r,m} - \log \Delta_m &= \boldsymbol{X}_{s,r}(t)'\boldsymbol{\beta} \\
        \log \mu_{s,r,m} &= \log \Delta_m + \boldsymbol{X}_{s,r}(t)'\boldsymbol{\beta} ,
    \end{aligned}
\end{equation}

which is a Poisson regression model that can be estimated using a binary indicator for the occurrence of an event between sender $s$ and receiver $r$, $y_{s_mr_m}$ as a dependent variable, and $\log \Delta_m$ as offset. We provide a practical tutorial on this reparameterization under \url{https://jonathankoop.eu/blog/rem-to-poisson/}.

\section{Variable Selection}
\label{app2}

\ref{app2} provides the evaluation of the variable selection on the generated dependent relational event history (REH) data. Below, true discovery rates, false discovery rates, distance metrics, and Matthews' Correlation Coefficients are reported as means across iterations in the simulation, together with Monte Carlo standard errors.

\begin{landscape}
\begin{table}[ht]
  \centering
  \caption{True Discovery Rates (TDR) from the generated dependent Relational Event History (REH) data, defined as the proportion of truly nonzero effects correctly identified, for endogenous and exogenous effects (weak, moderate, strong) across numbers of events \(M\) (100–6,400). Results are reported as \textit{mean (Monte Carlo standard error)} over the 100 iterations in the simulation. Thirteen selection methods are compared: (i) maximum likelihood estimation (MLE) via univariate selection at \(\alpha=0.05\); (ii) approximate Bayesian regularization (ABR) with Horseshoe (abbreviated as HS) and Ridge priors using posterior 95\% highest-density intervals (HDI) and 0.1 thresholding on posterior mean, median, and mode; and (iii) exact Bayesian regularization (EBR) with Horseshoe prior using analogous posterior summaries. Boldface highlights the highest TDR in each row. A dash (–) indicates that a given criterion was not assessed.}
  \scriptsize
  \setlength\tabcolsep{2pt}
  \resizebox{1.6\textwidth}{!}{%
        \begin{tabular}{l r *{4}{c} *{4}{c} *{4}{c} c}
      \toprule
      & & \multicolumn{4}{c}{\itshape ABR HS}
        & \multicolumn{4}{c}{\itshape ABR Ridge}
        & \multicolumn{4}{c}{\itshape EBR HS}
        & \itshape MLE \\
      \cmidrule(lr){3-6}\cmidrule(lr){7-10}\cmidrule(lr){11-14}
      Sample Size ($M$)
        & & $95\%$ HDI & $|\mathrm{Mean}|\ge0.1$ & $|\mathrm{Median}|\ge0.1$ & $|\mathrm{Mode}|\ge0.1$
        & $95\%$ HDI  & $|\mathrm{Mean}|\ge0.1$ & $|\mathrm{Median}|\ge0.1$ & $|\mathrm{Mode}|\ge0.1$
        & $95\%$ HDI & $|\mathrm{Mean}|\ge0.1$ & $|\mathrm{Median}|\ge0.1$ & $|\mathrm{Mode}|\ge0.1$
        & $\alpha=0.05$ \\
      \midrule
      \multicolumn{15}{l}{\textbf{Endogenous}}\\
      $M=100$    &   & 0.473 (0.025) & 0.860 (0.019) & 0.860 (0.019) & 0.609 (0.019)
                & 0.638 (0.020) & \textbf{0.939 (0.013)} & \textbf{0.939 (0.013)} & \textbf{0.939 (0.013)}
                & 0.595 (0.015) & 0.857 (0.019) & 0.803 (0.019) & 0.663 (0.012)
                & 0.548 (0.024) \\
      200        &   & 0.522 (0.021) & 0.838 (0.019) & 0.838 (0.019) & 0.623 (0.017)
                & 0.657 (0.015) & \textbf{0.926 (0.014)} & \textbf{0.926 (0.014)} & 0.916 (0.015)
                & 0.616 (0.012) & 0.838 (0.017) & 0.801 (0.017) & 0.677 (0.012)
                & 0.599 (0.023) \\
      400        &   & 0.590 (0.019) & 0.863 (0.018) & 0.863 (0.018) & 0.643 (0.019)
                & 0.680 (0.014) & \textbf{0.917 (0.015)} & \textbf{0.917 (0.015)} & \textbf{0.917 (0.015)}
                & – & – & – & –
                & 0.670 (0.022) \\
      800        &   & 0.610 (0.018) & 0.860 (0.019) & 0.860 (0.019) & 0.700 (0.021)
                & 0.690 (0.013) & \textbf{0.913 (0.016)} & \textbf{0.913 (0.016)} & 0.910 (0.016)
                & – & – & – & –
                & 0.677 (0.019) \\
      1600       &   & 0.647 (0.021) & 0.873 (0.020) & 0.873 (0.020) & 0.710 (0.025)
                & 0.713 (0.014) & \textbf{0.927 (0.015)} & \textbf{0.927 (0.015)} & \textbf{0.927 (0.015)}
                & – & – & – & –
                & 0.700 (0.018) \\
      3200       &   & 0.687 (0.020) & 0.890 (0.020) & 0.890 (0.020) & 0.767 (0.025)
                & 0.720 (0.014) & \textbf{0.923 (0.016)} & \textbf{0.923 (0.016)} & 0.910 (0.017)
                & – & – & – & –
                & 0.690 (0.017) \\
      6400       &   & 0.727 (0.019) & 0.933 (0.016) & 0.933 (0.016) & 0.823 (0.022)
                & 0.727 (0.015) & \textbf{0.940 (0.014)} & \textbf{0.940 (0.014)} & 0.937 (0.015)
                & – & – & – & –
                & 0.683 (0.017) \\
      \addlinespace
      \multicolumn{15}{l}{\textbf{Exogenous}}\\
      \multicolumn{15}{l}{\itshape weak}\\
      $M=100$    &   & 0.004 (0.004) & 0.297 (0.029) & 0.297 (0.029) & 0.036 (0.011)
                & 0.068 (0.014) & \textbf{0.713 (0.025)} & \textbf{0.713 (0.025)} & \textbf{0.713 (0.025)}
                & 0.004 (0.004) & 0.312 (0.027) & 0.197 (0.023) & 0.022 (0.009)
                & 0.161 (0.022) \\
      200        &   & 0.007 (0.005) & 0.343 (0.028) & 0.343 (0.028) & 0.047 (0.012)
                & 0.094 (0.017) & 0.684 (0.026) & 0.684 (0.026) & \textbf{0.687 (0.026)}
                & 0.003 (0.003) & 0.239 (0.023) & 0.165 (0.022) & 0.034 (0.010)
                & 0.168 (0.023) \\
      400        &   & 0.030 (0.010) & 0.363 (0.028) & 0.363 (0.028) & 0.097 (0.019)
                & 0.143 (0.022) & 0.670 (0.028) & 0.670 (0.028) & \textbf{0.673 (0.027)}
                & – & – & – & –
                & 0.237 (0.025) \\
      800        &   & 0.057 (0.013) & 0.463 (0.026) & 0.463 (0.026) & 0.237 (0.022)
                & 0.270 (0.026) & 0.693 (0.027) & 0.693 (0.027) & \textbf{0.700 (0.027)}
                & – & – & – & –
                & 0.360 (0.028) \\
      1600       &   & 0.253 (0.026) & 0.583 (0.029) & 0.583 (0.029) & 0.427 (0.030)
                & 0.447 (0.031) & \textbf{0.773 (0.024)} & \textbf{0.773 (0.024)} & 0.770 (0.024)
                & – & – & – & –
                & 0.543 (0.031) \\
      3200       &   & 0.510 (0.029) & 0.733 (0.026) & 0.733 (0.026) & 0.673 (0.027)
                & 0.690 (0.029) & \textbf{0.850 (0.021)} & \textbf{0.850 (0.021)} & \textbf{0.850 (0.021)}
                & – & – & – & –
                & 0.753 (0.025) \\
      6400       &   & 0.817 (0.022) & 0.877 (0.019) & 0.877 (0.019) & 0.893 (0.019)
                & 0.917 (0.017) & 0.957 (0.013) & 0.957 (0.013) & \textbf{0.960 (0.012)}
                & – & – & – & –
                & 0.930 (0.016) \\
      \addlinespace
      \multicolumn{15}{l}{\itshape moderate}\\
      $M=100$    &   & 0.050 (0.015) & 0.581 (0.031) & 0.581 (0.031) & 0.100 (0.021)
                & 0.244 (0.028) & 0.828 (0.021) & 0.828 (0.021) & \textbf{0.835 (0.021)}
                & 0.079 (0.019) & 0.631 (0.030) & 0.491 (0.034) & 0.172 (0.026)
                & 0.312 (0.030) \\
      200        &   & 0.138 (0.025) & 0.700 (0.029) & 0.700 (0.029) & 0.195 (0.031)
                & 0.357 (0.030) & \textbf{0.912 (0.016)} & \textbf{0.912 (0.016)} & \textbf{0.912 (0.016)}
                & 0.172 (0.028) & 0.771 (0.025) & 0.626 (0.031) & 0.276 (0.033)
                & 0.407 (0.033) \\
      400        &   & 0.307 (0.032) & 0.837 (0.020) & 0.837 (0.020) & 0.447 (0.035)
                & 0.593 (0.028) & \textbf{0.957 (0.011)} & \textbf{0.957 (0.011)} & 0.953 (0.012)
                & – & – & – & –
                & 0.607 (0.033) \\
      800        &   & 0.633 (0.033) & 0.953 (0.013) & 0.953 (0.013) & 0.757 (0.029)
                & 0.830 (0.024) & 0.993 (0.005) & 0.993 (0.005) & \textbf{0.997 (0.003)}
                & – & – & – & –
                & 0.853 (0.023) \\
      1600       &   & 0.850 (0.024) & 0.993 (0.005) & 0.993 (0.005) & 0.923 (0.017)
                & 0.973 (0.009) & \textbf{1.000 (0.000)} & \textbf{1.000 (0.000)} & \textbf{1.000 (0.000)}
                & – & – & – & –
                & 0.980 (0.008) \\
      3200       &   & \textbf{1.000 (0.000)} & \textbf{1.000 (0.000)} & \textbf{1.000 (0.000)} & \textbf{1.000 (0.000)}
                & \textbf{1.000 (0.000)} & \textbf{1.000 (0.000)} & \textbf{1.000 (0.000)} & \textbf{1.000 (0.000)}
                & – & – & – & –
                & \textbf{1.000 (0.000)} \\
      6400       &   & \textbf{1.000 (0.000)} & \textbf{1.000 (0.000)} & \textbf{1.000 (0.000)} & \textbf{1.000 (0.000)}
                & \textbf{1.000 (0.000)} & \textbf{1.000 (0.000)} & \textbf{1.000 (0.000)} & \textbf{1.000 (0.000)}
                & – & – & – & –
                & \textbf{1.000 (0.000)} \\
      \addlinespace
      \multicolumn{15}{l}{\itshape strong}\\
      $M=100$    &   & 0.326 (0.036) & 0.921 (0.016) & 0.921 (0.016) & 0.423 (0.037)
                & 0.591 (0.036) & \textbf{0.964 (0.011)} & \textbf{0.964 (0.011)} & 0.961 (0.011)
                & 0.466 (0.036) & 0.953 (0.012) & 0.889 (0.018) & 0.573 (0.032)
                & 0.620 (0.034) \\
      200        &   & 0.498 (0.032) & 0.939 (0.017) & 0.939 (0.017) & 0.636 (0.032)
                & 0.764 (0.029) & 0.973 (0.009) & 0.973 (0.009) & 0.973 (0.009)
                & 0.626 (0.029) & \textbf{0.976 (0.010)} & 0.943 (0.014) & 0.747 (0.028)
                & 0.781 (0.025) \\
      400        &   & 0.723 (0.027) & 0.963 (0.010) & 0.963 (0.010) & 0.807 (0.025)
                & 0.897 (0.019) & 0.997 (0.003) & 0.997 (0.003) & \textbf{1.000 (0.000)}
                & – & – & – & –
                & 0.897 (0.019) \\
      800        &   & 0.890 (0.019) & 0.977 (0.009) & 0.977 (0.009) & 0.917 (0.016)
                & 0.950 (0.012) & \textbf{0.997 (0.003)} & \textbf{0.997 (0.003)} & \textbf{0.997 (0.003)}
                & – & – & – & –
                & 0.963 (0.010) \\
      1600       &   & 0.933 (0.013) & 0.990 (0.006) & 0.990 (0.006) & 0.950 (0.012)
                & 0.960 (0.011) & \textbf{1.000 (0.000)} & \textbf{1.000 (0.000)} & \textbf{1.000 (0.000)}
                & – & – & – & –
                & 0.990 (0.006) \\
      3200       &   & 0.967 (0.010) & \textbf{1.000 (0.000)} & \textbf{1.000 (0.000)} & 0.973 (0.009)
                & \textbf{1.000 (0.000)} & \textbf{1.000 (0.000)} & \textbf{1.000 (0.000)} & \textbf{1.000 (0.000)}
                & – & – & – & –
                & \textbf{1.000 (0.000)} \\
      6400       &   & \textbf{1.000 (0.000)} & \textbf{1.000 (0.000)} & \textbf{1.000 (0.000)} & \textbf{1.000 (0.000)}
                & \textbf{1.000 (0.000)} & \textbf{1.000 (0.000)} & \textbf{1.000 (0.000)} & \textbf{1.000 (0.000)}
                & – & – & – & –
                & \textbf{1.000 (0.000)} \\
      \bottomrule
    \end{tabular}%
  }
  \label{tab:tdr}
\end{table}
\end{landscape}

\newpage

\begin{landscape}
\begin{table}[ht]
  \centering
  \caption{False Discovery Rates (FDR) from the generated dependent Relational Event History (REH) data, defined as the proportion of truly zero effects incorrectly identified as nonzero, for endogenous and exogenous effects across numbers of events \(M\) (100–6400). Results are reported as \textit{mean (Monte Carlo standard error)} over the 100 iterations in the simulation. Thirteen selection methods are compared: (i) maximum likelihood estimation (MLE) via univariate selection at \(\alpha=0.05\); (ii) approximate Bayesian regularization (ABR) with Horseshoe (abbreviated as HS) and Ridge priors using posterior 95\% highest-density intervals (HDI) and 0.1 thresholding on posterior mean, median, and mode; and (iii) exact Bayesian regularization (EBR) with Horseshoe prior using analogous posterior summaries. Boldface highlights the lowest FDR in each row. A dash (–) indicates that a given criterion was not assessed.}
  \scriptsize
  \setlength\tabcolsep{2pt}
  \resizebox{1.6\textwidth}{!}{%
        \begin{tabular}{l r *{4}{c} *{4}{c} *{4}{c} c}
      \toprule
      & & \multicolumn{4}{c}{\itshape ABR HS}
        & \multicolumn{4}{c}{\itshape ABR Ridge}
        & \multicolumn{4}{c}{\itshape EBR HS}
        & \itshape MLE \\
      \cmidrule(lr){3-6}\cmidrule(lr){7-10}\cmidrule(lr){11-14}
      Sample Size ($M$)
        & & $95\%$ HDI & $|\mathrm{Mean}|\ge0.1$ & $|\mathrm{Median}|\ge0.1$ & $|\mathrm{Mode}|\ge0.1$
        & $95\%$ HDI  & $|\mathrm{Mean}|\ge0.1$ & $|\mathrm{Median}|\ge0.1$ & $|\mathrm{Mode}|\ge0.1$
        & $95\%$ HDI & $|\mathrm{Mean}|\ge0.1$ & $|\mathrm{Median}|\ge0.1$ & $|\mathrm{Mode}|\ge0.1$
        & $\alpha=0.05$ \\
      \midrule
      \multicolumn{15}{l}{\textbf{Endogenous}}\\
      $M=100$    &   & 0.183 (0.012) & 0.520 (0.011) & 0.520 (0.011) & 0.232 (0.010)
                & 0.129 (0.005) & 0.703 (0.010) & 0.703 (0.010) & 0.696 (0.010)
                & \textbf{0.015 (0.003)} & 0.428 (0.013) & 0.223 (0.011) & 0.039 (0.005)
                & 0.381 (0.008) \\
      200        &   & 0.113 (0.010) & 0.434 (0.013) & 0.434 (0.013) & 0.172 (0.010)
                & 0.116 (0.004) & 0.644 (0.010) & 0.644 (0.010) & 0.638 (0.010)
                & \textbf{0.015 (0.003)} & 0.356 (0.011) & 0.193 (0.009) & 0.040 (0.004)
                & 0.390 (0.008) \\
      400        &   & \textbf{0.069 (0.008)} & 0.328 (0.015) & 0.328 (0.015) & 0.097 (0.010)
                & 0.111 (0.004) & 0.594 (0.011) & 0.594 (0.011) & 0.587 (0.011)
                & – & – & – & –
                & 0.388 (0.007) \\
      800        &   & \textbf{0.048 (0.006)} & 0.239 (0.013) & 0.239 (0.013) & 0.062 (0.006)
                & 0.109 (0.005) & 0.558 (0.009) & 0.558 (0.009) & 0.553 (0.009)
                & – & – & – & –
                & 0.377 (0.007) \\
      1600       &   & \textbf{0.037 (0.004)} & 0.172 (0.009) & 0.172 (0.009) & 0.046 (0.004)
                & 0.096 (0.005) & 0.548 (0.009) & 0.548 (0.009) & 0.536 (0.009)
                & – & – & – & –
                & 0.373 (0.007) \\
      3200       &   & \textbf{0.026 (0.003)} & 0.133 (0.006) & 0.133 (0.006) & 0.033 (0.003)
                & 0.066 (0.004) & 0.509 (0.009) & 0.509 (0.009) & 0.498 (0.009)
                & – & – & – & –
                & 0.363 (0.006) \\
      6400       &   & \textbf{0.019 (0.004)} & 0.118 (0.006) & 0.118 (0.006) & 0.027 (0.004)
                & 0.051 (0.005) & 0.478 (0.008) & 0.478 (0.008) & 0.467 (0.008)
                & – & – & – & –
                & 0.365 (0.006) \\
      \addlinespace
      \multicolumn{15}{l}{\textbf{Exogenous}}\\
      $M=100$    &   & \textbf{0.000 (0.000)} & 0.242 (0.014) & 0.242 (0.014) & 0.007 (0.002)
                & 0.026 (0.005) & 0.629 (0.015) & 0.629 (0.015) & 0.639 (0.015)
                & 0.001 (0.001) & 0.246 (0.014) & 0.141 (0.012) & 0.019 (0.004)
                & 0.079 (0.008) \\
      200        &   & \textbf{0.002 (0.001)} & 0.185 (0.012) & 0.185 (0.012) & 0.012 (0.003)
                & 0.025 (0.005) & 0.560 (0.015) & 0.560 (0.015) & 0.560 (0.016)
                & 0.003 (0.001) & 0.201 (0.013) & 0.121 (0.010) & 0.019 (0.004)
                & 0.074 (0.009) \\
      400        &   & \textbf{0.003 (0.001)} & 0.138 (0.011) & 0.138 (0.011) & 0.019 (0.004)
                & 0.037 (0.006) & 0.473 (0.015) & 0.473 (0.015) & 0.485 (0.015)
                & – & – & – & –
                & 0.081 (0.009) \\
      800        &   & \textbf{0.002 (0.001)} & 0.079 (0.008) & 0.079 (0.008) & 0.015 (0.003)
                & 0.034 (0.004) & 0.369 (0.014) & 0.369 (0.014) & 0.372 (0.014)
                & – & – & – & –
                & 0.061 (0.007) \\
      1600       &   & \textbf{0.007 (0.002)} & 0.049 (0.005) & 0.049 (0.005) & 0.017 (0.003)
                & 0.044 (0.005) & 0.265 (0.013) & 0.265 (0.013) & 0.263 (0.013)
                & – & – & – & –
                & 0.057 (0.006) \\
      3200       &   & \textbf{0.004 (0.002)} & 0.022 (0.004) & 0.022 (0.004) & 0.011 (0.003)
                & 0.041 (0.005) & 0.161 (0.011) & 0.161 (0.011) & 0.159 (0.010)
                & – & – & – & –
                & 0.055 (0.006) \\
      6400       &   & 0.005 (0.002) & 0.007 (0.002) & 0.007 (0.002) & \textbf{0.004 (0.002)}
                & 0.047 (0.006) & 0.073 (0.007) & 0.073 (0.007) & 0.075 (0.007)
                & – & – & – & –
                & 0.055 (0.006) \\
      \bottomrule
    \end{tabular}%

    \label{tab:fdr}
  }
\end{table}
\end{landscape}

\newpage

\begin{landscape}
\begin{table}[ht]
  \centering
  \caption{Distance metrics from the generated dependent Relational Event History (REH) data, quantifying the Euclidean distance from perfect classification (TDR = 1, FDR = 0) for endogenous and exogenous effects across numbers of events \(M\) (100–6400). Results are reported as \textit{mean (Monte Carlo standard error)} over the 100 iterations in the simulation. Thirteen selection methods are compared: (i) maximum likelihood estimation (MLE) via univariate selection at \(\alpha=0.05\); (ii) approximate Bayesian regularization (ABR) with Horseshoe (abbreviated as HS) and Ridge priors using posterior 95\% highest-density intervals (HDI) and 0.1 thresholding on posterior mean, median, and mode; and (iii) exact Bayesian regularization (EBR) with Horseshoe prior using analogous posterior summaries. Boldface highlights the lowest distance in each row. A dash (–) indicates that a given criterion was not assessed.}
  \scriptsize
  \setlength\tabcolsep{2pt}
  \resizebox{1.6\textwidth}{!}{%
        \begin{tabular}{l r *{4}{c} *{4}{c} *{4}{c} c}
      \toprule
      & & \multicolumn{4}{c}{\itshape ABR HS}
        & \multicolumn{4}{c}{\itshape ABR Ridge}
        & \multicolumn{4}{c}{\itshape EBR HS}
        & \itshape MLE \\
      \cmidrule(lr){3-6}\cmidrule(lr){7-10}\cmidrule(lr){11-14}
      Sample Size ($M$)
        & & $95\%$ HDI & $|\mathrm{Mean}|\ge0.1$ & $|\mathrm{Median}|\ge0.1$ & $|\mathrm{Mode}|\ge0.1$
        & $95\%$ HDI  & $|\mathrm{Mean}|\ge0.1$ & $|\mathrm{Median}|\ge0.1$ & $|\mathrm{Mode}|\ge0.1$
        & $95\%$ HDI & $|\mathrm{Mean}|\ge0.1$ & $|\mathrm{Median}|\ge0.1$ & $|\mathrm{Mode}|\ge0.1$
        & $\alpha=0.05$ \\
      \midrule
      \multicolumn{15}{l}{\textbf{Endogenous}}\\
      $M=100$    &   & 0.579 (0.022) & 0.567 (0.013) & 0.567 (0.013) & 0.478 (0.015)
                & 0.400 (0.017) & 0.717 (0.011) & 0.717 (0.011) & 0.710 (0.010)
                & 0.406 (0.015) & 0.483 (0.014) & \textbf{0.337 (0.014)} & 0.343 (0.012)
                & 0.619 (0.016) \\
      200        &   & 0.503 (0.021) & 0.496 (0.015) & 0.496 (0.015) & 0.435 (0.015)
                & 0.372 (0.014) & 0.661 (0.011) & 0.661 (0.011) & 0.658 (0.012)
                & 0.385 (0.012) & 0.426 (0.012) & \textbf{0.316 (0.012)} & 0.331 (0.011)
                & 0.588 (0.016) \\
      400        &   & 0.425 (0.019) & 0.390 (0.016) & 0.390 (0.016) & 0.385 (0.018)
                & \textbf{0.348 (0.012)} & 0.617 (0.012) & 0.617 (0.012) & 0.608 (0.012)
                & – & – & – & –
                & 0.545 (0.013) \\
      800        &   & 0.399 (0.017) & \textbf{0.318 (0.017)} & \textbf{0.318 (0.017)} & 0.321 (0.020)
                & 0.341 (0.010) & 0.584 (0.011) & 0.584 (0.011) & 0.580 (0.011)
                & – & – & – & –
                & 0.522 (0.011) \\
      1600       &   & 0.360 (0.021) & \textbf{0.252 (0.018)} & \textbf{0.252 (0.018)} & 0.305 (0.024)
                & 0.314 (0.012) & 0.570 (0.010) & 0.570 (0.010) & 0.559 (0.010)
                & – & – & – & –
                & 0.506 (0.010) \\
      3200       &   & 0.318 (0.020) & \textbf{0.208 (0.017)} & \textbf{0.208 (0.017)} & 0.245 (0.024)
                & 0.299 (0.012) & 0.533 (0.011) & 0.533 (0.011) & 0.529 (0.011)
                & – & – & – & –
                & 0.502 (0.010) \\
      6400       &   & 0.280 (0.019) & \textbf{0.163 (0.015)} & \textbf{0.163 (0.015)} & 0.190 (0.022)
                & 0.290 (0.014) & 0.497 (0.010) & 0.497 (0.010) & 0.489 (0.011)
                & – & – & – & –
                & 0.506 (0.009) \\
      \addlinespace
      \multicolumn{15}{l}{\textbf{Exogenous}}\\
      $M=100$    &   & 0.873 (0.016) & 0.490 (0.017) & 0.490 (0.017) & 0.814 (0.020)
                & 0.702 (0.020) & 0.659 (0.015) & 0.659 (0.015) & 0.669 (0.016)
                & 0.817 (0.016) & \textbf{0.465 (0.016)} & 0.509 (0.017) & 0.746 (0.018)
                & 0.647 (0.020) \\
      200        &   & 0.786 (0.017) & \textbf{0.408 (0.017)} & \textbf{0.408 (0.017)} & 0.708 (0.020)
                & 0.597 (0.018) & 0.589 (0.015) & 0.589 (0.015) & 0.587 (0.016)
                & 0.733 (0.017) & 0.411 (0.014) & 0.450 (0.016) & 0.649 (0.018)
                & 0.560 (0.019) \\
      400        &   & 0.647 (0.019) & \textbf{0.332 (0.013)} & \textbf{0.332 (0.013)} & 0.553 (0.021)
                & 0.461 (0.016) & 0.500 (0.015) & 0.500 (0.015) & 0.511 (0.015)
                & – & – & – & –
                & 0.437 (0.018) \\
      800        &   & 0.473 (0.017) & \textbf{0.234 (0.011)} & \textbf{0.234 (0.011)} & 0.365 (0.018)
                & 0.323 (0.015) & 0.395 (0.014) & 0.395 (0.014) & 0.397 (0.014)
                & – & – & – & –
                & 0.290 (0.015) \\
      1600       &   & 0.322 (0.017) & \textbf{0.168 (0.010)} & \textbf{0.168 (0.010)} & 0.238 (0.015)
                & 0.223 (0.012) & 0.288 (0.013) & 0.288 (0.013) & 0.286 (0.013)
                & – & – & – & –
                & 0.189 (0.010) \\
      3200       &   & 0.175 (0.011) & \textbf{0.101 (0.008)} & \textbf{0.101 (0.008)} & 0.123 (0.010)
                & 0.126 (0.009) & 0.181 (0.011) & 0.181 (0.011) & 0.178 (0.011)
                & – & – & – & –
                & 0.119 (0.008) \\
      6400       &   & 0.066 (0.007) & 0.046 (0.007) & 0.046 (0.007) & \textbf{0.040 (0.006)}
                & 0.068 (0.007) & 0.081 (0.007) & 0.081 (0.007) & 0.082 (0.008)
                & – & – & – & –
                & 0.073 (0.007) \\
      \bottomrule
    \end{tabular}%
    \label{tab:dist}
  }
\end{table}
\end{landscape}

\newpage

\begin{landscape}
\begin{table}[ht]
  \centering
  \caption{Matthews correlation coefficient (MCC) from the generated dependent Relational Event History (REH) data, quantifying the agreement between true and selected nonzero effects for endogenous and exogenous effects across numbers of events \(M\) (100–6400). Results are reported as \textit{mean (Monte Carlo standard error)} over the 100 iterations in the simulation. Thirteen selection methods are compared: (i) maximum likelihood estimation (MLE) via univariate selection at \(\alpha=0.05\); (ii) approximate Bayesian regularization (ABR) with Horseshoe (HS) and Ridge priors using posterior 95\% highest-density intervals (HDI) and 0.1 thresholding on posterior mode, median, and mean; and (iii) exact Bayesian regularization (EBR) with Horseshoe prior using analogous posterior summaries. Boldface highlights the highest MCC in each row. A dash (–) indicates that a given criterion was not assessed.}
  \scriptsize
  \setlength\tabcolsep{2pt}
  \resizebox{1.6\textwidth}{!}{%
        \begin{tabular}{l r *{4}{c} *{4}{c} *{4}{c} c}
      \toprule
      & & \multicolumn{4}{c}{\itshape ABR HS}
        & \multicolumn{4}{c}{\itshape ABR Ridge}
        & \multicolumn{4}{c}{\itshape EBR HS}
        & \itshape MLE \\
      \cmidrule(lr){3-6}\cmidrule(lr){7-10}\cmidrule(lr){11-14}\cmidrule(lr){15-15}
      Sample Size ($M$)
        & & $95\%$ HDI & $|\mathrm{Mode}|\ge0.1$ & $|\mathrm{Median}|\ge0.1$ & $|\mathrm{Mean}|\ge0.1$
        & $95\%$ HDI  & $|\mathrm{Mode}|\ge0.1$ & $|\mathrm{Median}|\ge0.1$ & $|\mathrm{Mean}|\ge0.1$
        & $95\%$ HDI & $|\mathrm{Mode}|\ge0.1$ & $|\mathrm{Median}|\ge0.1$ & $|\mathrm{Mean}|\ge0.1$
        & $\alpha=0.05$ \\
      \midrule
      \multicolumn{15}{l}{\textbf{Endogenous}}\\
      $M=100$    &   & 0.251 (0.023) & 0.283 (0.015) & 0.223 (0.014) & 0.223 (0.014)
                      & 0.423 (0.017) & 0.175 (0.011) & 0.172 (0.011) & 0.172 (0.011)
                      & \textbf{0.690 (0.015)} & 0.666 (0.017) & 0.429 (0.017) & 0.283 (0.015)
                      & 0.110 (0.015) \\
      200        &   & 0.395 (0.025) & 0.373 (0.018) & 0.267 (0.016) & 0.267 (0.016)
                      & 0.457 (0.014) & 0.188 (0.014) & 0.192 (0.012) & 0.192 (0.012)
                      & \textbf{0.705 (0.013)} & 0.666 (0.015) & 0.460 (0.016) & 0.322 (0.014)
                      & 0.136 (0.015) \\
      400        &   & \textbf{0.537 (0.023)} & 0.517 (0.022) & 0.369 (0.018) & 0.369 (0.018)
                      & 0.486 (0.014) & 0.219 (0.013) & 0.215 (0.013) & 0.215 (0.013)
                      & – & – & – & –
                      & 0.183 (0.013) \\
      800        &   & 0.591 (0.020) & \textbf{0.616 (0.021)} & 0.451 (0.019) & 0.451 (0.019)
                      & 0.500 (0.013) & 0.234 (0.013) & 0.233 (0.013) & 0.233 (0.013)
                      & – & – & – & –
                      & 0.196 (0.011) \\
      1600       &   & 0.637 (0.022) & \textbf{0.650 (0.024)} & 0.535 (0.020) & 0.535 (0.020)
                      & 0.541 (0.015) & 0.256 (0.011) & 0.248 (0.011) & 0.248 (0.011)
                      & – & – & – & –
                      & 0.214 (0.011) \\
      3200       &   & 0.702 (0.019) & \textbf{0.724 (0.023)} & 0.597 (0.019) & 0.597 (0.019)
                      & 0.616 (0.014) & 0.267 (0.013) & 0.269 (0.012) & 0.269 (0.012)
                      & – & – & – & –
                      & 0.215 (0.010) \\
      6400       &   & 0.763 (0.017) & \textbf{0.795 (0.020)} & 0.654 (0.018) & 0.654 (0.018)
                      & 0.666 (0.016) & 0.303 (0.012) & 0.298 (0.011) & 0.298 (0.011)
                      & – & – & – & –
                      & 0.209 (0.010) \\
      \addlinespace
      \multicolumn{15}{l}{\textbf{Exogenous}}\\
      $M=100$    &   & 0.373 (0.013) & 0.383 (0.018) & 0.368 (0.023) & 0.368 (0.023)
                      & 0.412 (0.020) & 0.208 (0.023) & 0.216 (0.023) & 0.216 (0.023)
                      & 0.381 (0.015) & 0.380 (0.016) & \textbf{0.424 (0.021)} & 0.394 (0.022)
                      & 0.367 (0.023) \\
      200        &   & 0.399 (0.015) & 0.429 (0.018) & 0.490 (0.020) & 0.490 (0.020)
                      & \textbf{0.496 (0.019)} & 0.308 (0.020) & 0.309 (0.019) & 0.309 (0.019)
                      & 0.422 (0.015) & 0.460 (0.015) & 0.494 (0.021) & 0.472 (0.020)
                      & 0.454 (0.022) \\
      400        &   & 0.490 (0.016) & 0.539 (0.018) & \textbf{0.602 (0.017)} & \textbf{0.602 (0.017)}
                      & 0.589 (0.016) & 0.396 (0.018) & 0.405 (0.018) & 0.405 (0.018)
                      & – & – & – & –
                      & 0.555 (0.021) \\
      800        &   & 0.632 (0.014) & 0.697 (0.015) & \textbf{0.738 (0.014)} & \textbf{0.738 (0.014)}
                      & 0.705 (0.014) & 0.520 (0.017) & 0.520 (0.017) & 0.520 (0.017)
                      & – & – & – & –
                      & 0.700 (0.016) \\
      1600       &   & 0.742 (0.014) & 0.795 (0.013) & \textbf{0.823 (0.011)} & \textbf{0.823 (0.011)}
                      & 0.779 (0.012) & 0.648 (0.015) & 0.647 (0.016) & 0.647 (0.016)
                      & – & – & – & –
                      & 0.798 (0.012) \\
      3200       &   & 0.859 (0.009) & 0.893 (0.009) & \textbf{0.902 (0.009)} & \textbf{0.902 (0.009)}
                      & 0.866 (0.010) & 0.777 (0.014) & 0.776 (0.014) & 0.776 (0.014)
                      & – & – & – & –
                      & 0.866 (0.010) \\
      6400       &   & 0.946 (0.006) & \textbf{0.967 (0.005)} & 0.959 (0.006) & 0.959 (0.006)
                      & 0.920 (0.008) & 0.900 (0.009) & 0.900 (0.009) & 0.900 (0.009)
                      & – & – & – & –
                      & 0.914 (0.008) \\
      \bottomrule
    \end{tabular}%
    \label{tab:mcc}
  }
\end{table}
\end{landscape}

\section{Bias and Variance} 
\label{app3}

\begin{table}[ht]
    \centering
    \caption{Bias and variance of parameter estimates from the generated dependent Relational Event History (REH) data, where bias is the average deviation of estimates from the true effects and variance is the variance of this deviation over the 100 iterations in the simulation, for sample sizes \(M=100\)–6400 after the removal of participation shifts. Results are reported as \textit{bias (variance)}. Four estimation methods are compared: maximum likelihood estimation (MLE), approximate Bayesian regularization (ABR) with Horseshoe (HS) and Ridge priors, and exact Bayesian regularization (EBR) with Horseshoe prior. A dash (–) indicates that the method was not assessed at the given sample size.}
    \begin{tabular}{rcccc}
\toprule
Sample Size ($M$) & \itshape MLE & \itshape ABR HS & \itshape ABR Ridge & \itshape EBR HS \\
\midrule
100 & -0.135 (2.39) & 0.044 (1.13) & 0.188 (1.76) & 0.165 (2.59)\\
200 & -0.096 (1.11) & 0.068 (1.37) & 0.187 (1.70) & 0.165 (2.35)\\
400 & -0.075 (0.65) & 0.129 (1.40) & 0.186 (1.64) & –\\
800 & -0.072 (0.38) & 0.155 (1.52) & 0.183 (1.58) & –\\
1600 & -0.068 (0.31) & 0.173 (1.48) & 0.181 (1.51) & –\\
3200 & -0.065 (0.28) & 0.179 (1.44) & 0.180 (1.44) & –\\
6400 & -0.064 (0.26) & 0.178 (1.40) & 0.180 (1.37) & –\\
\bottomrule
\end{tabular}

    \label{tab:bias_variance}
\end{table}

\newpage

\begin{figure}[H]
    \centering
    \includegraphics[width=\textwidth,height=0.85\textheight,keepaspectratio]{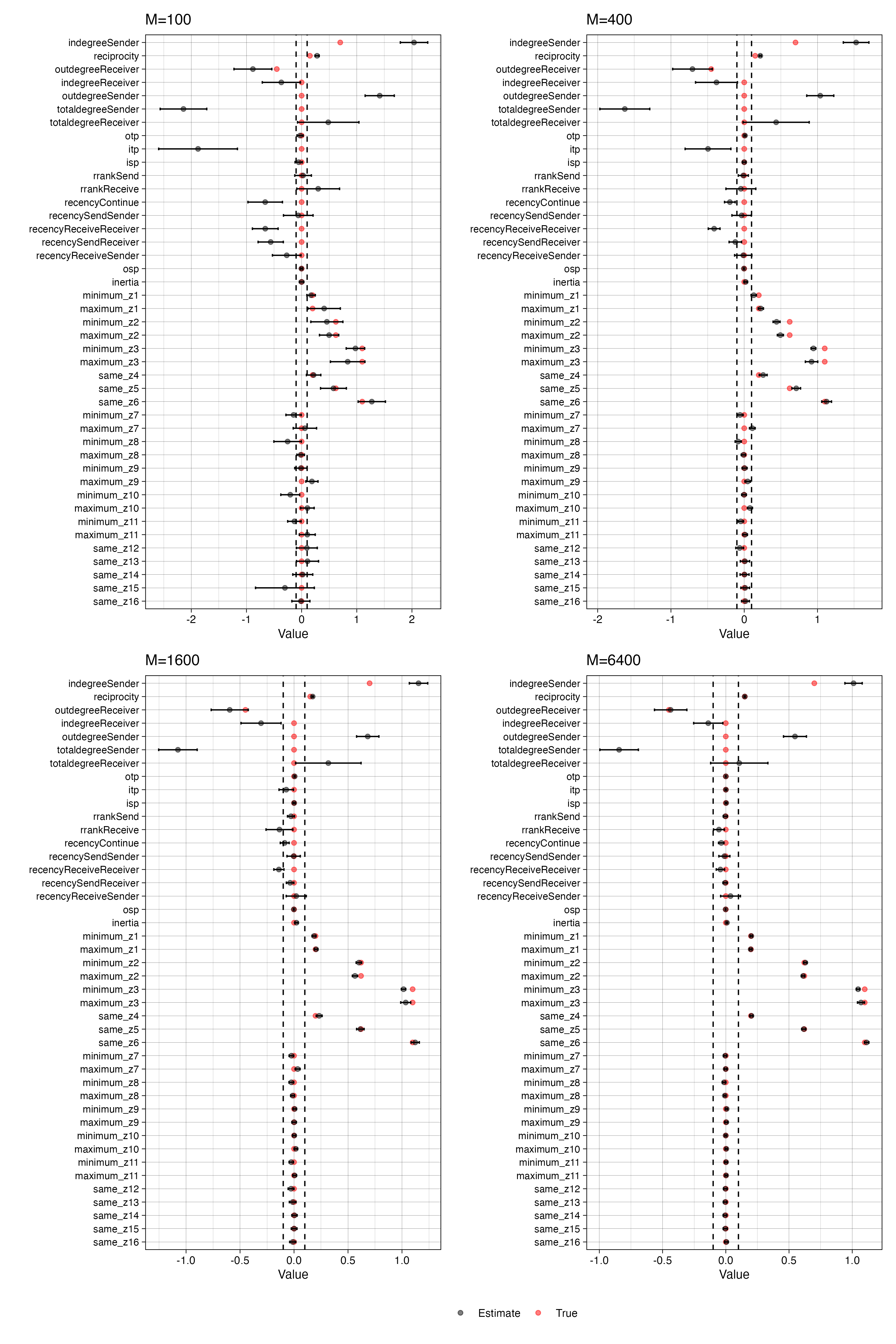}
    \caption{\textbf{Mean Maximum Likelihood Estimates for All Coefficients Excluding Participation Shifts Across Selected Sample Sizes} Mean parameter estimates for all coefficients excluding participation shifts averaged over all iterations under varying numbers of events $M$. The four panels correspond to selected sample sizes: $M=100$ (top-left), $M=400$ (top-right), $M=1600$ (bottom-left), and $M=6400$ (bottom-right). Vertical axes list the individual predictor variables evaluated in the model. Horizontal axes display the corresponding parameter values. Mean parameter estimates obtained via Maximum Likelihood Estimation (MLE) are represented by black dots, with horizontal error bars indicating the 95\% confidence intervals across all iterations for each $M$. True parameter values are overlaid as red dots for direct comparison. Vertical dashed lines are positioned at 0.1 to denote thresholds for the corresponding selection criteria.}
    \label{fig:plots_mle}
\end{figure}

\begin{figure}[H]
    \centering
    \includegraphics[width=\textwidth]{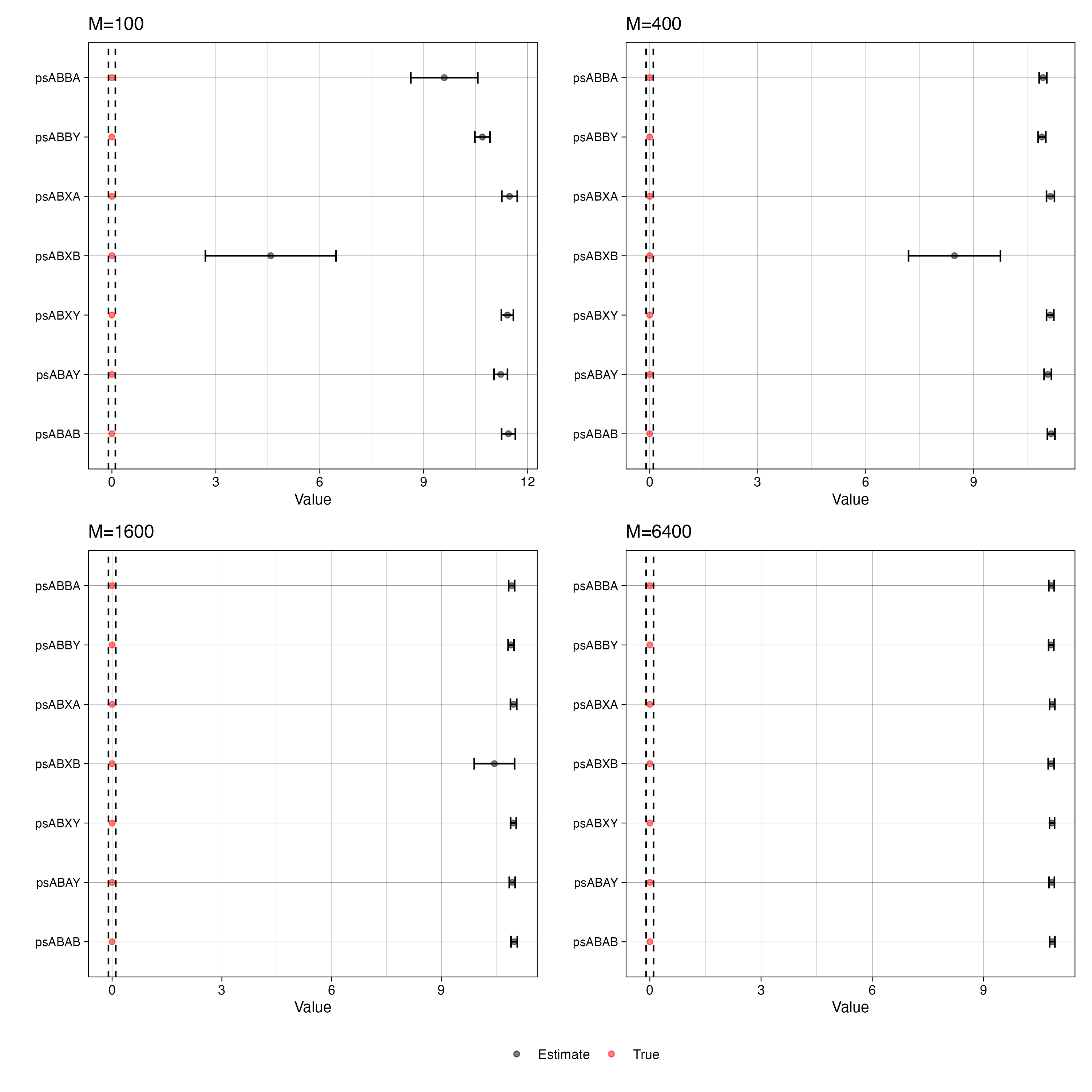}
     \caption{\textbf{Mean Maximum Likelihood Estimates for Participation Shifts Across Selected Sample Sizes} Mean parameter estimates for all participation shifts averaged over all iterations under varying numbers of events $M$. The four panels correspond to selected sample sizes: $M=100$ (top-left), $M=400$ (top-right), $M=1600$ (bottom-left), and $M=6400$ (bottom-right). Vertical axes list the individual predictor variables evaluated in the model. Horizontal axes display the corresponding parameter values. Mean parameter estimates obtained via Maximum Likelihood Estimation (MLE) are represented by black dots, with horizontal error bars indicating the 95\% confidence intervals across all iterations for each $M$. True parameter values are overlaid as red dots for direct comparison. Vertical dashed lines are positioned at 0.1 to denote thresholds for the corresponding selection criteria.}
    \label{fig:plots_mle_ps}
\end{figure}

\newpage

\section{Predictive Performance}
\label{app4}

\begin{table}[ht]
  \centering
  \caption{In-sample predictive performance of the full Relational Event Models from the generated dependent Relational Event History (REH) data, quantified as the proportion of true events captured within the top 20\%, 10\%, and 5\% highest predicted-probability events for training sample sizes \(M=100\)–6400. Event probabilities were calculated at each time \(t\) on the training set, and coverage of the actual events by the highest-probability subsets was evaluated. Four estimation approaches are compared: maximum likelihood estimation (MLE), approximate Bayesian regularization with Ridge (ABR Ridge) and Horseshoe (ABR HS) priors, and exact Bayesian regularization with Horseshoe prior (EBR HS). Results are reported as \textit{mean (Monte Carlo standard error)} over the 100 iterations in the simulation. Boldface highlights the best performance per threshold and sample size. A dash (–) indicates that a given configuration was not assessed.}
    \begin{tabular}{l c c c c}
    \toprule
    Sample Size ($M$)
      & \itshape MLE
      & \itshape ABR Ridge
      & \itshape ABR HS
      & \itshape EBR HS \\
    \midrule
    \multicolumn{5}{l}{\textbf{Top 20\%}}\\
    $M=100$    & \textbf{0.975 (0.002)} & 0.970 (0.002) & 0.938 (0.011) & 0.965 (0.002) \\
    200        & \textbf{0.984 (0.001)} & 0.982 (0.001) & 0.967 (0.004) & 0.978 (0.001) \\
    400        & \textbf{0.990 (0.001)} & 0.989 (0.001) & 0.974 (0.009) & – \\
    800        & \textbf{0.992 (0.001)} & \textbf{0.992 (0.001)} & 0.988 (0.001) & – \\
    1600       & \textbf{0.993 (0.000)} & \textbf{0.993 (0.000)} & 0.991 (0.001) & – \\
    3200       & \textbf{0.994 (0.000)} & \textbf{0.994 (0.000)} & 0.992 (0.001) & – \\
    6400       & \textbf{0.994 (0.000)} & \textbf{0.994 (0.000)} & \textbf{0.994 (0.000)} & – \\

    \addlinespace
    \multicolumn{5}{l}{\textbf{Top 10\%}}\\
    $M=100$    & \textbf{0.950 (0.004)} & 0.942 (0.004) & 0.905 (0.011) & 0.935 (0.004) \\
    200        & \textbf{0.967 (0.003)} & 0.963 (0.003) & 0.944 (0.005) & 0.959 (0.003) \\
    400        & \textbf{0.977 (0.002)} & 0.976 (0.002) & 0.957 (0.010) & – \\
    800        & \textbf{0.983 (0.001)} & 0.982 (0.001) & 0.976 (0.002) & – \\
    1600       & \textbf{0.985 (0.001)} & \textbf{0.985 (0.001)} & 0.980 (0.001) & – \\
    3200       & \textbf{0.987 (0.001)} & 0.986 (0.001) & 0.983 (0.001) & – \\
    6400       & \textbf{0.987 (0.001)} & \textbf{0.987 (0.001)} & 0.986 (0.001) & – \\

    \addlinespace
    \multicolumn{5}{l}{\textbf{Top 5\%}}\\
    $M=100$    & \textbf{0.905 (0.011)} & \textbf{0.905 (0.006)} & 0.860 (0.012) & 0.893 (0.007) \\
    200        & \textbf{0.942 (0.005)} & 0.934 (0.005) & 0.910 (0.007) & 0.928 (0.005) \\
    400        & \textbf{0.957 (0.004)} & 0.954 (0.004) & 0.932 (0.010) & – \\
    800        & \textbf{0.966 (0.002)} & 0.964 (0.002) & 0.957 (0.003) & – \\
    1600       & \textbf{0.971 (0.002)} & 0.970 (0.002) & 0.964 (0.002) & – \\
    3200       & \textbf{0.974 (0.001)} & \textbf{0.974 (0.001)} & 0.970 (0.002) & – \\
    6400       & \textbf{0.976 (0.001)} & \textbf{0.976 (0.001)} & 0.973 (0.001) & – \\

    \bottomrule
  \end{tabular}%
  \label{tab:pp_is}
\end{table}

\begin{table}[ht]
  \centering
  \caption{Out‐of‐sample predictive performance of the full Relational Event Models from the generated dependent Relational Event History (REH) data, quantified as the proportion of true events captured within the top 20\%, 10\%, and 5\% highest predicted‐probability events for training sample sizes \(M=100\)–6400. Predictive probabilities were calculated at each time \(t\) on an out-of-sample set of 1000 events generated beyond the training data, and coverage of the actual events by the highest‐probability subsets was evaluated. Four estimation approaches are compared: maximum likelihood estimation (MLE), approximate Bayesian regularization with Ridge (ABR Ridge) and Horseshoe (ABR HS) priors, and exact Bayesian regularization with Horseshoe prior (EBR HS). Results are reported as \textit{mean (Monte Carlo standard error)} over the 100 iterations in the simulation. Boldface highlights the best performance per threshold and sample size. A dash (–) indicates that a given configuration was not assessed.}

    \begin{tabular}{l c c c c}
    \toprule
    Sample Size ($M$)
      & \itshape MLE
      & \itshape ABR Ridge
      & \itshape ABR HS
      & \itshape EBR HS \\
    \midrule
    \multicolumn{5}{l}{\textbf{Top 20\%}}\\
    $M=100$    & 0.936 (0.013) & 0.988 (0.001) & 0.944 (0.014) & \textbf{0.989 (0.001)} \\
    200        & 0.983 (0.003) & \textbf{0.992 (0.000)} & 0.977 (0.004) & 0.990 (0.001) \\
    400        & 0.989 (0.002) & \textbf{0.992 (0.000)} & 0.978 (0.010) & – \\
    800        & 0.993 (0.000) & \textbf{0.994 (0.000)} & 0.990 (0.001) & – \\
    1600       & \textbf{0.994 (0.000)} & \textbf{0.994 (0.000)} & 0.992 (0.001) & – \\
    3200       & \textbf{0.994 (0.000)} & \textbf{0.994 (0.000)} & 0.993 (0.001) & – \\
    6400       & \textbf{0.995 (0.000)} & \textbf{0.995 (0.000)} & 0.994 (0.000) & – \\

    \addlinespace
    \multicolumn{5}{l}{\textbf{Top 10\%}}\\
    $M=100$    & 0.916 (0.015) & \textbf{0.978 (0.002)} & 0.928 (0.014) & \textbf{0.978 (0.002)} \\
    200        & 0.974 (0.003) & \textbf{0.983 (0.001)} & 0.965 (0.004) & 0.982 (0.002) \\
    400        & 0.981 (0.003) & \textbf{0.985 (0.001)} & 0.967 (0.010) & – \\
    800        & 0.986 (0.001) & \textbf{0.987 (0.001)} & 0.981 (0.002) & – \\
    1600       & \textbf{0.987 (0.001)} & \textbf{0.987 (0.001)} & 0.983 (0.001) & – \\
    3200       & \textbf{0.987 (0.001)} & \textbf{0.987 (0.001)} & 0.984 (0.001) & – \\
    6400       & \textbf{0.988 (0.001)} & \textbf{0.988 (0.001)} & 0.987 (0.001) & – \\

    \addlinespace
    \multicolumn{5}{l}{\textbf{Top 5\%}}\\
    $M=100$    & 0.884 (0.019) & \textbf{0.963 (0.003)} & 0.911 (0.015) & \textbf{0.963 (0.003)} \\
    200        & 0.960 (0.004) & \textbf{0.970 (0.002)} & 0.950 (0.005) & 0.968 (0.002) \\
    400        & 0.969 (0.003) & \textbf{0.973 (0.002)} & 0.952 (0.010) & – \\
    800        & \textbf{0.975 (0.001)} & \textbf{0.975 (0.001)} & 0.968 (0.002) & – \\
    1600       & \textbf{0.976 (0.001)} & 0.975 (0.001) & 0.969 (0.002) & – \\
    3200       & \textbf{0.976 (0.001)} & 0.975 (0.001) & 0.971 (0.002) & – \\
    6400       & \textbf{0.977 (0.001)} & \textbf{0.977 (0.001)} & 0.975 (0.001) & – \\

    \bottomrule
  \end{tabular}%
  \label{tab:pp_oos}
\end{table}

\begin{landscape}
\begin{table}[ht]
  \centering
  \caption{In‐sample predictive performance of sparse Relational Event Models from the generated dependent Relational Event History (REH) data including only predictors selected by the assessed criteria. Performance is defined as the proportion of true events included within the top 20\%, 10\%, and 5\% highest-predicted-probability events computed on all training events for sample sizes \(M=100\)–6400. Results are reported as \textit{mean (Monte Carlo standard error)} across 100 simulation iterations. 13 strategies are compared: (i) maximum likelihood estimation (MLE) via univariate selection at \(\alpha=0.05\); (ii) approximate Bayesian regularization (ABR) with Horseshoe (HS) and Ridge priors using posterior 95\% highest-density intervals (HDI) and 0.1 thresholding on posterior mode, median, and mean; and (iii) exact Bayesian regularization (EBR) with Horseshoe prior using analogous posterior summaries. Boldface highlights the best performance per threshold and sample size. A dash (–) indicates that a given configuration was not assessed.}

  \scriptsize
  \setlength\tabcolsep{2pt}
  \resizebox{1.6\textwidth}{!}{%
        \begin{tabular}{l r *{4}{c} *{4}{c} *{4}{c} c}
      \toprule
      & & \multicolumn{4}{c}{\itshape ABR HS}
        & \multicolumn{4}{c}{\itshape ABR Ridge}
        & \multicolumn{4}{c}{\itshape EBR HS}
        & \itshape MLE \\
      \cmidrule(lr){3-6}\cmidrule(lr){7-10}\cmidrule(lr){11-14}\cmidrule(lr){15-15}
      Sample Size ($M$)
        & & $95\%$ HDI & $|\mathrm{Mode}|\ge0.1$ & $|\mathrm{Median}|\ge0.1$ & $|\mathrm{Mean}|\ge0.1$
        & $95\%$ HDI  & $|\mathrm{Mode}|\ge0.1$ & $|\mathrm{Median}|\ge0.1$ & $|\mathrm{Mean}|\ge0.1$
        & $95\%$ HDI & $|\mathrm{Mode}|\ge0.1$ & $|\mathrm{Median}|\ge0.1$ & $|\mathrm{Mean}|\ge0.1$
        & $\alpha=0.05$ \\
      \midrule
      \multicolumn{15}{l}{\textbf{Top 20\%}}\\
      $M=100$    &   & 0.869 (0.018) & 0.923 (0.012) & 0.938 (0.011) & 0.938 (0.011)
                & \textbf{0.970 (0.002)} & \textbf{0.970 (0.002)} & \textbf{0.970 (0.002)} & \textbf{0.970 (0.002)}
                & 0.938 (0.007) & 0.962 (0.002) & 0.964 (0.002) & 0.965 (0.002)
                & 0.928 (0.013) \\
      200        &   & 0.922 (0.012) & 0.964 (0.004) & 0.967 (0.004) & 0.967 (0.004)
                & \textbf{0.982 (0.001)} & \textbf{0.982 (0.001)} & \textbf{0.982 (0.001)} & \textbf{0.982 (0.001)}
                & 0.970 (0.003) & 0.977 (0.002) & 0.978 (0.001) & 0.978 (0.001)
                & 0.948 (0.014) \\
      400        &   & 0.968 (0.010) & 0.973 (0.010) & 0.974 (0.009) & 0.974 (0.009)
                & \textbf{0.989 (0.001)} & \textbf{0.989 (0.001)} & \textbf{0.989 (0.001)} & \textbf{0.989 (0.001)}
                & – & – & – & –
                & 0.978 (0.008) \\
      800        &   & 0.976 (0.009) & 0.988 (0.001) & 0.988 (0.001) & 0.988 (0.001)
                & \textbf{0.992 (0.001)} & \textbf{0.992 (0.001)} & \textbf{0.992 (0.001)} & \textbf{0.992 (0.001)}
                & – & – & – & –
                & 0.990 (0.001) \\
      1600       &   & 0.988 (0.001) & 0.991 (0.001) & 0.991 (0.001) & 0.991 (0.001)
                & \textbf{0.993 (0.000)} & \textbf{0.993 (0.000)} & \textbf{0.993 (0.000)} & \textbf{0.993 (0.000)}
                & – & – & – & –
                & 0.977 (0.010) \\
      3200       &   & 0.991 (0.001) & 0.992 (0.001) & 0.992 (0.001) & 0.992 (0.001)
                & \textbf{0.994 (0.000)} & \textbf{0.994 (0.000)} & \textbf{0.994 (0.000)} & \textbf{0.994 (0.000)}
                & – & – & – & –
                & 0.981 (0.008) \\
      6400       &   & 0.993 (0.000) & \textbf{0.994 (0.000)} & \textbf{0.994 (0.000)} & \textbf{0.994 (0.000)}
                & \textbf{0.994 (0.000)} & \textbf{0.994 (0.000)} & \textbf{0.994 (0.000)} & \textbf{0.994 (0.000)}
                & – & – & – & –
                & 0.984 (0.008) \\
      \addlinespace
      \multicolumn{15}{l}{\textbf{Top 10\%}}\\
      $M=100$    &   & 0.835 (0.018) & 0.890 (0.012) & 0.906 (0.011) & 0.906 (0.011)
                & \textbf{0.942 (0.004)} & \textbf{0.942 (0.004)} & \textbf{0.942 (0.004)} & \textbf{0.942 (0.004)}
                & 0.907 (0.008) & 0.932 (0.004) & 0.935 (0.004) & 0.935 (0.004)
                & 0.878 (0.017) \\
      200        &   & 0.892 (0.012) & 0.941 (0.005) & 0.944 (0.005) & 0.944 (0.005)
                & \textbf{0.963 (0.003)} & \textbf{0.963 (0.003)} & \textbf{0.963 (0.003)} & \textbf{0.963 (0.003)}
                & 0.947 (0.004) & 0.958 (0.003) & 0.959 (0.003) & 0.959 (0.003)
                & 0.924 (0.014) \\
      400        &   & 0.948 (0.010) & 0.956 (0.010) & 0.957 (0.010) & 0.957 (0.010)
                & \textbf{0.976 (0.002)} & \textbf{0.976 (0.002)} & \textbf{0.976 (0.002)} & \textbf{0.976 (0.002)}
                & – & – & – & –
                & 0.962 (0.008) \\
      800        &   & 0.960 (0.009) & 0.975 (0.002) & 0.976 (0.002) & 0.976 (0.002)
                & \textbf{0.982 (0.001)} & \textbf{0.982 (0.001)} & \textbf{0.982 (0.001)} & \textbf{0.982 (0.001)}
                & – & – & – & –
                & 0.976 (0.002) \\
      1600       &   & 0.976 (0.002) & 0.980 (0.001) & 0.980 (0.001) & 0.980 (0.001)
                & \textbf{0.985 (0.001)} & \textbf{0.985 (0.001)} & \textbf{0.985 (0.001)} & \textbf{0.985 (0.001)}
                & – & – & – & –
                & 0.967 (0.011) \\
      3200       &   & 0.980 (0.001) & 0.983 (0.001) & 0.983 (0.001) & 0.983 (0.001)
                & \textbf{0.986 (0.001)} & \textbf{0.986 (0.001)} & \textbf{0.986 (0.001)} & \textbf{0.986 (0.001)}
                & – & – & – & –
                & 0.968 (0.011) \\
      6400       &   & 0.985 (0.001) & 0.986 (0.001) & 0.986 (0.001) & 0.986 (0.001)
                & \textbf{0.987 (0.001)} & \textbf{0.987 (0.001)} & \textbf{0.987 (0.001)} & \textbf{0.987 (0.001)}
                & – & – & – & –
                & 0.974 (0.008) \\
      \addlinespace
      \multicolumn{15}{l}{\textbf{Top 5\%}}\\
      $M=100$    &   & 0.786 (0.018) & 0.844 (0.012) & 0.860 (0.012) & 0.860 (0.012)
                & \textbf{0.905 (0.006)} & \textbf{0.905 (0.006)} & \textbf{0.905 (0.006)} & \textbf{0.905 (0.006)}
                & 0.862 (0.009) & 0.888 (0.007) & 0.893 (0.007) & 0.893 (0.007)
                & 0.817 (0.021) \\
      200        &   & 0.850 (0.013) & 0.907 (0.007) & 0.911 (0.007) & 0.911 (0.007)
                & \textbf{0.934 (0.005)} & \textbf{0.934 (0.005)} & \textbf{0.934 (0.005)} & \textbf{0.934 (0.005)}
                & 0.914 (0.006) & 0.927 (0.005) & 0.928 (0.005) & 0.927 (0.005)
                & 0.880 (0.017) \\
      400        &   & 0.921 (0.011) & 0.931 (0.010) & 0.932 (0.010) & 0.932 (0.010)
                & \textbf{0.954 (0.004)} & \textbf{0.954 (0.004)} & \textbf{0.954 (0.004)} & \textbf{0.954 (0.004)}
                & – & – & – & –
                & 0.936 (0.009) \\
      800        &   & 0.939 (0.010) & 0.956 (0.003) & 0.957 (0.003) & 0.957 (0.003)
                & \textbf{0.964 (0.002)} & \textbf{0.964 (0.002)} & \textbf{0.964 (0.002)} & \textbf{0.964 (0.002)}
                & – & – & – & –
                & 0.947 (0.009) \\
      1600       &   & 0.959 (0.002) & 0.964 (0.002) & 0.964 (0.002) & 0.964 (0.002)
                & \textbf{0.970 (0.002)} & \textbf{0.970 (0.002)} & \textbf{0.970 (0.002)} & \textbf{0.970 (0.002)}
                & – & – & – & –
                & 0.949 (0.011) \\
      3200       &   & 0.966 (0.002) & 0.969 (0.002) & 0.970 (0.002) & 0.970 (0.002)
                & \textbf{0.974 (0.001)} & \textbf{0.974 (0.001)} & \textbf{0.974 (0.001)} & \textbf{0.974 (0.001)}
                & – & – & – & –
                & 0.946 (0.013) \\
      6400       &   & 0.971 (0.001) & 0.973 (0.001) & 0.973 (0.001) & 0.973 (0.001)
                & \textbf{0.976 (0.001)} & \textbf{0.976 (0.001)} & \textbf{0.976 (0.001)} & \textbf{0.976 (0.001)}
                & – & – & – & –
                & 0.955 (0.010) \\
      \bottomrule
    \end{tabular}%
  }
  \label{tab:pp_is_sparse}
\end{table}
\end{landscape}

\begin{landscape}
\begin{table}[ht]
  \centering
  \caption{Out‐of‐sample predictive performance of sparse Relational Event Models from the generated dependent Relational Event History (REH) data including only predictors selected by the assessed criteria. Performance is defined as the proportion of true events included within the top 20\%, 10\%, and 5\% highest‐predicted‐probability events computed on an out-of-sample set of 1000 events for training sample sizes \(M=100\)–6400. Results are reported as \textit{mean (Monte Carlo standard error)} across 100 simulation iterations. Thirteen strategies are compared: (i) maximum likelihood estimation (MLE) via univariate selection at \(\alpha=0.05\); (ii) approximate Bayesian regularization (ABR) with Horseshoe (abbreviated as HS) and Ridge priors using posterior 95\% highest-density intervals (HDI) and 0.1 thresholding on posterior mode, median, and mean; and (iii) exact Bayesian regularization (EBR) with Horseshoe prior using analogous posterior summaries. Boldface highlights the best performance per threshold and sample size. A dash (–) indicates that a given configuration was not assessed.}

  \scriptsize
  \setlength\tabcolsep{2pt}
  \resizebox{1.6\textwidth}{!}{%
        \begin{tabular}{l r *{4}{c} *{4}{c} *{4}{c} c}
      \toprule
      & & \multicolumn{4}{c}{\itshape ABR HS}
        & \multicolumn{4}{c}{\itshape ABR Ridge}
        & \multicolumn{4}{c}{\itshape EBR HS}
        & \itshape MLE \\
      \cmidrule(lr){3-6}\cmidrule(lr){7-10}\cmidrule(lr){11-14}\cmidrule(lr){15-15}
      Sample Size ($M$)
        & & $95\%$ HDI & $|\mathrm{Mode}|\ge0.1$ & $|\mathrm{Median}|\ge0.1$ & $|\mathrm{Mean}|\ge0.1$
        & $95\%$ HDI  & $|\mathrm{Mode}|\ge0.1$ & $|\mathrm{Median}|\ge0.1$ & $|\mathrm{Mean}|\ge0.1$
        & $95\%$ HDI & $|\mathrm{Mode}|\ge0.1$ & $|\mathrm{Median}|\ge0.1$ & $|\mathrm{Mean}|\ge0.1$
        & $\alpha=0.05$ \\
      \midrule
      \multicolumn{15}{l}{\textbf{Top 20\%}}\\
      $M=100$    &   & 0.893 (0.021) & 0.940 (0.014) & 0.946 (0.014) & 0.946 (0.014)
                & 0.988 (0.001) & 0.988 (0.001) & 0.988 (0.001) & 0.988 (0.001)
                & 0.973 (0.005) & 0.988 (0.001) & \textbf{0.989 (0.001)} & \textbf{0.989 (0.001)}
                & 0.905 (0.020) \\
      200        &   & 0.933 (0.012) & 0.975 (0.004) & 0.977 (0.004) & 0.977 (0.004)
                & \textbf{0.992 (0.000)} & \textbf{0.992 (0.000)} & \textbf{0.992 (0.000)} & \textbf{0.992 (0.000)}
                & 0.984 (0.002) & 0.990 (0.001) & 0.991 (0.001) & 0.991 (0.001)
                & 0.954 (0.015) \\
      400        &   & 0.972 (0.010) & 0.977 (0.010) & 0.978 (0.010) & 0.978 (0.010)
                & \textbf{0.992 (0.000)} & \textbf{0.992 (0.000)} & \textbf{0.992 (0.000)} & \textbf{0.992 (0.000)}
                & – & – & – & –
                & 0.979 (0.009) \\
      800        &   & 0.977 (0.009) & 0.990 (0.001) & 0.990 (0.001) & 0.990 (0.001)
                & \textbf{0.994 (0.000)} & \textbf{0.994 (0.000)} & \textbf{0.994 (0.000)} & \textbf{0.994 (0.000)}
                & – & – & – & –
                & 0.991 (0.001) \\
      1600       &   & 0.988 (0.001) & 0.991 (0.001) & 0.992 (0.001) & 0.992 (0.001)
                & \textbf{0.994 (0.000)} & \textbf{0.994 (0.000)} & \textbf{0.994 (0.000)} & \textbf{0.994 (0.000)}
                & – & – & – & –
                & 0.979 (0.010) \\
      3200       &   & 0.991 (0.001) & 0.992 (0.001) & 0.993 (0.001) & 0.993 (0.001)
                & \textbf{0.994 (0.000)} & \textbf{0.994 (0.000)} & \textbf{0.994 (0.000)} & \textbf{0.994 (0.000)}
                & – & – & – & –
                & 0.982 (0.008) \\
      6400       &   & 0.994 (0.000) & 0.994 (0.000) & 0.994 (0.000) & 0.994 (0.000)
                & \textbf{0.995 (0.000)} & \textbf{0.995 (0.000)} & \textbf{0.995 (0.000)} & \textbf{0.995 (0.000)}
                & – & – & – & –
                & 0.984 (0.008) \\
      \addlinespace
      \multicolumn{15}{l}{\textbf{Top 10\%}}\\
      $M=100$    &   & 0.879 (0.021) & 0.925 (0.015) & 0.931 (0.014) & 0.931 (0.014)
                & \textbf{0.978 (0.002)} & \textbf{0.978 (0.002)} & \textbf{0.978 (0.002)} & \textbf{0.978 (0.002)}
                & 0.962 (0.006) & 0.977 (0.002) & \textbf{0.978 (0.002)} & \textbf{0.978 (0.002)}
                & 0.869 (0.025) \\
      200        &   & 0.917 (0.012) & 0.963 (0.005) & 0.965 (0.004) & 0.965 (0.004)
                & \textbf{0.983 (0.001)} & \textbf{0.983 (0.001)} & \textbf{0.983 (0.001)} & \textbf{0.983 (0.001)}
                & 0.972 (0.003) & 0.981 (0.002) & 0.981 (0.002) & 0.981 (0.002)
                & 0.937 (0.016) \\
      400        &   & 0.956 (0.011) & 0.966 (0.010) & 0.967 (0.010) & 0.967 (0.010)
                & \textbf{0.985 (0.001)} & \textbf{0.985 (0.001)} & \textbf{0.985 (0.001)} & \textbf{0.985 (0.001)}
                & – & – & – & –
                & 0.969 (0.009) \\
      800        &   & 0.965 (0.009) & 0.980 (0.002) & 0.981 (0.002) & 0.981 (0.002)
                & \textbf{0.987 (0.001)} & \textbf{0.987 (0.001)} & \textbf{0.987 (0.001)} & \textbf{0.987 (0.001)}
                & – & – & – & –
                & 0.981 (0.002) \\
      1600       &   & 0.977 (0.002) & 0.982 (0.001) & 0.983 (0.001) & 0.983 (0.001)
                & \textbf{0.987 (0.001)} & \textbf{0.987 (0.001)} & \textbf{0.987 (0.001)} & \textbf{0.987 (0.001)}
                & – & – & – & –
                & 0.969 (0.011) \\
      3200       &   & 0.981 (0.001) & 0.984 (0.001) & 0.984 (0.001) & 0.984 (0.001)
                & \textbf{0.987 (0.001)} & \textbf{0.987 (0.001)} & \textbf{0.987 (0.001)} & \textbf{0.987 (0.001)}
                & – & – & – & –
                & 0.968 (0.011) \\
      6400       &   & 0.985 (0.001) & 0.986 (0.001) & 0.987 (0.001) & 0.987 (0.001)
                & \textbf{0.988 (0.001)} & \textbf{0.988 (0.001)} & \textbf{0.988 (0.001)} & \textbf{0.988 (0.001)}
                & – & – & – & –
                & 0.974 (0.009) \\
      \addlinespace
      \multicolumn{15}{l}{\textbf{Top 5\%}}\\
      $M=100$    &   & 0.858 (0.021) & 0.907 (0.015) & 0.914 (0.014) & 0.914 (0.014)
                & \textbf{0.963 (0.003)} & \textbf{0.963 (0.003)} & \textbf{0.963 (0.003)} & \textbf{0.963 (0.003)}
                & 0.947 (0.006) & \textbf{0.963 (0.003)} & \textbf{0.963 (0.003)} & \textbf{0.963 (0.003)}
                & 0.840 (0.027) \\
      200        &   & 0.888 (0.013) & 0.948 (0.005) & 0.950 (0.005) & 0.950 (0.005)
                & \textbf{0.970 (0.002)} & \textbf{0.970 (0.002)} & \textbf{0.970 (0.002)} & \textbf{0.970 (0.002)}
                & 0.957 (0.004) & 0.967 (0.003) & 0.968 (0.003) & 0.968 (0.002)
                & 0.910 (0.018) \\
      400        &   & 0.940 (0.011) & 0.951 (0.010) & 0.952 (0.010) & 0.952 (0.010)
                & \textbf{0.973 (0.002)} & \textbf{0.973 (0.002)} & \textbf{0.973 (0.002)} & \textbf{0.973 (0.002)}
                & – & – & – & –
                & 0.953 (0.009) \\
      800        &   & 0.950 (0.010) & 0.968 (0.002) & 0.968 (0.002) & 0.968 (0.002)
                & \textbf{0.975 (0.001)} & \textbf{0.975 (0.001)} & \textbf{0.975 (0.001)} & \textbf{0.975 (0.001)}
                & – & – & – & –
                & 0.958 (0.009) \\
      1600       &   & 0.963 (0.003) & 0.969 (0.002) & 0.969 (0.002) & 0.969 (0.002)
                & \textbf{0.975 (0.001)} & \textbf{0.975 (0.001)} & \textbf{0.975 (0.001)} & \textbf{0.975 (0.001)}
                & – & – & – & –
                & 0.954 (0.011) \\
      3200       &   & 0.967 (0.002) & 0.971 (0.002) & 0.971 (0.002) & 0.971 (0.002)
                & \textbf{0.975 (0.001)} & \textbf{0.975 (0.001)} & \textbf{0.975 (0.001)} & \textbf{0.975 (0.001)}
                & – & – & – & –
                & 0.947 (0.013) \\
      6400       &   & 0.973 (0.002) & 0.975 (0.001) & 0.975 (0.001) & 0.975 (0.001)
                & \textbf{0.977 (0.001)} & \textbf{0.977 (0.001)} & \textbf{0.977 (0.001)} & \textbf{0.977 (0.001)}
                & – & – & – & –
                & 0.956 (0.010) \\
      \bottomrule
    \end{tabular}%
  }
  \label{tab:pp_oos_sparse}
\end{table}
\end{landscape}

\section{Independent Analyses} \label{app5}

\ref{app5} reports the results of the analyses on independently generated Relational Event History (REH) data. Here, instead of generating a large data set and consequently subsetting to the needed sample sizes, data sets are generated independently. This serves as a robustness check for the results from the dependent data.

\subsection{Variable Selection} \label{varsel_ind}

\ref{varsel_ind} provides the evaluation of the variable selection on the generated independent relational event history (REH) data. Below, true discovery rates, false discovery rates, distance metrics, and Matthews' Correlation Coefficients are reported as means across iterations in the simulation, together with Monte Carlo standard errors.

\begin{landscape}
\begin{table}[ht]
  \centering
  \caption{True Discovery Rates (TDR) from the generated independent Relational Event History (REH) data, defined as the proportion of truly nonzero effects correctly identified, for endogenous and exogenous effects (weak, moderate, strong) across numbers of events \(M\) (100–6400). Results are reported as \textit{mean (Monte Carlo standard error)} over the 100 iterations in the simulation. Nine selection methods are compared: (i) maximum likelihood estimation (MLE) via univariate selection at \(\alpha=0.05\) and (ii) approximate Bayesian regularization (ABR) with Horseshoe (abbreviated as HS) and Ridge priors using posterior 95\% highest-density intervals (HDI) and 0.1 thresholding on posterior mean, median, and mode. Boldface highlights the highest TDR in each row.}
  \scriptsize
  \setlength\tabcolsep{2pt}
  \resizebox{1.6\textwidth}{!}{%
        \begin{tabular}{l *{9}{c}}
      \toprule
      \multicolumn{1}{c}{} 
        & \multicolumn{4}{c}{\itshape ABR HS} 
        & \multicolumn{4}{c}{\itshape ABR Ridge} 
        & \multicolumn{1}{c}{\itshape MLE} \\
      \cmidrule(lr){2-5}\cmidrule(lr){6-9}
      Sample Size ($M$)
        & $95\%$ HDI 
        & $|\mathrm{Mean}|\ge 0.1$ 
        & $|\mathrm{Median}|\ge 0.1$ 
        & $|\mathrm{Mode}|\ge 0.1$
        & $95\%$ HDI 
        & $|\mathrm{Mean}|\ge 0.1$ 
        & $|\mathrm{Median}|\ge 0.1$ 
        & $|\mathrm{Mode}|\ge 0.1$ 
        & $\alpha=0.05$ \\
      \midrule
      \multicolumn{10}{l}{\textbf{Endogenous}}\\
      $M=100$    & 0.470 (0.024) & 0.853 (0.020) & 0.853 (0.020) & 0.572 (0.019)
                & 0.670 (0.019) & \textbf{0.954 (0.012)} & \textbf{0.954 (0.012)} & 0.947 (0.013)
                & 0.547 (0.024)\\
      200        & 0.563 (0.020) & 0.867 (0.019) & 0.867 (0.019) & 0.653 (0.020)
                & 0.683 (0.017) & \textbf{0.950 (0.012)} & \textbf{0.950 (0.012)} & 0.947 (0.012)
                & 0.623 (0.018)\\
      400        & 0.590 (0.021) & 0.867 (0.018) & 0.867 (0.018) & 0.677 (0.017)
                & 0.683 (0.014) & 0.947 (0.012) & 0.947 (0.012) & \textbf{0.957 (0.011)}
                & 0.657 (0.016)\\
      800        & 0.620 (0.018) & 0.870 (0.018) & 0.870 (0.018) & 0.693 (0.020)
                & 0.707 (0.014) & \textbf{0.943 (0.013)} & \textbf{0.943 (0.013)} & 0.940 (0.013)
                & 0.697 (0.018)\\
      1600       & 0.680 (0.016) & 0.907 (0.018) & 0.907 (0.018) & 0.760 (0.020)
                & 0.717 (0.013) & \textbf{0.953 (0.013)} & \textbf{0.953 (0.013)} & \textbf{0.953 (0.013)}
                & 0.703 (0.016)\\
      3200       & 0.727 (0.019) & 0.957 (0.011) & 0.957 (0.011) & 0.820 (0.019)
                & 0.760 (0.015) & 0.967 (0.010) & 0.967 (0.010) & \textbf{0.973 (0.009)}
                & 0.737 (0.017)\\
      6400       & 0.743 (0.018) & 0.950 (0.013) & 0.950 (0.013) & 0.817 (0.020)
                & 0.743 (0.015) & 0.960 (0.011) & 0.960 (0.011) & \textbf{0.963 (0.010)}
                & 0.703 (0.015)\\
      \addlinespace
      \multicolumn{10}{l}{\textbf{Exogenous}}\\
      \multicolumn{10}{l}{\itshape weak}\\
      $M=100$    & 0.000 (0.000) & 0.319 (0.029) & 0.319 (0.029) & 0.025 (0.009)
                & 0.056 (0.013) & \textbf{0.677 (0.023)} & \textbf{0.677 (0.023)} & 0.667 (0.025)
                & 0.144 (0.022)\\
      200        & 0.020 (0.008) & 0.333 (0.026) & 0.333 (0.026) & 0.047 (0.015)
                & 0.093 (0.018) & \textbf{0.690 (0.025)} & \textbf{0.690 (0.025)} & 0.683 (0.024)
                & 0.177 (0.024)\\
      400        & 0.033 (0.011) & 0.350 (0.029) & 0.350 (0.029) & 0.087 (0.017)
                & 0.147 (0.021) & 0.683 (0.024) & 0.683 (0.024) & \textbf{0.687 (0.025)}
                & 0.213 (0.025)\\
      800        & 0.097 (0.018) & 0.460 (0.027) & 0.460 (0.027) & 0.213 (0.021)
                & 0.240 (0.023) & \textbf{0.700 (0.024)} & \textbf{0.700 (0.024)} & \textbf{0.700 (0.023)}
                & 0.323 (0.025)\\
      1600       & 0.210 (0.025) & 0.530 (0.029) & 0.530 (0.029) & 0.387 (0.029)
                & 0.347 (0.030) & \textbf{0.753 (0.027)} & \textbf{0.753 (0.027)} & 0.750 (0.027)
                & 0.450 (0.031)\\
      3200       & 0.537 (0.028) & 0.727 (0.025) & 0.727 (0.025) & 0.677 (0.025)
                & 0.680 (0.027) & \textbf{0.857 (0.017)} & \textbf{0.857 (0.017)} & \textbf{0.857 (0.018)}
                & 0.753 (0.023)\\
      6400       & 0.803 (0.025) & 0.860 (0.020) & 0.860 (0.020) & 0.877 (0.019)
                & 0.887 (0.018) & 0.943 (0.013) & 0.943 (0.013) & \textbf{0.953 (0.012)}
                & 0.927 (0.015)\\
      \addlinespace
      \multicolumn{10}{l}{\itshape moderate}\\
      $M=100$    & 0.032 (0.010) & 0.530 (0.034) & 0.530 (0.034) & 0.109 (0.020)
                & 0.200 (0.025) & \textbf{0.884 (0.019)} & \textbf{0.884 (0.019)} & 0.877 (0.018)
                & 0.305 (0.029)\\
      200        & 0.107 (0.019) & 0.683 (0.029) & 0.683 (0.029) & 0.200 (0.029)
                & 0.353 (0.026) & \textbf{0.910 (0.017)} & \textbf{0.910 (0.017)} & 0.903 (0.017)
                & 0.383 (0.032)\\
      400        & 0.243 (0.028) & 0.893 (0.020) & 0.893 (0.020) & 0.403 (0.035)
                & 0.570 (0.028) & 0.963 (0.012) & 0.963 (0.012) & \textbf{0.967 (0.010)}
                & 0.593 (0.032)\\
      800        & 0.620 (0.034) & 0.947 (0.014) & 0.947 (0.014) & 0.720 (0.031)
                & 0.833 (0.023) & \textbf{0.993 (0.005)} & \textbf{0.993 (0.005)} & \textbf{0.993 (0.005)}
                & 0.867 (0.022)\\
      1600       & 0.903 (0.020) & 0.993 (0.005) & 0.993 (0.005) & 0.937 (0.016)
                & 0.970 (0.010) & \textbf{1.000 (0.000)} & \textbf{1.000 (0.000)} & \textbf{1.000 (0.000)}
                & 0.980 (0.008)\\
      3200       & 0.993 (0.005) & \textbf{1.000 (0.000)} & \textbf{1.000 (0.000)} & 0.997 (0.003)
                & 0.997 (0.003) & \textbf{1.000 (0.000)} & \textbf{1.000 (0.000)} & \textbf{1.000 (0.000)}
                & 0.997 (0.003)\\
      6400       & \textbf{1.000 (0.000)} & \textbf{1.000 (0.000)} & \textbf{1.000 (0.000)} & \textbf{1.000 (0.000)}
                & \textbf{1.000 (0.000)} & \textbf{1.000 (0.000)} & \textbf{1.000 (0.000)} & \textbf{1.000 (0.000)}
                & \textbf{1.000 (0.000)}\\
      \addlinespace
      \multicolumn{10}{l}{\itshape strong}\\
      $M=100$    & 0.347 (0.034) & 0.898 (0.022) & 0.898 (0.022) & 0.446 (0.034)
                & 0.604 (0.038) & \textbf{0.975 (0.009)} & \textbf{0.975 (0.009)} & 0.972 (0.010)
                & 0.674 (0.032)\\
      200        & 0.520 (0.034) & 0.967 (0.010) & 0.967 (0.010) & 0.623 (0.031)
                & 0.783 (0.026) & \textbf{0.983 (0.007)} & \textbf{0.983 (0.007)} & 0.980 (0.008)
                & 0.773 (0.025)\\
      400        & 0.693 (0.026) & 0.977 (0.009) & 0.977 (0.009) & 0.777 (0.026)
                & 0.890 (0.020) & \textbf{0.990 (0.006)} & \textbf{0.990 (0.006)} & \textbf{0.990 (0.006)}
                & 0.890 (0.020)\\
      800        & 0.883 (0.018) & 0.993 (0.005) & 0.993 (0.005) & 0.897 (0.018)
                & 0.950 (0.012) & \textbf{1.000 (0.000)} & \textbf{1.000 (0.000)} & \textbf{1.000 (0.000)}
                & 0.977 (0.009)\\
      1600       & 0.963 (0.010) & \textbf{1.000 (0.000)} & \textbf{1.000 (0.000)} & 0.973 (0.009)
                & 0.990 (0.006) & \textbf{1.000 (0.000)} & \textbf{1.000 (0.000)} & \textbf{1.000 (0.000)}
                & 0.997 (0.003)\\
      3200       & 0.980 (0.008) & \textbf{1.000 (0.000)} & \textbf{1.000 (0.000)} & 0.987 (0.007)
                & \textbf{1.000 (0.000)} & \textbf{1.000 (0.000)} & \textbf{1.000 (0.000)} & \textbf{1.000 (0.000)}
                & \textbf{1.000 (0.000)}\\
      6400       & \textbf{1.000 (0.000)} & \textbf{1.000 (0.000)} & \textbf{1.000 (0.000)} & \textbf{1.000 (0.000)}
                & \textbf{1.000 (0.000)} & \textbf{1.000 (0.000)} & \textbf{1.000 (0.000)} & \textbf{1.000 (0.000)}
                & \textbf{1.000 (0.000)}\\
      \bottomrule
    \end{tabular}%
  }
  \label{tab:tdr_ind}
\end{table}
\end{landscape}

\begin{landscape}
\begin{table}[ht]
  \centering
  \caption{False Discovery Rates (FDR) from the generated independent Relational Event History (REH) data, defined as the proportion of truly zero effects incorrectly identified as nonzero, for endogenous and exogenous effects across numbers of events \(M\) (100–6400). Results are reported as \textit{mean (Monte Carlo standard error)} over the 100 iterations in the simulation. Nine selection methods are compared: (i) maximum likelihood estimation (MLE) via univariate selection at \(\alpha=0.05\) and (ii) approximate Bayesian regularization (ABR) with Horseshoe (abbreviated as HS) and Ridge priors using posterior 95\% highest-density intervals (HDI) and 0.1 thresholding on posterior mean, median, and mode. Boldface highlights the lowest FDR in each row.}
  \scriptsize
  \setlength\tabcolsep{2pt}
  \resizebox{1.6\textwidth}{!}{%
        \begin{tabular}{l r *{4}{c} *{4}{c} c}
      \toprule
      \multicolumn{1}{c}{} 
        & \multicolumn{1}{c}{} 
        & \multicolumn{4}{c}{\itshape ABR HS} 
        & \multicolumn{4}{c}{\itshape ABR Ridge} 
        & \multicolumn{1}{c}{\itshape MLE} \\
      \cmidrule(lr){3-6}\cmidrule(lr){7-10}
      Sample Size ($M$) 
        & & $95\%$ HDI 
            & $|\mathrm{Mean}|\ge0.1$ 
            & $|\mathrm{Median}|\ge0.1$ 
            & $|\mathrm{Mode}|\ge0.1$
        & $95\%$ HDI 
            & $|\mathrm{Mean}|\ge0.1$ 
            & $|\mathrm{Median}|\ge0.1$ 
            & $|\mathrm{Mode}|\ge0.1$
        & $\alpha=0.05$ \\
      \midrule
      \multicolumn{11}{l}{\textbf{Endogenous}}\\
      $M=100$    &  & 0.154 (0.012) & 0.480 (0.013) & 0.480 (0.013) & 0.203 (0.012)
                & \textbf{0.134 (0.005)} & 0.708 (0.011) & 0.708 (0.011) & 0.705 (0.011)
                & 0.384 (0.007) \\
      200        &  & \textbf{0.097 (0.010)} & 0.406 (0.015) & 0.406 (0.015) & 0.140 (0.010)
                & 0.128 (0.005) & 0.643 (0.010) & 0.643 (0.010) & 0.639 (0.010)
                & 0.378 (0.007) \\
      400        &  & \textbf{0.079 (0.009)} & 0.336 (0.013) & 0.336 (0.013) & 0.100 (0.009)
                & 0.118 (0.005) & 0.616 (0.010) & 0.616 (0.010) & 0.612 (0.009)
                & 0.383 (0.007) \\
      800        &  & \textbf{0.037 (0.003)} & 0.212 (0.011) & 0.212 (0.011) & 0.055 (0.004)
                & 0.100 (0.004) & 0.570 (0.008) & 0.570 (0.008) & 0.554 (0.009)
                & 0.373 (0.006) \\
      1600       &  & \textbf{0.031 (0.003)} & 0.181 (0.007) & 0.181 (0.007) & 0.043 (0.003)
                & 0.083 (0.005) & 0.539 (0.007) & 0.539 (0.007) & 0.528 (0.008)
                & 0.371 (0.006) \\
      3200       &  & \textbf{0.020 (0.003)} & 0.128 (0.006) & 0.128 (0.006) & 0.029 (0.003)
                & 0.063 (0.004) & 0.496 (0.009) & 0.496 (0.009) & 0.478 (0.008)
                & 0.363 (0.005) \\
      6400       &  & \textbf{0.017 (0.002)} & 0.110 (0.006) & 0.110 (0.006) & 0.020 (0.003)
                & 0.048 (0.005) & 0.471 (0.008) & 0.471 (0.008) & 0.452 (0.008)
                & 0.364 (0.006) \\
      \addlinespace
      \multicolumn{11}{l}{\textbf{Exogenous}}\\
      $M=100$    &  & \textbf{0.001 (0.001)} & 0.217 (0.012) & 0.217 (0.012) & 0.006 (0.002)
                & 0.020 (0.004) & 0.620 (0.015) & 0.620 (0.015) & 0.624 (0.015)
                & 0.106 (0.010) \\
      200        &  & \textbf{0.002 (0.001)} & 0.173 (0.012) & 0.173 (0.012) & 0.012 (0.003)
                & 0.032 (0.004) & 0.559 (0.014) & 0.559 (0.014) & 0.557 (0.013)
                & 0.079 (0.009) \\
      400        &  & \textbf{0.003 (0.001)} & 0.139 (0.010) & 0.139 (0.010) & 0.013 (0.003)
                & 0.029 (0.005) & 0.484 (0.012) & 0.484 (0.012) & 0.481 (0.013)
                & 0.066 (0.007) \\
      800        &  & \textbf{0.003 (0.001)} & 0.089 (0.008) & 0.089 (0.008) & 0.014 (0.003)
                & 0.039 (0.005) & 0.388 (0.013) & 0.388 (0.013) & 0.387 (0.012)
                & 0.069 (0.006) \\
      1600       &  & \textbf{0.004 (0.002)} & 0.051 (0.006) & 0.051 (0.006) & 0.015 (0.003)
                & 0.031 (0.005) & 0.260 (0.014) & 0.260 (0.014) & 0.254 (0.014)
                & 0.053 (0.006) \\
      3200       &  & \textbf{0.005 (0.002)} & 0.021 (0.004) & 0.021 (0.004) & 0.011 (0.003)
                & 0.044 (0.006) & 0.150 (0.011) & 0.150 (0.011) & 0.151 (0.011)
                & 0.067 (0.007) \\
      6400       &  & \textbf{0.004 (0.002)} & 0.009 (0.003) & 0.009 (0.003) & 0.006 (0.002)
                & 0.039 (0.005) & 0.069 (0.007) & 0.069 (0.007) & 0.071 (0.007)
                & 0.051 (0.006) \\
      \bottomrule
    \end{tabular}%
  }
  \label{tab:fdr_ind}
\end{table}
\end{landscape}

\begin{landscape}
\begin{table}[ht]
  \centering
  \caption{Distance metrics from the generated independent Relational Event History (REH) data, quantifying the Euclidean distance from perfect classification (TDR = 1, FDR = 0) for endogenous and exogenous effects across numbers of events \(M\) (100–6400). Results are reported as \textit{mean (Monte Carlo standard error)} over the 100 iterations in the simulation. Nine selection methods are compared: (i) maximum likelihood estimation (MLE) via univariate selection at \(\alpha=0.05\) and (ii) approximate Bayesian regularization (ABR) with Horseshoe (abbreviated as HS) and Ridge priors using posterior 95\% highest-density intervals (HDI) and 0.1 thresholding on posterior mean, median, and mode. Boldface highlights the lowest distance in each row.}
  \scriptsize
  \setlength\tabcolsep{2pt}
  \resizebox{1.6\textwidth}{!}{%
        \begin{tabular}{l r *{4}{c} *{4}{c} c}
      \toprule
      & & \multicolumn{4}{c}{\itshape ABR HS}
        & \multicolumn{4}{c}{\itshape ABR Ridge}
        & \itshape MLE \\
      \cmidrule(lr){3-6}\cmidrule(lr){7-10}
      Sample Size ($M$)
        & & $95\%$ HDI 
            & $|\mathrm{Mean}|\ge0.1$ 
            & $|\mathrm{Median}|\ge0.1$ 
            & $|\mathrm{Mode}|\ge0.1$
        & $95\%$ HDI  
            & $|\mathrm{Mean}|\ge0.1$ 
            & $|\mathrm{Median}|\ge0.1$ 
            & $|\mathrm{Mode}|\ge0.1$
        & $\alpha=0.05$ \\
      \midrule
      \multicolumn{11}{l}{\textbf{Endogenous}}\\
      $M=100$    &   & 0.570 (0.023) & 0.536 (0.013) & 0.536 (0.013) & 0.499 (0.016)
                & \textbf{0.371 (0.016)} & 0.718 (0.011) & 0.718 (0.011) & 0.717 (0.011)
                & 0.620 (0.017) \\
      200        &   & 0.462 (0.019) & 0.462 (0.016) & 0.462 (0.016) & 0.401 (0.017)
                & \textbf{0.359 (0.014)} & 0.655 (0.010) & 0.655 (0.010) & 0.653 (0.010)
                & 0.551 (0.014) \\
      400        &   & 0.428 (0.021) & 0.395 (0.016) & 0.395 (0.016) & 0.354 (0.017)
                & \textbf{0.351 (0.011)} & 0.629 (0.010) & 0.629 (0.010) & 0.623 (0.010)
                & 0.532 (0.011) \\
      800        &   & 0.386 (0.017) & \textbf{0.285 (0.015)} & \textbf{0.285 (0.015)} & 0.320 (0.019)
                & 0.322 (0.011) & 0.585 (0.009) & 0.585 (0.009) & 0.571 (0.010)
                & 0.505 (0.010) \\
      1600       &   & 0.325 (0.016) & \textbf{0.238 (0.015)} & \textbf{0.238 (0.015)} & 0.256 (0.019)
                & 0.304 (0.012) & 0.554 (0.008) & 0.554 (0.008) & 0.543 (0.009)
                & 0.498 (0.009) \\
      3200       &   & 0.279 (0.019) & \textbf{0.155 (0.010)} & \textbf{0.155 (0.010)} & 0.194 (0.018)
                & 0.266 (0.012) & 0.505 (0.010) & 0.505 (0.010) & 0.485 (0.009)
                & 0.476 (0.008) \\
      6400       &   & 0.261 (0.018) & \textbf{0.143 (0.012)} & \textbf{0.143 (0.012)} & 0.193 (0.019)
                & 0.272 (0.014) & 0.484 (0.009) & 0.484 (0.009) & 0.463 (0.009)
                & 0.488 (0.009) \\
      \addlinespace
      \multicolumn{11}{l}{\textbf{Exogenous}}\\
      $M=100$    &   & 0.874 (0.013) & \textbf{0.493 (0.017)} & \textbf{0.493 (0.017)} & 0.807 (0.017)
                & 0.715 (0.019) & 0.647 (0.015) & 0.647 (0.015) & 0.652 (0.015)
                & 0.641 (0.018) \\
      200        &   & 0.785 (0.017) & \textbf{0.403 (0.014)} & \textbf{0.403 (0.014)} & 0.711 (0.020)
                & 0.593 (0.017) & 0.585 (0.014) & 0.585 (0.014) & 0.585 (0.014)
                & 0.569 (0.019) \\
      400        &   & 0.677 (0.017) & \textbf{0.316 (0.013)} & \textbf{0.316 (0.013)} & 0.579 (0.020)
                & 0.469 (0.015) & 0.507 (0.012) & 0.507 (0.012) & 0.504 (0.013)
                & 0.447 (0.016) \\
      800        &   & 0.467 (0.019) & \textbf{0.233 (0.011)} & \textbf{0.233 (0.011)} & 0.391 (0.019)
                & 0.332 (0.015) & 0.410 (0.013) & 0.410 (0.013) & 0.409 (0.012)
                & 0.297 (0.012) \\
      1600       &   & 0.308 (0.013) & \textbf{0.181 (0.010)} & \textbf{0.181 (0.010)} & 0.238 (0.014)
                & 0.240 (0.011) & 0.286 (0.014) & 0.286 (0.014) & 0.280 (0.014)
                & 0.212 (0.011) \\
      3200       &   & 0.165 (0.010) & \textbf{0.102 (0.008)} & \textbf{0.102 (0.008)} & 0.116 (0.009)
                & 0.130 (0.009) & 0.167 (0.011) & 0.167 (0.011) & 0.168 (0.011)
                & 0.126 (0.008) \\
      6400       &   & 0.067 (0.008) & 0.053 (0.007) & 0.053 (0.007) & \textbf{0.045 (0.006)}
                & 0.068 (0.007) & 0.082 (0.007) & 0.082 (0.007) & 0.082 (0.007)
                & 0.068 (0.007) \\
      \bottomrule
    \end{tabular}%
  }
  \label{tab:dist_ind}
\end{table}
\end{landscape}

\begin{landscape}
\begin{table}[ht]
  \centering
  \caption{Matthews correlation coefficient (MCC) from the generated independent Relational Event History (REH) data, quantifying the agreement between true and selected nonzero effects for endogenous and exogenous effects across numbers of events \(M\) (100–6400). Results are reported as \textit{mean (Monte Carlo standard error)} over the 100 iterations in the simulation. Nine selection methods are compared: (i) maximum likelihood estimation (MLE) via univariate selection at \(\alpha=0.05\) and (ii) approximate Bayesian regularization (ABR) with Horseshoe (abbreviated as HS) and Ridge priors using posterior 95\% highest-density intervals (HDI) and 0.1 thresholding on posterior mean, median, and mode. Boldface highlights the highest MCC in each row.}
  \scriptsize
  \setlength\tabcolsep{2pt}
  \resizebox{1.6\textwidth}{!}{%
        \begin{tabular}{l r *{4}{c} *{4}{c} c}
      \toprule
      & & \multicolumn{4}{c}{\itshape ABR HS}
        & \multicolumn{4}{c}{\itshape ABR Ridge}
        & \itshape MLE \\
      \cmidrule(lr){3-6}\cmidrule(lr){7-10}
      Sample Size ($M$)
        & & $95\%$ HDI
            & $|\mathrm{Mean}|\ge0.1$
            & $|\mathrm{Median}|\ge0.1$
            & $|\mathrm{Mode}|\ge0.1$
        & $95\%$ HDI
            & $|\mathrm{Mean}|\ge0.1$
            & $|\mathrm{Median}|\ge0.1$
            & $|\mathrm{Mode}|\ge0.1$
        & $\alpha=0.05$ \\
      \midrule
      \multicolumn{11}{l}{\textbf{Endogenous}}\\
      $M=100$    &   & 0.299 (0.024) & 0.244 (0.015) & 0.244 (0.015) & 0.305 (0.018)
                & \textbf{0.439 (0.017)} & 0.179 (0.011) & 0.179 (0.011) & 0.175 (0.011)
                & 0.106 (0.016) \\
      200        &   & \textbf{0.466 (0.024)} & 0.309 (0.017) & 0.309 (0.017) & 0.445 (0.021)
                & 0.457 (0.014) & 0.212 (0.010) & 0.212 (0.010) & 0.212 (0.010)
                & 0.161 (0.013) \\
      400        &   & 0.526 (0.024) & 0.362 (0.017) & 0.362 (0.017) & \textbf{0.534 (0.021)}
                & 0.477 (0.012) & 0.223 (0.010) & 0.223 (0.010) & 0.232 (0.010)
                & 0.178 (0.011) \\
      800        &   & \textbf{0.619 (0.017)} & 0.485 (0.018) & 0.485 (0.018) & 0.617 (0.019)
                & 0.526 (0.013) & 0.246 (0.010) & 0.246 (0.010) & 0.253 (0.011)
                & 0.211 (0.011) \\
      1600       &   & 0.683 (0.015) & 0.539 (0.017) & 0.539 (0.017) & \textbf{0.697 (0.018)}
                & 0.573 (0.016) & 0.271 (0.009) & 0.271 (0.009) & 0.277 (0.009)
                & 0.217 (0.010) \\
      3200       &   & 0.752 (0.017) & 0.649 (0.014) & 0.649 (0.014) & \textbf{0.785 (0.017)}
                & 0.653 (0.013) & 0.305 (0.010) & 0.305 (0.010) & 0.320 (0.009)
                & 0.244 (0.010) \\
      6400       &   & 0.779 (0.016) & 0.682 (0.016) & 0.682 (0.016) & \textbf{0.814 (0.017)}
                & 0.689 (0.017) & 0.315 (0.009) & 0.315 (0.009) & 0.330 (0.009)
                & 0.222 (0.010) \\
      \addlinespace
      \multicolumn{11}{l}{\textbf{Exogenous}}\\
      $M=100$    &   & 0.361 (0.009) & 0.379 (0.021) & 0.379 (0.021) & 0.389 (0.014)
                & \textbf{0.426 (0.018)} & 0.235 (0.020) & 0.235 (0.020) & 0.225 (0.019)
                & 0.330 (0.024) \\
      200        &   & 0.401 (0.014) & \textbf{0.505 (0.017)} & \textbf{0.505 (0.017)} & 0.427 (0.017)
                & 0.481 (0.016) & 0.312 (0.018) & 0.312 (0.018) & 0.308 (0.018)
                & 0.433 (0.023) \\
      400        &   & 0.462 (0.014) & \textbf{0.614 (0.017)} & \textbf{0.614 (0.017)} & 0.526 (0.016)
                & 0.596 (0.014) & 0.400 (0.014) & 0.400 (0.014) & 0.404 (0.015)
                & 0.561 (0.018) \\
      800        &   & 0.635 (0.015) & \textbf{0.725 (0.016)} & \textbf{0.725 (0.016)} & 0.677 (0.015)
                & 0.691 (0.014) & 0.505 (0.015) & 0.505 (0.015) & 0.505 (0.014)
                & 0.688 (0.013) \\
      1600       &   & 0.758 (0.011) & \textbf{0.809 (0.012)} & \textbf{0.809 (0.012)} & 0.798 (0.013)
                & 0.779 (0.011) & 0.646 (0.018) & 0.646 (0.018) & 0.651 (0.018)
                & 0.780 (0.013) \\
      3200       &   & 0.867 (0.008) & \textbf{0.901 (0.009)} & \textbf{0.901 (0.009)} & 0.896 (0.008)
                & 0.858 (0.011) & 0.788 (0.014) & 0.788 (0.014) & 0.787 (0.014)
                & 0.850 (0.011) \\
      6400       &   & 0.944 (0.007) & 0.951 (0.006) & 0.951 (0.006) & \textbf{0.960 (0.006)}
                & 0.922 (0.008) & 0.902 (0.008) & 0.902 (0.008) & 0.902 (0.008)
                & 0.917 (0.008) \\
      \bottomrule
    \end{tabular}
  }
  \label{tab:mcc_ind}
\end{table}
\end{landscape}

\subsection{Bias and Variance}

\begin{table}[ht]
    \centering
    \caption{Bias and variance of parameter estimates from the generated independent Relational Event History (REH) data, where bias is the average deviation of estimates from the true effects and variance is the variance of this deviation over the 100 iterations in the simulation, for sample sizes \(M=100\)–6400  after the removal of participation shifts. Results are reported as \textit{bias (variance)}. Three estimation methods are compared: maximum likelihood estimation (MLE) and approximate Bayesian regularization (ABR) with Horseshoe (HS) and Ridge priors. A dash (–) indicates that the method was not assessed at the given sample size.}
\begin{tabular}{rccc}
\toprule
Sample Size ($M$) & \itshape MLE & \itshape ABR HS & \itshape ABR Ridge\\
\midrule
100 & -0.117 (3.50) & 0.072 (1.28) & 0.194 (1.78)\\
200 & -0.088 (0.74) & 0.094 (1.43) & 0.187 (1.70)\\
400 & -0.084 (0.59) & 0.111 (1.51) & 0.185 (1.65)\\
800 & -0.064 (0.33) & 0.167 (1.49) & 0.186 (1.58)\\
1600 & -0.062 (0.25) & 0.176 (1.48) & 0.181 (1.53)\\
3200 & -0.062 (0.23) & 0.179 (1.41) & 0.180 (1.43)\\
6400 & -0.064 (0.25) & 0.181 (1.40) & 0.180 (1.37)\\
\bottomrule
\end{tabular}
    \label{tab:biasvar_independent}
\end{table}

\newpage

\subsubsection{Predictive Performance}

\begin{table}[ht]
  \centering
  \caption{In-sample predictive performance of the full Relational Event Models from the generated independent Relational Event History (REH) data, quantified as the proportion of true events captured within the top 20\%, 10\%, and 5\% highest predicted-probability events for training sample sizes \(M=100\)–6400. Event probabilities were calculated at each time \(t\) on the training set, and coverage of the actual events by the highest-probability subsets was evaluated. Three estimation approaches are compared: maximum likelihood estimation (MLE), approximate Bayesian regularization with Ridge (ABR Ridge) and Horseshoe (ABR HS) priors. Results are reported as \textit{mean (Monte Carlo standard error)} over the 100 iterations in the simulation. Boldface highlights the best performance per threshold and sample size.}
    \begin{tabular}{l c c c}
    \toprule
    \textbf{Threshold / \(M\)}
      & \itshape MLE
      & \itshape ABR Ridge
      & \itshape ABR HS \\
    \midrule
    \multicolumn{4}{l}{\textbf{Top 20\%}}\\
    \(M=100\)        & \(\mathbf{0.961 (0.003)}\) & \(0.945 (0.003)\) & \(0.897 (0.013)\) \\
    200              & \(\mathbf{0.971 (0.002)}\) & \(0.966 (0.002)\) & \(0.918 (0.013)\) \\
    400              & \(\mathbf{0.984 (0.001)}\) & \(0.982 (0.001)\) & \(0.957 (0.013)\) \\
    800              & \(\mathbf{0.989 (0.001)}\) & \(0.988 (0.001)\) & \(0.984 (0.001)\) \\
    1600             & \(\mathbf{0.992 (0.000)}\) & \(0.991 (0.000)\) & \(0.990 (0.001)\) \\
    3200             & \(\mathbf{0.993 (0.000)}\) & \(\mathbf{0.993 (0.000)}\) & \(\mathbf{0.993 (0.000)}\) \\
    6400             & \(\mathbf{0.995 (0.000)}\) & \(\mathbf{0.995 (0.000)}\) & \(0.994 (0.000)\) \\

    \addlinespace
    \multicolumn{4}{l}{\textbf{Top 10\%}}\\
    \(M=100\)        & \(\mathbf{0.920 (0.006)}\) & \(0.902 (0.006)\) & \(0.835 (0.013)\) \\
    200              & \(\mathbf{0.943 (0.004)}\) & \(0.938 (0.004)\) & \(0.875 (0.016)\) \\
    400              & \(\mathbf{0.969 (0.002)}\) & \(0.967 (0.002)\) & \(0.938 (0.013)\) \\
    800              & \(\mathbf{0.978 (0.001)}\) & \(\mathbf{0.978 (0.001)}\) & \(0.966 (0.005)\) \\
    1600             & \(\mathbf{0.983 (0.001)}\) & \(0.982 (0.001)\) & \(0.980 (0.001)\) \\
    3200             & \(\mathbf{0.986 (0.001)}\) & \(0.985 (0.001)\) & \(0.984 (0.001)\) \\
    6400             & \(\mathbf{0.988 (0.001)}\) & \(\mathbf{0.988 (0.001)}\) & \(\mathbf{0.988 (0.001)}\) \\

    \addlinespace
    \multicolumn{4}{l}{\textbf{Top 5\%}}\\
    \(M=100\)        & \(\mathbf{0.859 (0.009)}\) & \(0.834 (0.008)\) & \(0.747 (0.014)\) \\
    200              & \(\mathbf{0.903 (0.007)}\) & \(0.891 (0.007)\) & \(0.808 (0.019)\) \\
    400              & \(\mathbf{0.944 (0.002)}\) & \(0.940 (0.002)\) & \(0.904 (0.013)\) \\
    800              & \(\mathbf{0.960 (0.002)}\) & \(0.959 (0.002)\) & \(0.940 (0.009)\) \\
    1600             & \(\mathbf{0.967 (0.002)}\) & \(\mathbf{0.967 (0.002)}\) & \(0.963 (0.002)\) \\
    3200             & \(\mathbf{0.973 (0.002)}\) & \(\mathbf{0.973 (0.002)}\) & \(0.971 (0.002)\) \\
    6400             & \(\mathbf{0.978 (0.001)}\) & \(0.977 (0.001)\) & \(0.977 (0.001)\) \\

    \bottomrule
  \end{tabular}
\end{table}

\begin{table}[ht]
  \centering
  \caption{Out‐of‐sample predictive performance of the full Relational Event Models from the generated independent Relational Event History (REH) data, quantified as the proportion of true events captured within the top 20\%, 10\%, and 5\% highest predicted‐probability events for training sample sizes \(M=100\)–6400. Predictive probabilities were calculated at each time \(t\) on an out-of-sample set of 1000 events generated beyond the training data, and coverage of the actual events by the highest‐probability subsets was evaluated. Three estimation approaches are compared: maximum likelihood estimation (MLE), approximate Bayesian regularization with Ridge (ABR Ridge) and Horseshoe (ABR HS) priors. Results are reported as \textit{mean (Monte Carlo standard error)} over the 100 iterations in the simulation. Boldface highlights the best performance per threshold and sample size.}
    \begin{tabular}{l c c c}
    \toprule
    Sample Size ($M$)
      & \itshape MLE
      & \itshape ABR Ridge
      & \itshape ABR HS \\
    \midrule
    \multicolumn{4}{l}{\textbf{Top 20\%}}\\
    $M=100$    & 0.946 (0.011) & \textbf{0.988 (0.001)} & 0.959 (0.011) \\
    200        & 0.986 (0.001) & \textbf{0.989 (0.001)} & 0.970 (0.005) \\
    400        & 0.990 (0.001) & \textbf{0.991 (0.000)} & 0.976 (0.010) \\
    800        & \textbf{0.993 (0.000)} & \textbf{0.993 (0.000)} & 0.989 (0.001) \\
    1600       & \textbf{0.994 (0.000)} & \textbf{0.994 (0.000)} & 0.993 (0.001) \\
    3200       & \textbf{0.994 (0.000)} & \textbf{0.994 (0.000)} & 0.993 (0.000) \\
    6400       & \textbf{0.995 (0.000)} & \textbf{0.995 (0.000)} & \textbf{0.995 (0.000)} \\

    \addlinespace
    \multicolumn{4}{l}{\textbf{Top 10\%}}\\
    $M=100$    & 0.919 (0.017) & \textbf{0.979 (0.001)} & 0.947 (0.011) \\
    200        & 0.978 (0.001) & \textbf{0.981 (0.001)} & 0.952 (0.011) \\
    400        & 0.982 (0.001) & \textbf{0.983 (0.001)} & 0.961 (0.011) \\
    800        & \textbf{0.986 (0.001)} & \textbf{0.986 (0.001)} & 0.979 (0.002) \\
    1600       & 0.987 (0.001) & \textbf{0.988 (0.001)} & 0.986 (0.001) \\
    3200       & \textbf{0.986 (0.001)} & \textbf{0.986 (0.001)} & \textbf{0.986 (0.001)} \\
    6400       & \textbf{0.988 (0.001)} & \textbf{0.988 (0.001)} & \textbf{0.988 (0.001)} \\

    \addlinespace
    \multicolumn{4}{l}{\textbf{Top 5\%}}\\
    $M=100$    & 0.886 (0.019) & \textbf{0.964 (0.002)} & 0.925 (0.012) \\
    200        & 0.964 (0.002) & \textbf{0.968 (0.002)} & 0.930 (0.013) \\
    400        & 0.968 (0.002) & \textbf{0.969 (0.002)} & 0.940 (0.014) \\
    800        & \textbf{0.974 (0.001)} & \textbf{0.974 (0.001)} & 0.964 (0.003) \\
    1600       & \textbf{0.976 (0.002)} & \textbf{0.976 (0.002)} & 0.973 (0.002) \\
    3200       & \textbf{0.975 (0.002)} & \textbf{0.975 (0.002)} & 0.973 (0.002) \\
    6400       & \textbf{0.978 (0.001)} & \textbf{0.978 (0.002)} & \textbf{0.978 (0.001)} \\

    \bottomrule
  \end{tabular}
\end{table}

\begin{landscape}
\begin{table}[ht]
  \centering
  \caption{In‐sample predictive performance of sparse Relational Event Models from the generated independent Relational Event History (REH) data including only predictors selected by the assessed criteria. Performance is defined as the proportion of true events included within the top 20\%, 10\%, and 5\% highest-predicted-probability events computed on all training events for sample sizes \(M=100\)–6400. Results are reported as \textit{mean (Monte Carlo standard error)} across 100 simulation iterations. Nine strategies are compared: (i) maximum likelihood estimation (MLE) via univariate selection at \(\alpha=0.05\); and (ii) approximate Bayesian regularization (ABR) with Horseshoe (HS) and Ridge priors using posterior 95\% highest-density intervals (HDI) and 0.1 thresholding on posterior mode, median, and mean. Boldface highlights the best performance per threshold and sample size.}
  \scriptsize
  \setlength\tabcolsep{2pt}
  \resizebox{1.6\textwidth}{!}{%
        \begin{tabular}{l r *{4}{c} *{4}{c} c}
  \toprule
  & & \multicolumn{4}{c}{\itshape ABR HS}
    & \multicolumn{4}{c}{\itshape ABR Ridge}
    & \itshape MLE \\
  \cmidrule(lr){3-6}\cmidrule(lr){7-10}\cmidrule(lr){11-11}
  Sample Size ($M$)
    & & $95\%$ HDI
    & $|\mathrm{Mean}|\ge0.1$ & $|\mathrm{Median}|\ge0.1$ & $|\mathrm{Mode}|\ge0.1$
    & $95\%$ HDI
    & $|\mathrm{Mean}|\ge0.1$ & $|\mathrm{Median}|\ge0.1$ & $|\mathrm{Mode}|\ge0.1$
    & $\alpha=0.05$ \\
  \midrule
  \multicolumn{11}{l}{\textbf{Top 20\%}}\\
  $M=100$    &   & 0.782 (0.027) & 0.897 (0.013) & 0.897 (0.013) & 0.886 (0.014)
              & \textbf{0.945 (0.003)} & \textbf{0.945 (0.003)} & \textbf{0.945 (0.003)} & \textbf{0.945 (0.003)}
              & 0.894 (0.013) \\
  200        &   & 0.858 (0.020) & 0.918 (0.013) & 0.918 (0.013) & 0.906 (0.015)
              & \textbf{0.966 (0.002)} & \textbf{0.966 (0.002)} & \textbf{0.966 (0.002)} & \textbf{0.966 (0.002)}
              & 0.917 (0.016) \\
  400        &   & 0.930 (0.018) & 0.957 (0.013) & 0.957 (0.013) & 0.956 (0.013)
              & \textbf{0.982 (0.001)} & \textbf{0.982 (0.001)} & \textbf{0.982 (0.001)} & \textbf{0.982 (0.001)}
              & 0.965 (0.010) \\
  800        &   & 0.978 (0.002) & 0.984 (0.001) & 0.984 (0.001) & 0.983 (0.001)
              & \textbf{0.988 (0.001)} & \textbf{0.988 (0.001)} & \textbf{0.988 (0.001)} & \textbf{0.988 (0.001)}
              & 0.975 (0.008) \\
  1600       &   & 0.988 (0.001) & 0.990 (0.001) & 0.990 (0.001) & 0.990 (0.001)
              & \textbf{0.991 (0.000)} & \textbf{0.991 (0.000)} & \textbf{0.991 (0.000)} & \textbf{0.991 (0.000)}
              & 0.986 (0.003) \\
  3200       &   & 0.992 (0.001) & \textbf{0.993 (0.000)} & \textbf{0.993 (0.000)} & \textbf{0.993 (0.000)}
              & \textbf{0.993 (0.000)} & \textbf{0.993 (0.000)} & \textbf{0.993 (0.000)} & \textbf{0.993 (0.000)}
              & 0.992 (0.001) \\
  6400       &   & 0.994 (0.000) & 0.994 (0.000) & 0.994 (0.000) & 0.994 (0.000)
              & \textbf{0.995 (0.000)} & \textbf{0.995 (0.000)} & \textbf{0.995 (0.000)} & \textbf{0.995 (0.000)}
              & 0.989 (0.005) \\
  \addlinespace
  \multicolumn{11}{l}{\textbf{Top 10\%}}\\
  $M=100$    &   & 0.696 (0.026) & 0.837 (0.013) & 0.837 (0.013) & 0.811 (0.014)
              & \textbf{0.902 (0.006)} & \textbf{0.902 (0.006)} & \textbf{0.902 (0.006)} & \textbf{0.902 (0.006)}
              & 0.806 (0.017) \\
  200        &   & 0.813 (0.021) & 0.874 (0.016) & 0.874 (0.016) & 0.862 (0.018)
              & \textbf{0.938 (0.004)} & \textbf{0.938 (0.004)} & \textbf{0.938 (0.004)} & \textbf{0.938 (0.004)}
              & 0.877 (0.017) \\
  400        &   & 0.904 (0.018) & 0.938 (0.013) & 0.938 (0.013) & 0.936 (0.013)
              & \textbf{0.967 (0.002)} & \textbf{0.967 (0.002)} & \textbf{0.967 (0.002)} & \textbf{0.967 (0.002)}
              & 0.942 (0.012) \\
  800        &   & 0.955 (0.008) & 0.965 (0.006) & 0.965 (0.006) & 0.964 (0.006)
              & \textbf{0.978 (0.001)} & \textbf{0.978 (0.001)} & \textbf{0.978 (0.001)} & \textbf{0.978 (0.001)}
              & 0.958 (0.010) \\
  1600       &   & 0.976 (0.002) & 0.980 (0.001) & 0.980 (0.001) & 0.980 (0.001)
              & \textbf{0.982 (0.001)} & \textbf{0.982 (0.001)} & \textbf{0.982 (0.001)} & \textbf{0.982 (0.001)}
              & 0.969 (0.007) \\
  3200       &   & 0.982 (0.001) & 0.984 (0.001) & 0.984 (0.001) & 0.984 (0.001)
              & \textbf{0.985 (0.001)} & \textbf{0.985 (0.001)} & \textbf{0.985 (0.001)} & \textbf{0.985 (0.001)}
              & 0.982 (0.001) \\
  6400       &   & 0.986 (0.001) & \textbf{0.988 (0.001)} & \textbf{0.988 (0.001)} & \textbf{0.988 (0.001)}
              & \textbf{0.988 (0.001)} & \textbf{0.988 (0.001)} & \textbf{0.988 (0.001)} & \textbf{0.988 (0.001)}
              & 0.981 (0.006) \\
  \addlinespace
  \multicolumn{11}{l}{\textbf{Top 5\%}}\\
  $M=100$    &   & 0.591 (0.025) & 0.749 (0.014) & 0.749 (0.014) & 0.705 (0.018)
              & \textbf{0.834 (0.008)} & \textbf{0.834 (0.008)} & \textbf{0.834 (0.008)} & \textbf{0.834 (0.008)}
              & 0.693 (0.020) \\
  200        &   & 0.763 (0.020) & 0.807 (0.019) & 0.807 (0.019) & 0.796 (0.020)
              & \textbf{0.891 (0.007)} & \textbf{0.891 (0.007)} & \textbf{0.891 (0.007)} & \textbf{0.891 (0.007)}
              & 0.805 (0.020) \\
  400        &   & 0.869 (0.018) & 0.904 (0.013) & 0.904 (0.013) & 0.903 (0.013)
              & \textbf{0.940 (0.002)} & \textbf{0.940 (0.002)} & \textbf{0.940 (0.002)} & \textbf{0.940 (0.002)}
              & 0.908 (0.012) \\
  800        &   & 0.929 (0.009) & 0.940 (0.009) & 0.940 (0.009) & 0.939 (0.009)
              & \textbf{0.959 (0.002)} & \textbf{0.959 (0.002)} & \textbf{0.959 (0.002)} & \textbf{0.959 (0.002)}
              & 0.929 (0.012) \\
  1600       &   & 0.957 (0.003) & 0.963 (0.002) & 0.963 (0.002) & 0.962 (0.002)
              & \textbf{0.967 (0.002)} & \textbf{0.967 (0.002)} & \textbf{0.967 (0.002)} & \textbf{0.967 (0.002)}
              & 0.943 (0.011) \\
  3200       &   & 0.967 (0.002) & 0.970 (0.002) & 0.970 (0.002) & 0.970 (0.002)
              & \textbf{0.973 (0.002)} & \textbf{0.973 (0.002)} & \textbf{0.973 (0.002)} & \textbf{0.973 (0.002)}
              & 0.967 (0.002) \\
  6400       &   & 0.974 (0.002) & 0.976 (0.001) & 0.976 (0.001) & 0.976 (0.001)
              & \textbf{0.977 (0.001)} & \textbf{0.977 (0.001)} & \textbf{0.977 (0.001)} & \textbf{0.977 (0.001)}
              & 0.968 (0.007) \\
  \bottomrule
    \end{tabular}%
  }
  \label{tab:pp_is_independent}
\end{table}
\end{landscape}

\begin{landscape}
\begin{table}[ht]
  \centering
  \caption{Out‐of‐sample predictive performance of sparse Relational Event Models from the generated independent Relational Event History (REH) data including only predictors selected by the assessed criteria. Performance is defined as the proportion of true events included within the top 20\%, 10\%, and 5\% highest‐predicted‐probability events computed on an out-of-sample set of 1000 events for training sample sizes \(M=100\)–6400. Results are reported as \textit{mean (Monte Carlo standard error)} across 100 simulation iterations. Nine strategies are compared: (i) maximum likelihood estimation (MLE) via univariate selection at \(\alpha=0.05\); and (ii) approximate Bayesian regularization (ABR) with Horseshoe (abbreviated as HS) and Ridge priors using posterior 95\% highest-density intervals (HDI) and 0.1 thresholding on posterior mode, median, and mean. Boldface highlights the best performance per threshold and sample size.}
  \scriptsize
  \setlength\tabcolsep{2pt}
  \resizebox{1.6\textwidth}{!}{%
        \begin{tabular}{l r *{4}{c} *{4}{c} c}
  \toprule
  & & \multicolumn{4}{c}{\itshape ABR HS}
    & \multicolumn{4}{c}{\itshape ABR Ridge}
    & \itshape MLE \\
  \cmidrule(lr){3-6}\cmidrule(lr){7-10}\cmidrule(lr){11-11}
  Sample Size ($M$)
    & & $95\%$ HDI
    & $|\mathrm{Mean}|\ge0.1$ & $|\mathrm{Median}|\ge0.1$ & $|\mathrm{Mode}|\ge0.1$
    & $95\%$ HDI
    & $|\mathrm{Mean}|\ge0.1$ & $|\mathrm{Median}|\ge0.1$ & $|\mathrm{Mode}|\ge0.1$
    & $\alpha=0.05$ \\
  \midrule
  \multicolumn{11}{l}{\textbf{Top 20\%}}\\
  $M=100$    &   & 0.870 (0.026) & 0.958 (0.011) & 0.958 (0.011) & 0.952 (0.012)
              & \textbf{0.988 (0.001)} & \textbf{0.988 (0.001)} & \textbf{0.988 (0.001)} & \textbf{0.988 (0.001)}
              & 0.932 (0.015) \\
  200        &   & 0.929 (0.014) & 0.970 (0.005) & 0.970 (0.005) & 0.955 (0.012)
              & \textbf{0.989 (0.001)} & \textbf{0.989 (0.001)} & \textbf{0.989 (0.001)} & \textbf{0.989 (0.001)}
              & 0.944 (0.017) \\
  400        &   & 0.958 (0.014) & 0.976 (0.010) & 0.976 (0.010) & 0.975 (0.010)
              & \textbf{0.991 (0.000)} & \textbf{0.991 (0.000)} & \textbf{0.991 (0.000)} & \textbf{0.991 (0.000)}
              & 0.977 (0.010) \\
  800        &   & 0.984 (0.002) & 0.989 (0.001) & 0.989 (0.001) & 0.989 (0.001)
              & \textbf{0.993 (0.000)} & \textbf{0.993 (0.000)} & \textbf{0.993 (0.000)} & \textbf{0.993 (0.000)}
              & 0.986 (0.004) \\
  1600       &   & 0.991 (0.001) & 0.993 (0.001) & 0.993 (0.001) & 0.993 (0.001)
              & \textbf{0.994 (0.000)} & \textbf{0.994 (0.000)} & \textbf{0.994 (0.000)} & \textbf{0.994 (0.000)}
              & 0.990 (0.002) \\
  3200       &   & 0.992 (0.001) & 0.993 (0.000) & 0.993 (0.000) & 0.993 (0.000)
              & \textbf{0.994 (0.000)} & \textbf{0.994 (0.000)} & \textbf{0.994 (0.000)} & \textbf{0.994 (0.000)}
              & 0.993 (0.000) \\
  6400       &   & 0.994 (0.000) & \textbf{0.995 (0.000)} & \textbf{0.995 (0.000)} & \textbf{0.995 (0.000)}
              & \textbf{0.995 (0.000)} & \textbf{0.995 (0.000)} & \textbf{0.995 (0.000)} & \textbf{0.995 (0.000)}
              & 0.989 (0.005) \\
  \addlinespace
  \multicolumn{11}{l}{\textbf{Top 10\%}}\\
  $M=100$    &   & 0.855 (0.026) & 0.947 (0.012) & 0.947 (0.012) & 0.940 (0.012)
              & \textbf{0.979 (0.001)} & \textbf{0.979 (0.001)} & \textbf{0.979 (0.001)} & \textbf{0.979 (0.001)}
              & 0.898 (0.018) \\
  200        &   & 0.917 (0.014) & 0.952 (0.011) & 0.952 (0.011) & 0.944 (0.012)
              & \textbf{0.981 (0.001)} & \textbf{0.981 (0.001)} & \textbf{0.981 (0.001)} & \textbf{0.981 (0.001)}
              & 0.928 (0.018) \\
  400        &   & 0.935 (0.016) & 0.959 (0.012) & 0.959 (0.012) & 0.960 (0.012)
              & \textbf{0.983 (0.001)} & \textbf{0.983 (0.001)} & \textbf{0.983 (0.001)} & \textbf{0.983 (0.001)}
              & 0.965 (0.010) \\
  800        &   & 0.972 (0.003) & 0.979 (0.002) & 0.979 (0.002) & 0.979 (0.002)
              & \textbf{0.986 (0.001)} & \textbf{0.986 (0.001)} & \textbf{0.986 (0.001)} & \textbf{0.986 (0.001)}
              & 0.967 (0.010) \\
  1600       &   & 0.983 (0.002) & 0.986 (0.001) & 0.986 (0.001) & 0.986 (0.001)
              & \textbf{0.988 (0.001)} & \textbf{0.988 (0.001)} & \textbf{0.988 (0.001)} & \textbf{0.988 (0.001)}
              & 0.975 (0.008) \\
  3200       &   & 0.984 (0.001) & \textbf{0.986 (0.001)} & \textbf{0.986 (0.001)} & \textbf{0.986 (0.001)}
              & \textbf{0.986 (0.001)} & \textbf{0.986 (0.001)} & \textbf{0.986 (0.001)} & \textbf{0.986 (0.001)}
              & 0.984 (0.001) \\
  6400       &   & 0.987 (0.001) & \textbf{0.988 (0.001)} & \textbf{0.988 (0.001)} & \textbf{0.988 (0.001)}
              & \textbf{0.988 (0.001)} & \textbf{0.988 (0.001)} & \textbf{0.988 (0.001)} & \textbf{0.988 (0.001)}
              & 0.982 (0.005) \\
  \addlinespace
  \multicolumn{11}{l}{\textbf{Top 5\%}}\\
  $M=100$    &   & 0.803 (0.026) & 0.925 (0.012) & 0.925 (0.012) & 0.901 (0.014)
              & \textbf{0.964 (0.002)} & \textbf{0.964 (0.002)} & \textbf{0.964 (0.002)} & \textbf{0.964 (0.002)}
              & 0.824 (0.027) \\
  200        &   & 0.891 (0.014) & 0.930 (0.013) & 0.930 (0.013) & 0.921 (0.014)
              & \textbf{0.968 (0.002)} & \textbf{0.968 (0.002)} & \textbf{0.968 (0.002)} & \textbf{0.968 (0.002)}
              & 0.908 (0.019) \\
  400        &   & 0.915 (0.017) & 0.940 (0.014) & 0.940 (0.014) & 0.939 (0.014)
              & \textbf{0.969 (0.002)} & \textbf{0.969 (0.002)} & \textbf{0.969 (0.002)} & \textbf{0.969 (0.002)}
              & 0.949 (0.010) \\
  800        &   & 0.949 (0.009) & 0.963 (0.004) & 0.963 (0.004) & 0.963 (0.004)
              & \textbf{0.974 (0.001)} & \textbf{0.974 (0.001)} & \textbf{0.974 (0.001)} & \textbf{0.974 (0.001)}
              & 0.948 (0.012) \\
  1600       &   & 0.968 (0.002) & 0.973 (0.002) & 0.973 (0.002) & 0.973 (0.002)
              & \textbf{0.976 (0.002)} & \textbf{0.976 (0.002)} & \textbf{0.976 (0.002)} & \textbf{0.976 (0.002)}
              & 0.953 (0.011) \\
  3200       &   & 0.970 (0.002) & 0.973 (0.002) & 0.973 (0.002) & 0.973 (0.002)
              & \textbf{0.975 (0.002)} & \textbf{0.975 (0.002)} & \textbf{0.975 (0.002)} & \textbf{0.975 (0.002)}
              & 0.970 (0.002) \\
  6400       &   & 0.975 (0.002) & 0.977 (0.002) & 0.977 (0.002) & 0.977 (0.002)
              & \textbf{0.978 (0.002)} & \textbf{0.978 (0.002)} & \textbf{0.978 (0.002)} & \textbf{0.978 (0.002)}
              & 0.970 (0.006) \\
  \bottomrule
    \end{tabular}%
  }
  \label{tab:pp_oos_independent}
\end{table}
\end{landscape}

\end{document}